\documentclass[a4paper,11pt]{article}
\usepackage{jheppub} 
\usepackage{lineno}
\usepackage{physics}
\usepackage{amsmath,amssymb}
\usepackage{xcolor}

\title{\boldmath On the Structure Constants of ${\cal N}=2$ Liouville Theory}

\author[a]{
Jingxin Tu}
\affiliation[a]{University of Amsterdam,\\Amsterdam, the Netherlands}

\emailAdd{j.tu2@uva.nl}

\abstract{We show the structure constants of $\mathcal{N}=2$ spacelike Liouville theory. The absence of self-duality under $b\leftrightarrow b^{-1}$, where $b$ is the Liouville coupling, prevents a straightforward application of the standard bootstrap methods used in $\mathcal{N}=0$ and $\mathcal{N}=1$ Liouville theory. Motivated by the fermionic FZZ duality with the $SL(2,\mathbb{R})/U(1)$ supercoset theory, we establish an algebraic correspondence between the two theories, including a matching of their null vectors. We then derive the three-point structure constants in superspace. The neutral sector structure constants are associated with the spectral-flow-preserving structure constant of the $SL(2,\mathbb{R})$ WZW model, while the two charged sectors correspond to its spectral-flow-violating structure constants. Finally, we give the correct relative normalization between the neutral and charged sectors and test crossing symmetry numerically.}

\begin{document}
\maketitle
\flushbottom

\section{Introduction}
\label{sec:intro}

Liouville theory, as an irrational and non-compact conformal field theory with exact solvable results, has played an essential role in theoretical physics. It is a central example of an interacting conformal field theory, with important applications to two-dimensional quantum gravity \cite{David:1988hj,Distler:1988jt,Knizhnik:1988ak} and string theory \cite{Polyakov:1981rd,Ginsparg:1993is}, as well as close connections to matrix models \cite{Brezin:1990rb,Douglas:1989ve,Gross:1989vs,DiFrancesco:1993cyw}. Supersymmetric extensions of Liouville theory extend the framework to fermionic strings \cite{Polyakov:1981re} and provide building blocks for the noncritical superstring backgrounds \cite{Murthy:2003es,Kutasov:1990ua}.

Of our particular interest is the $\mathcal{N}=2$ Liouville theory, whose equivalence to the supersymmetric coset $SL(2,\mathbb{R})/U(1)$ theory provides a dual worldsheet description of strings propagating on the two-dimensional fermionic black hole, with its characteristic cigar geometry. This correspondence, known as the fermionic FZZ duality, was derived as a mirror symmetry by Hori and Kapustin in \cite{Hori:2001ax}. Although the duality was established at the level of the action, there remains a lack of understanding for the $\mathcal{N}=2$ Liouville theory itself. Such an understanding is essential for exploring a potential $\mathcal{N}=2$ matrix model, complementing the bosonic construction in \cite{Kazakov:2000pm}. The first task is to determine the structure constants, which encode the dynamical information of the theory, motivating a detailed study of correlation functions on both sides of the duality.

The three-point structure constant of bosonic $\mathcal{N}=0$ Liouville theory is given by the \emph{DOZZ} formula, proposed in \cite{Dorn:1994xn,Zamolodchikov_1996} and subsequently derived through a bootstrap approach by Teschner in \cite{Teschner_1995}. A rigorous probabilistic construction of the Liouville path integral was given in \cite{David:2014aha}. The bootstrap approach uses the null vector decoupling conditions to constrain four-point functions containing one insertion of degenerate state, and determine the corresponding conformal blocks. Crossing symmetry then yields a shift relation for the structure constant. The self-duality $b\rightarrow b^{-1}$ supplies a second, dual shift relation, and the two together with suitable analyticity assumptions determine the structure constant up to normalization. The same approach applies to $\mathcal{N}=1$ Liouville theory, which is also self-dual \cite{Poghossian:1996agj,Rashkov:1996np,Muhlmann:2025ngz}. In $\mathcal{N}=2$ Liouville theory this symmetry is absent, so the second shift relation does not follow from the first and has to be obtained from an independent degenerate field.

Alternative approaches have provided substantial information about $\mathcal{N}=2$ Liouville theory on the other hand. As reviewed in \cite{NAKAYAMA_2004}, its fusion rules and correlation functions can be studied using the free-field path integrals. Hosomichi computed the three-point structure constants with the Coulomb-gas method and found agreement with the $SL(2,\mathbb{R})/U(1)$ coset \cite{Hosomichi:2004ph}. On the coset side, an algebraic bootstrap approach has been used to determine the structure constant of the $H_3^+$ WZNW model \cite{Teschner:1997ft}; there the two shift relations come from two independent degenerate fields rather than from a self-duality. This suggests that an algebraic approach should also be possible for $\mathcal{N}=2$ Liouville.

Motivated by the fermionic FZZ duality, we compare the representation spectrum and corresponding characters in the two theories to show their agreement. In particular, we examine the null vector conditions of $\mathcal{N}=2$ Liouville theory derived in \cite{Ahn_2004_ZZ} and verify explicitly, at levels 1 and 2, that they map to those of the $SL(2,\mathbb{R})$ WZW model. This correspondence establishes the algebraic constraints for the structure constants. We then analyze three-point functions of $\mathcal{N}=2$ Liouville in superspace and identify the neutral and charged structure constants with the spectral-flow-preserving and spectral-flow-violating WZW structure constants, respectively. The neutral and charged sectors, together with their associated structure constants, will be defined precisely below. Finally, with the constraints from superconformal Ward identities, we can give the consistent results of the neutral sector structure constants. With numerical tests of crossing symmetry, we confirm the relative normalization between the neutral and charged sectors, finally determining the three-point structure constants in superspace. Similar results were recently obtained in \cite{Muhlmann:2026pdc} from a different perspective: there the structure constants of the superprimary components, rather than of the superfields, are derived from the supercoset and tested semiclassically against the Liouville path integral to one-loop order.

The rest of the paper is organized as follows. In Section~\ref{sec:N=2 Liouville}, we introduce $\mathcal{N}=2$ super-Liouville theory, and review the $\mathcal{N}=2$ superconformal algebra and its representation spectrum. In the end of this section, we will define the vertex operators and set up the problem we aim to address. In Section~\ref{sec:Relation to SL(2,R) WZW}, we review the coset construction of $\mathcal{N}=2$ superconformal algebra and show the correspondence between the representation spectra, obtaining the parameter dictionary between the two theories. In Section~\ref{sec:Identification of duality}, we match the level 1 null vectors of $\mathcal{N}=2$ Liouville theory with their $SL(2,\mathbb{R})$ counterparts, and derive the associated BPZ-type equations. In Section~\ref{sec:correlators in superspace}, we present our structure constants satisfying constraints from superconformal Ward identities in superspace, together with the numerical checks for crossing symmetry. Finally, in Section~\ref{sec:discussion}, we summarize our results and discuss potential applications and directions for the future work.

\section{\texorpdfstring{$\mathcal{N}=2$}{N=2} super-Liouville Theory}
\label{sec:N=2 Liouville}

The $\mathcal N=2$ super-Liouville theory is the supersymmetric extension of ordinary Liouville theory which preserves two supercharges and carries a natural $U(1)$ symmetry. It is best understood as a non-compact superconformal model in which the exponential interaction and the background charge deform the free theory into a genuinely interacting system. The $\mathcal{N}=2$ superconformal algebra constrains the form of its correlation functions, while as a non-compact CFT, determining the structure constants and dynamic information requires a detailed analysis of the spectrum.

In this section, we will give the definition of ${\cal N}=2$ Liouville theory starting from the free field realization. Then we will review the $\mathcal{N}=2$ superconformal algebra and its representation spectra. After giving the definition for the vertex operators in $\mathcal{N}=2$ Liouville theory, we will bring up the setting of the problem.

\subsection{Definition of \texorpdfstring{$\mathcal{N}=2$}{N=2} Liouville Theory}
\label{subsec:def N=2 Liouville}

A convenient way to formulate the theory is in terms of a superfield description. Let $\Phi$ denote a chiral superfield, and let $\bar{\Phi}$ be its anti-chiral conjugate. The $\mathcal N=2$ Liouville action can be written in the form
\begin{equation}
\begin{aligned}
    S_{\mathrm{L}}&=S_0+S_c+S_{nc},\\
    S_0&=\frac{1}{8\pi}\int d^2z\, d\theta^+ d\bar\theta^+d\theta^- d\bar \theta^- \Phi\bar{\Phi}\\
    S_c&=\frac{\mu}{2\pi} \int d^2z\, d\theta^- d\bar\theta^-e^{\frac{b}{\sqrt{2}}\Phi}+ \frac{\bar\mu}{2\pi} \int d^2z\, d\theta^+ d\bar \theta^+ e^{\frac{b}{\sqrt{2}}\bar{\Phi}}\\
    S_{nc}&=\frac{\tilde\mu}{4\pi}\int d^2zd\theta^+ d\bar\theta^+d\theta^- d\bar \theta^-  e^{\frac{1}{\sqrt{2}b}(\Phi+\bar\Phi)}
\end{aligned}
\end{equation}
where $b$ is the Liouville coupling parameter and $\mu,\bar\mu,\tilde\mu$ are the cosmological constants. The bosonic coordinates $(z,\bar z)$ span the flat Euclidean worldsheet. The $(2,2)$ superspace has four anti-commuting coordinates $\theta^{\pm}$ and $\bar\theta^{\pm}$ which are related by complex conjugation: $(\theta^{\pm})^{\dagger}=\bar\theta^{\mp}$. $S_0$ denotes the kinetic term and $S_c,S_{nc}$ denote chiral and non-chiral interaction terms respectively without ruining the $\mathcal{N}=2$ supersymmetry.

Using the $\theta$-expansions for $\Phi$ and $\bar\Phi$:
\begin{equation}
\begin{aligned}
   \Phi&=\phi+iY+2i\theta^-\psi^-+2i\bar{\theta}^-\bar{\psi}^-+2\theta^-\bar{\theta}^- F+\cdots\\
   \bar\Phi&=\phi-iY+2i\theta^{+}\psi^++2i\bar{\theta}^+\bar{\psi}^++2\theta^+\bar{\theta}^+F^*+\cdots,
\end{aligned}
\end{equation}
the action can be written in terms of component fields:
\begin{equation}
    S_0=\frac{1}{2\pi}\int d^2z(\partial\phi\bar{\partial}\phi+\partial Y\bar{\partial}Y
    +\psi^+\bar{\partial}\psi^-+\psi^-\bar{\partial}\psi^++\bar{\psi}^+\partial\bar{\psi}^-+\bar{\psi}^-\partial\bar{\psi}^+)
\end{equation}
The chiral interaction part is given by:
\begin{equation}
    S_c\propto\frac{b^2}{\pi}\int d^2z\left(\mu\psi^-\bar{\psi}^- e^{\frac{b}{\sqrt{2}}(\phi+iY)}+\bar{\mu}\psi^+\bar{\psi}^+e^{\frac{b}{\sqrt{2}}(\phi-iY)}\right)
\end{equation}
and the non-chiral interaction is given by
\begin{equation}
    S_{nc}\propto \frac{\tilde{\mu}}{2\pi b^2}\int d^2z(\partial\phi+i\partial Y-\frac{\sqrt{2}}{b}\psi^+\psi^-)(\bar{\partial}\phi+i\bar{\partial}Y-\frac{\sqrt{2}}{b}\bar{\psi}^+\bar{\psi}^-)e^{\frac{\sqrt{2}}{b}\phi}
\end{equation}
In components, this theory contains a complex scalar field $\phi+iY$ and two fermions $\psi^{\pm}$, and the interaction is generated by exponential screening operators\footnote{We drop the auxiliary fields $F$, $F^*$; the contact terms they generate, as well as those in the expansion of the non-chiral interaction, are neglected.}. These fields satisfy the canonical OPEs:
\begin{equation}
    \phi(z)\phi(w)\sim Y(z)Y(w)\sim -\ln|z-w|^2,\, \quad \psi^+(z)\psi^-(w)\sim \frac{1}{z-w}
\end{equation}
With the background charge $Q_L=\frac{1}{b}$ for $\phi$, the central charge is given by $c=3+6Q_L^2=3+\frac{6}{b^2}$. One of the most significant features of the $\mathcal{N}=2$ theory is the nonrenormalization theorem. In this case, the background charge $Q_L$ is not renormalized quantum mechanically \cite{Distler:1989nt}. We can easily observe that there is no $b\leftrightarrow b^{-1}$ symmetry as in $\mathcal{N}=0$ and $\mathcal{N}=1$ Liouville theory. Because of this property, the central charge of $\mathcal{N}=2$ Liouville theory is monotonic and makes sense for arbitrary real $b$. Consequently, the central charge can become small enough to introduce other transverse dimensions, so that we can embed the $\mathcal{N}=2$ Liouville theory into the superstring theory propagating in the nontrivial background whose dimension is larger than two. We will comment on this further in the discussion section.

The exponential interaction preserves the superconformal structure while introducing the nontrivial Liouville potential. Therefore, the flat target-space free theory is deformed into a non-compact interacting model whose spectrum is organized by the representation theory of the $\mathcal N=2$ superconformal algebra, generated by the stress tensor $T(z)$, the fermionic generators $G^{\pm}(z)$, and the $U(1)$ current $J(z)$:
\begin{equation}
\begin{split}
   T(z)&=-\frac{1}{2}(\partial\phi\partial\phi+\partial Y\partial Y)-\frac{1}{2}(\psi^+\partial\psi^-+\psi^-\partial\psi^+)+\frac{1}{\sqrt{2}b}\partial^2\phi\\
   G^+(z)&=i\psi^+(\partial\phi+i\partial Y)-i\frac{\sqrt{2}}{b}\partial\psi^+\\
   G^-(z)&=i\psi^-(\partial\phi-i\partial Y)-i\frac{\sqrt{2}}{b}\partial\psi^-\\
   J(z)&=\psi^+\psi^-+i\frac{\sqrt{2}}{b}\partial Y
\end{split}
\end{equation}

Their operator product expansions encode the superconformal structure of the model. The $\mathcal{N}=2$ algebras are given by different modes of the generators, corresponding to different ways of choosing boundary conditions. The NS algebra has integer modes for $J_m$ and half-integer modes for $G_{r}^{\pm}$, the algebra in R sector has integer modes for both $J_m$ and $G_{r}^{\pm}$. If we define their modes as follows:
\begin{equation}
    T(z)=\sum_{m\in\mathbb{Z}} L_mz^{-m-2}\, \quad G^{\pm}(z)=\sum_{r\in\mathbb{Z}\pm\alpha+1/2} G_{r}^{\pm}z^{-r-3/2}\, \quad J(z)=\sum_{m\in\mathbb{Z}} J_mz^{-m-1}
\end{equation}
where NS and R algebras are labeled by $\alpha=0$ and $\alpha=\frac{1}{2}$, then they obey the (anti-)commutation relations:
\begin{equation}
\begin{aligned}
    \relax [L_m,L_n]&=(m-n)L_{m+n}+\frac{c}{12}(m^3-m)\delta_{m,-n}\\
    [L_m,G_r^{\pm}]&=(\frac{m}{2}-r)G_{m+r}^{\pm} \quad [L_m,J_n]=-nJ_{m+n}\\
    \{G_r^+,G_s^-\}&=2L_{r+s}+(r-s)J_{r+s}+\frac{c}{3}(r^2-\frac{1}{4})\delta_{r,-s}\\
    \{G_r^{\pm},G_{s}^{\pm}\}&=0 \quad [J_n,G_r^{\pm}]=\pm G_{r+n}^{\pm} \quad [J_m,J_n]=\frac{c}{3}m\delta_{m,-n}
    \label{eq:N=2 algebra}
\end{aligned}
\end{equation}
The first two lines of~\eqref{eq:N=2 algebra} state the usual Virasoro algebra and last two relations specify a $U(1)$ current algebra and that $G^{\pm}(z)$ has charge $\pm1$. 

Theories with $\mathcal{N}=2$ supersymmetry have distinctive new features which are not present for those with $\mathcal{N}=0$ or $\mathcal{N}=1$ supersymmetry \cite{blumenhagen2009introduction}. These are the chiral ring and the spectral flow, which will be briefly explained in the following paragraphs.

\paragraph{Chiral Ring}

A central feature of $\mathcal N=2$ superconformal theory is the existence of a chiral ring. A state $\ket{h,Q}$ labeled by conformal dimension $h$ and $U(1)$ charge $Q$ in the NS sector of the Hilbert space of an $\mathcal{N}=2$ SCFT is called chiral (anti-chiral) state if it satisfies $G^+_{-1/2}\ket{h,Q}=0$ ($G^-_{-1/2}\ket{h,Q}=0$)\footnote{Here we focus on the holomorphic sector.}. We can also easily derive that the chiral primary state $\ket{h,Q}$ satisfies $h_{\mathrm{chiral}}=\frac{Q}{2}$ with the anti-commutation relations of superconformal algebra. Similarly, for anti-chiral primary state we have $h_{\mathrm{anti-chiral}}=-\frac{Q}{2}$. In the compact case, chiral primaries furnish a closed algebra under operator product expansion, since the fusion of two chiral primary states gives another chiral primary state.

In $\mathcal{N}=2$ Liouville theory, the chiral and anti-chiral primary states are both BPS states, and they generate a shorter representation. More explicitly, starting from a primary state $\ket{h,Q}$, in general we have four components contributing to one $\mathcal{N}=2$ super primary field:
\begin{equation*}
   \ket{h,Q}\,\quad G_{-1/2}^+\ket{h,Q}\, \quad G_{-1/2}^-\ket{h,Q}\, \quad G_{-1/2}^+G_{-1/2}^-\ket{h,Q}
\end{equation*}
In case of a chiral field, we have $G^+_{-1/2}\ket{h,Q}=0$ so that the super primary consists of only two components $\ket{h,Q}$ and $G_{-1/2}^-\ket{h,Q}$. Such supermultiplets are shorter since the zero-norm states decouple from the physical representation and such states are also null states at level $1/2$.

\paragraph{Spectral Flow}

Now we turn to the second characteristic feature of $\mathcal{N}=2$ superconformal field theories. In last paragraph, we focus on the NS algebras, but R algebras are related to them by spectral flow:
\begin{equation}
\begin{aligned}
    U_{\alpha}L_mU_{\alpha}^{\dagger}&=L_m+\alpha J_m + \frac{\alpha^2}{6}c\delta_{m,0}\\
    U_{\alpha}J_mU_{\alpha}^{\dagger}&=J_m+\frac{c}{3}\alpha\delta_{m,0}\\
    U_{\alpha}G_r^{\pm}U_{\alpha}^{\dagger}&=G_{r\pm\alpha}^{\pm}
\end{aligned}
\end{equation}
where $U_{\alpha}$ is a unitary operator and $\alpha$ is a continuous parameter. The spectral flowed operators still satisfy the algebra~\eqref{eq:N=2 algebra}. In particular, for $\alpha\in\mathbb{Z}+\frac{1}{2}$, the flow brings NS(R) algebras to R(NS) algebras, while the flows labeled by $\alpha\in\mathbb{Z}$ are automorphisms of these algebras.

\subsection{Spectrum}
\label{subsec: spectrum}

Having introduced the definition of $\mathcal{N}=2$ superconformal algebra, now we show the spectrum for this algebra. We will focus on the representations for NS algebras, since those of R algebras can be obtained by half-integer spectral flow on NS algebras as we mentioned above. 

We consider highest weight representations of the $\mathcal{N}=2$ algebras, which are generated from the \textit{highest weight vector (hwv)} labeled by conformal dimension $h$ and $U(1)$ charge $Q$. The largest possible representation generated from hwv is a quotient of the Verma module, whose basis is a maximal independent set of descendant states generated from hwv with ordered monomials of generators. Different level, relative charge\footnote{A descendant state generated from hwv can be characterized by level $n$ and relative charge $m$ with $L_0$ eigenvalue $h+n$ and $J_0$ eigenvalue $Q+m$.} and fermion parity decompose the Verma module into orthogonal subspaces. The unitarity conditions and the null vector conditions are both encoded in determinant formula denoted by $\det M_{m,n}^{NS}(c,h,Q)$, which is a polynomial expression in $c,h,Q$ for the matrix determinant of inner products of a basis for a given subspace. A unitary representation requires that $\det M_{m,n}^{NS}(c,h,Q)\geq0$, which sets a lower bound for $h$. The vanishing surfaces of the determinant formula describe the Verma modules which contain null vectors \cite{Boucher:1986bh}.

The character formulas for unitary representations have been derived in \cite{Eguchi:1988af}. Giving the determinant formula for NS algebras:
\begin{equation}
    \det M_{m,n}^{NS}(\tilde{c},h,Q)=\prod_{1\leq rs\leq 2n,s \hspace{2pt}even}(f_{r,s}^{NS})^{P_{NS}(n-rs/2,m)}\prod_{k\in\mathbb{Z}+1/2}(g_k^{NS})^{\tilde{P}_{NS}(n-|k|,m-sgn(k);k)}
\end{equation}
where 
\begin{equation}
\begin{split}
    f_{r,s}^{NS}(\tilde{c},h,Q)&=2(\tilde{c}-1)h-Q^2-\frac{1}{4}(\tilde{c}-1)^2+\frac{1}{4}((\tilde{c}-1)r+s)^2\\
    g_{k}^{NS}(\tilde{c},h,Q)&=2h-2kQ+(\tilde{c}-1)(k^2-\frac{1}{4})
\end{split}
\label{eq:determinant formula}
\end{equation}
with $\tilde{c}\equiv \frac{c}{3}$ ($\tilde{c}\geq1$) and $r,s\in\mathbb{Z}^+$, the $\mathcal{N}=2$ Liouville theory has three types of unitary representations\footnote{The convention here is $\theta_3(\tau;z)=\prod_{m=1}^{\infty}(1-q^m)(1+yq^{m-1/2})(1+y^{-1}q^{m-1/2})$, $\eta(\tau)=q^{1/24}\prod_{n=1}^{\infty}(1-q^n)$ where $q\equiv e^{2\pi i\tau}$ and $y\equiv e^{2\pi iz}$. $P_{NS}$ and $\tilde{P}_{NS}$ are partition functions counting the states with relative level $n$ and relative charge $m$.}:
\begin{enumerate}
    \item \textbf{Massive representation}

    They are unitary representations for highest weight vector with $(\tilde{c},h,Q)$ satisfying certain conditions such that $g_{n}^{NS}>0$ for $\forall n\in\mathbb{Z}+1/2$. Plugging in~\eqref{eq:determinant formula}, the unitary condition sets a lower bound for $h$:
    \begin{equation}
        h>nQ-\frac{1}{2}(\tilde{c}-1)(n^2-\frac{1}{4})
    \end{equation}
    And the irreducible character for massive representation is given by:
    \begin{equation}
        \text{ch}^{NS}(h,Q;z)=q^{h-(\tilde{c}-1)/8}y^Q\frac{\theta_3(\tau;z)}{\eta(\tau)^3}
    \end{equation}
        \item \textbf{Massless representation}

    They are unitary representations lying on the vanishing surface $g_{n}^{NS}=0$ for some $n\in\mathbb{Z}+1/2$ \cite{Boucher:1986bh}. For $n=\pm\frac{1}{2}$ they are the chiral and anti-chiral primaries, $h=\frac{|Q|}{2}$ with $0<|Q|<\tilde{c}$, whose irreducible character is
    \begin{equation}
        \text{ch}^{NS}(\tfrac{|Q|}{2},Q;z)=q^{\frac{|Q|}{2}-(\tilde{c}-1)/8}\,\frac{y^{Q}}{1+y^{\mathrm{sgn}(Q)}q^{1/2}}\,\frac{\theta_3(\tau;z)}{\eta(\tau)^3}
    \end{equation}
    The representations with other values of $n$ are obtained from these by integer spectral flow.

\item \textbf{Graviton representation}

There is a special case when $h=Q=0$, which obviously lies on the vanishing surface and is the endpoint $Q=0$ of the massless family. The irreducible character is given by:
\begin{equation}
    \text{ch}^{NS}(\tau;z)=q^{-(\tilde{c}-1)/8}\frac{1-q}{(1+yq^{1/2})(1+y^{-1}q^{1/2})}\frac{\theta_3(\tau;z)}{\eta(\tau)^3}
\end{equation}

\end{enumerate}

There are also non-unitary representations, which are given by the vanishing surface of the determinant formula $f_{r,s}^{NS}(\tilde{c},h,Q)=0$. Such representations are generated by a series of degenerate states labeled by two integers $r,s$. And these null states play an essential role in the derivation of structure constant and correlation function of the conformal field theory. We can easily get the solution:
\begin{equation}
    h=\frac{1}{2(\tilde{c}-1)}Q^2+\frac{1}{8}(\tilde{c}-1)-\frac{1}{8(\tilde{c}-1)}((\tilde{c}-1)r+s)^2
    \label{eq:non-unitary repr N=2}
\end{equation}

Hereby we summarized the spectrum for NS algebras in $\mathcal{N}=2$ Liouville theory. With $m\rightarrow m+1/2$ ($y\rightarrow yq^{1/2}$), which can be considered as $1/2$ spectral flow, we get the character formulas for R algebras $\textrm{ch}^R(\tau,z)=q^{\tilde c/8}y^{\tilde c/2}\textrm{ch}^{NS}(\tau,z+\tau/2)$.

\subsection{Vertex Observables}
\label{subsec:vertex observables}

Now we turn to the structure constants of $\mathcal{N}=2$ Liouville theory, which provide the essential dynamical data needed to construct the correlation functions. First of all, we have to clarify the definition of observables. 

A particularly important class of observables consists of the vertex operators, which are also primary states \cite{Ahn_2004_ZZ,Ahn_2004}:
\begin{equation}
\begin{aligned}
N_{\alpha,\tilde{\alpha}}&=e^{\frac{1}{\sqrt{2}}(\alpha+\tilde{\alpha})\phi+\frac{i}{\sqrt{2}}(\alpha-\tilde{\alpha})Y} \quad \quad \text{for NS algebras}\\
R^{\pm}_{\alpha,\tilde{\alpha}}&=\sigma^{\pm}e^{\frac{1}{\sqrt{2}}(\alpha+\tilde{\alpha})\phi+\frac{i}{\sqrt{2}}(\alpha-\tilde{\alpha})Y} \quad \text{for R algebras}
\label{eq:NS and R operators}
\end{aligned}
\end{equation}
where $\sigma^{\pm}$ are spin operators and $\phi,Y$ are both free bosons. The conformal weight of these operators is determined by the parameters $\alpha,\tilde \alpha$ via:
\begin{equation}
\begin{split}
    \Delta_N&=-\alpha\tilde{\alpha}+\frac{1}{2b}(\alpha+\tilde{\alpha})\\
    \Delta_R&=-\alpha\tilde{\alpha}+\frac{1}{2b}(\alpha+\tilde{\alpha})+\frac{1}{8}
\end{split}
\end{equation}
and $U(1)$ charges are given by:
\begin{equation}
    Q_N=\frac{1}{b}(\alpha-\tilde{\alpha})\equiv Q\, \quad \quad   Q_R^{\pm}=Q\pm\frac{1}{2}
\end{equation}
These operators furnish the basic building blocks of the spectrum and play the role of the analogue of ordinary primary fields in the bosonic Liouville theory. These primary vertex operators can also be uniformly written as a general vertex operator labeled by $\alpha,\tilde\alpha,\eta$:
\begin{equation}
    V_{\alpha,\tilde\alpha}^{\eta}(z)=e^{\frac{1}{\sqrt{2}}(\alpha+\tilde\alpha)\phi(z)+\frac{i}{\sqrt{2}}(\alpha-\tilde\alpha)Y(z)+i\eta K(z)}
\end{equation}
with $h=-\alpha\tilde\alpha+\frac{1}{2b}(\alpha+\tilde\alpha)+\frac{\eta^2}{2}$ and $Q=\frac{1}{b}(\alpha-\tilde\alpha)+\eta$. $K(z)$ is the bosonization of the fermions, with $K(z)K(w)\sim -\ln(z-w)$ and $\psi^{\pm}=e^{\pm iK}$. For NS primaries, $\eta\in\mathbb{Z}$, and for R primaries, $\eta\in\mathbb{Z}+\frac{1}{2}$. In our definition above, $\eta=0$ for NS primary states and $\eta=\pm\frac{1}{2}$ for R primary states.

The superfield associated to the superprimary states in superspace is defined by the \emph{supertranslation} operators. In the NS sector, it is given by\footnote{Here we choose the order of highest component to be $\theta^+\theta^-$.}:
\begin{equation}
\begin{aligned}
    \mathcal{V}_{\alpha,\tilde\alpha}^{\eta=0}(z,\theta^+,\theta^-)&= e^{zL_{-1}+\theta^+G_{-1/2}^++\theta^-G_{-1/2}^-}N_{\alpha,\tilde\alpha}(0)e^{-zL_{-1}-\theta^+G_{-1/2}^+-\theta^-G_{-1/2}^-}\\
    &=N_{\alpha,\tilde\alpha}(z)+\theta^+G^{+}_{-1/2}N_{\alpha,\tilde\alpha}(z)+\theta^-G^{-}_{-1/2}N_{\alpha,\tilde\alpha}(z)\\
    &\qquad  +\theta^+\theta^-(L_{-1}-G^+_{-1/2}G^-_{-1/2})N_{\alpha,\tilde\alpha}(z)
    \label{eq: NS multiplet}
\end{aligned}
\end{equation}
The lowest component is the NS primary state defined above, and other components are descendant states generated from it. $L_{-1},G^+_{-1/2}$ and $G^-_{-1/2}$ form a closed algebra in superspace, and the operator $zL_{-1}+\theta^+G_{-1/2}^++\theta^-G_{-1/2}^-$ supertranslates the superprimary field from the origin of the superspace to $(z,\theta^+,\theta^-)$. For each NS vertex operator we have defined, there is a corresponding NS multiplet in the superspace $(z,\theta^+,\theta^-)$. The R multiplets are defined in the similar way. $\mathcal{V}^{1/2}_{\alpha,\tilde\alpha}(z)$ is defined by applying $\frac{1}{2}$-spectral flow for each generator in~\eqref{eq: NS multiplet} and replacing the lowest component by $R^+_{\alpha,\tilde\alpha}(z)$. Correspondingly, $\mathcal{V}^{-1/2{}}_{\alpha,\tilde\alpha}(z)$ is defined with $-\frac{1}{2}$-spectral flowed generators and R primary states $R^-_{\alpha,\tilde\alpha}(z)$.

\subsubsection*{Setting of the Problem}

The goal of our work is to figure out the three-point function in $\mathcal{N}=2$ Liouville theory, whose holomorphic part is written in superspace as\footnote{For anti-holomorphic part, we just replace the set of coordinates by $(\bar z_1,\bar z_2,\bar z_3,\bar\theta_1^{\pm}, \bar\theta_2^{\pm},\bar\theta_3^{\pm})$ and the label of states by $\bar{\alpha},\bar{\tilde{\alpha}}.$}:
\begin{equation*}
    G_3(z_1,z_2,z_3;\theta_1^{\pm},\theta_2^{\pm},\theta_3^{\pm})=\langle \mathcal{V}_{\alpha_1,\tilde\alpha_1}^{\eta=0}(z_1;\theta_1^{\pm})\mathcal{V}_{\alpha_2,\tilde\alpha_2}^{\eta=0}(z_2;\theta_2^{\pm})\mathcal{V}_{\alpha_3,\tilde\alpha_3}^{\eta=0}(z_3;\theta_3^{\pm}) \rangle
\end{equation*}
We restrict the problem in NS sector. The global $sl(2|1)$ Ward identities (Appendix~\ref{app:superconformal Ward identities}) fix the coordinate dependence completely. Bosonic coordinates are fixed to insertions on $(1,0,\infty)$ by $L_{-1},L_0,L_1$. $G^{\pm}_{1/2}$ and $G^{\pm}_{-1/2}$ fix a gauge for Grassmann variables. $J_0$ imposes the charge conservation. Therefore, with the gauge fixing $\theta_1^{\pm}=\theta_3^{\pm}=0$ and $\theta^{\pm}\equiv\theta_2^{\pm}$, the holomorphic part of the structure constant can be written as:
\begin{equation}
    C(\theta^+,\theta^-)=C_0+\theta^+C_++\theta^-C_-+\theta^+\theta^-\big((h_3-h_1-h_2)C_0-C_t\big)
    \label{eq:structure constant in superspace}
\end{equation}
where $C_0$, $C_{\pm}$ and $C_t$ are the structure constants of the lowest component $\phi_2$ and of its descendants $G^{\pm}_{-1/2}\phi_2$ and $G^{+}_{-1/2}G^{-}_{-1/2}\phi_2$ respectively (see Appendix~\ref{app:superconformal Ward identities}), and $C_{a}=C_a(\alpha_1,\tilde\alpha_1,\alpha_2,\tilde\alpha_2,\alpha_3,\tilde\alpha_3)$ for $a=\{0,+,-,t\}$. For readers interested mainly in the results: $C_0$ and $C_t$ are nonvanishing only for $\sum_iQ_i=0$ (the neutral sector) and $C_\pm$ only for $\sum_iQ_i=\mp1$ (the charged sectors). The combination with the anti-holomorphic sector is given at the beginning of Section~\ref{subsec:N=2 Liouville structure constant}, the neutral structure constant in~\eqref{eq:neutral-structure-constant} and~\eqref{eq:F-neutral}, and the charged ones in~\eqref{eq:charged-structure-constant} and~\eqref{eq:Fpm}.

Solving the null-vector equations for four-point functions directly in superspace does not give the structure constants straightforwardly: the solutions of the corresponding differential equations are not of hypergeometric type, since the relevant constraints are naturally expressed in a certain $U(1)$ charge sector. On the other hand, correlation functions in fixed charge sector can be obtained by transferring to $m$-basis $SL(2,\mathbb{R})$ WZW model \cite{Ribault:2005wp,Ribault:2005ms}, which introduces additional quantum numbers and selection rules, whose structure is closely related to the coset model $SL(2,R)/U(1)$. We therefore proceed indirectly. In Section~\ref{sec:Relation to SL(2,R) WZW} we recall the realization of the $\mathcal{N}=2$ algebra by the $SL(2,\mathbb{R})_k/U(1)$ supercoset, and in Section~\ref{sec:Identification of duality} we show, at levels 1 and 2, that the bosonic components of the Liouville null vectors coincide with the null vectors of the $SL(2,\mathbb{R})$ current algebra, whose bootstrap~\cite{Teschner:1997ft} is reviewed in Appendix~\ref{app: structure constant of SL(2,R)}. In Section~\ref{sec:correlators in superspace} we express the components $C_a$ of~\eqref{eq:structure constant in superspace} through the spectral-flow-preserving and spectral-flow-violating structure constants of the coset and test the result.

\section{Relation to \texorpdfstring{$SL(2,\mathbb{R})$}{SL(2,R)} WZW Model}
\label{sec:Relation to SL(2,R) WZW}

We now turn to the dual description of $\mathcal{N}=2$ Liouville theory. The $SL(2,\mathbb{R})$ WZW model is the worldsheet theory of bosonic strings on $\mathrm{AdS}_3\times\mathcal{M}$ with NS--NS flux, where $\mathcal{M}$ is a compact space~\cite{Maldacena:2000hw,Maldacena:2000kv,Maldacena:2001km}. Its symmetry is the affine algebra $\widehat{\mathfrak{sl}}(2,\mathbb{R})_L\times\widehat{\mathfrak{sl}}(2,\mathbb{R})_R$, and its states fall into representations of the universal cover of $SL(2,\mathbb{R})$ together with their affine descendants and spectral-flow images. Gauging a $U(1)$ subgroup and adding fermions gives the supersymmetric coset $SL(2,\mathbb{R})/U(1)$, the CFT of the two-dimensional cigar geometry.

In this section we review the fermionic FZZ duality of Hori and Kapustin, which identifies $\mathcal{N}=2$ Liouville theory with this coset, summarize the spectrum of the coset, and give the map of parameters and observables between the two theories.

\subsection{Fermionic FZZ Duality}
\label{subsec:fermionic FZZ duality}

The relation between supercoset $SL(2,R)/U(1)$, which is the background theory of a string propagating on a fermionic cigar black hole, and $\mathcal{N}=2$ Liouville theory is the supersymmetric version of the FZZ duality. In its original bosonic version, FZZ duality states that the $SL(2,\mathbb{R})/U(1)$ coset CFT is equivalent to a sine-Liouville theory. The geometric description is a semi-infinite cigar, which can be thought of as a Euclidean version of a 1+1 dimensional black hole. It is the solution of the equation of motion of two dimensional string theory \cite{Witten:1991yr,Dijkgraaf:1991ba}, which can be described by the metric and dilaton:
\begin{equation}
\begin{aligned}
    &ds^2=k[dr^2+\tanh^2rd\theta^2]\\
    &\Phi-\Phi_0=-2\log\cosh r
    \label{eq:metric}
\end{aligned}
\end{equation}
More precisely, the Minkowski version corresponds to the coset $SL(2,R)/U(1)$, while the Euclidean version is $\frac{SL(2,C)}{SU(2)}/U(1)$. The parameter $k$ in the metric~\eqref{eq:metric} is the level of $SL(2)$ current algebra in the construction, and the mass of the black hole $M$ is encoded in the value of the string coupling at the tip of the cigar, $M\propto\exp(-2\Phi_0)$. The dual description is sine-Liouville theory: a linear-dilaton field together with a compact scalar and a winding-one perturbation. In the asymptotic region, the metric~\eqref{eq:metric} describes a cylinder far from the tip, which is same as the geometry far from the potential wall in sine-Liouville CFT. While turning on the potential term, the two CFTs are equivalent via strong-weak coupling duality on the worldsheet. In both theories, the momentum is conserved, while the winding number is broken because of the existence of the tip and the potential wall. A matrix model was constructed for studying the two dimensional string theory in the Euclidean black hole background using this bosonic correspondence in \cite{Kazakov:2000pm}.

The fermionic generalization replaces the bosonic coset by the Kazama-Suzuki supercoset model, which is a fermionic generalization of 2d black hole. The fermionic FZZ duality states that \cite{Hori:2001ax}:
\begin{equation}
    \frac{SL(2,\mathbb{R})_k}{U(1)}^{SUSY}\quad \leftrightarrow \quad \text{$\mathcal{N}=2$ Liouville}
\end{equation}
After gauging the supersymmetric $U(1)$, the exact central charge is
\begin{equation}
    c_{\mathrm{coset}}=3\left(1+\frac{2}{k}\right).
\end{equation}
Since our convention in $\mathcal{N}=2$ Liouville theory gives $c_{\mathrm{L}}=3+\frac{6}{b^2}$, equality of the central charge gives the parameter identification
\begin{equation}
    k=b^2, \quad  Q_L=\frac{1}{b}=\frac{1}{\sqrt{k}}
    \label{eq:FZZ-parameter-map}
\end{equation}

Hori and Kapustin established the fermionic duality as an example of mirror symmetry in \cite{Hori:2001ax}. They begin with a two-dimensional $(2,2)$ gauged linear sigma-model containing two chiral superfields $\Phi$ and $P$ and a vector superfield $V$. The action of the system is:
\begin{equation}
 S=\frac{1}{2\pi}\int d^2x d^4\theta\left(\bar\Phi e^V\Phi+\frac{k}{4}(P+\bar P+V)^2
 -\frac{1}{2e^2}|\Sigma|^2\right).
\end{equation}
At energy much lower than $e\sqrt{k}$, the vector multiplet can be integrated out. But the classical background associated to the low energy theory is not yet the black hole metric. In the infrared limit, a modified, anomaly-free $R$-current is constructed with a $P$-dependent improvement. Such construction fixes the central charge of $\mathcal{N}=2$ superconformal algebra, and the resulting sigma model flows to the supersymmetric $SL(2,\mathbb{R})/U(1)$ coset. The resulting target space is a smooth cigar with an asymptotic cylinder, a linear dilaton, and an $\mathcal{N}=2$ superconformal symmetry.

To obtain the mirror, one T-dualizes the phase of $\Phi$ and the imaginary part of $P$. The lowest components of dual chiral fields $Y$ and $Y_P$ have period $2\pi i$, and the twisted superpotential is
\begin{equation}
 \widetilde W=\Sigma(Y+Y_P)+\mu e^{-Y}.
 \label{eq:FZZ-mirror-superpotential}
\end{equation}
The first term is classical, while $e^{-Y}$ is generated by vortex-instantons of the charged field $\Phi$. There is no corresponding $P$-vortex term. In the infrared limit, $\Sigma$ can be integrated out and this imposes $Y+Y_P=0$, leaving a single twisted chiral superfield $Y$ with the twisted potential $\widetilde{W}=\mu e^{-Y}$. The dual theory flows precisely to the $\mathcal N=2$ Liouville theory with the same central charge $3+\frac{6}{k}$ in IR limit.

This derivation establishes the equivalence between supercoset $SL(2,\mathbb{R})/U(1)$ and $\mathcal{N}=2$ Liouville at the level of action and symmetry. 
In particular, the claim is an equivalence of full $\mathcal N=2$ conformal field theories, not merely of semiclassical target spaces. Therefore, we would expect that their central charges, spectra and correlation functions also agree after applying~\eqref{eq:FZZ-parameter-map}, which we examine in the following sections.

\subsection{Supercoset \texorpdfstring{$SL(2,\mathbb{R})_k/U(1)$}{SL(2,R)k/U(1)} Model}
\label{subsec:supercoset}

The fermionic FZZ duality reviewed above implies the algebraic description for the duality, so that the $\mathcal{N}=2$ superconformal algebra can be equally well organized in terms of the supersymmetric coset model $SL(2,\mathbb{R})/U(1)$. We make this statement precise now. We first realize the $\mathcal{N}=2$ superconformal generators in terms of the $SL(2,\mathbb{R})$ currents and free fermions of the Kazama-Suzuki coset and review the irreducible representations of $SL(2,\mathbb{R})$ current algebra. Then we decompose representations of the parent theory with respect to the gauged $U(1)$ symmetry, and the resulting branching functions will be identified with the $\mathcal{N}=2$ characters reviewed in Section~\ref{subsec: spectrum}.

The coset CFT has supersymmetric level $k$. Equivalently, the parent theory consists of a bosonic $\widehat{SL(2,R)}_{k+2}$ algebra and adjoint free fermions, whose bilinear currents have level $-2$ effectively. After gauging the $U(1)$ symmetry, the Kazama-Suzuki supercoset model for $SL(2,R)_k/U(1)$ is defined as coset CFT with central charge $\tilde{c}=\frac{c}{3}=1+\frac{2}{k}$. The bosonic chiral currents $j^{\pm}(z),j^3(z)$ obey:
\begin{equation}
\begin{split}
    j^3(z)j^3(0)\sim -\frac{(k+2)/2}{z^2}\\
    j^3(z)j^{\pm}(0)\sim \frac{\pm1}{z}j^{\pm}(0)\\
    j^{+}(z)j^-(0)\sim \frac{(k+2)}{z^2}-\frac{2}{z}j^3(0)
\end{split}
\end{equation}
and the pair of free fermions satisfy the OPE's: $\psi^+(z)\psi^-(0)\sim\frac{1}{z}$, $\psi^{\pm}(z)\psi^{\pm}(0)\sim0$. In terms of these fields, the generators that commute with the gauged current take the form:
\begin{equation}
\begin{split}
    T(z)&=\frac{1}{2k}(j^+j^-+j^-j^+-2j^3j^3)+\frac{1}{k}J^3J^3-\frac{1}{2}(\psi^+\partial\psi^--\partial\psi^+\psi^-)\\
    J&=\psi^+\psi^-+\frac{2}{k}J^3, \quad G^{\pm}=\sqrt{\frac{2}{k}}\psi^{\pm}j^{\mp}
\end{split}
\end{equation}
where $J^3\equiv j^3+\psi^+\psi^-$ is the unique $U(1)$-current being removed by the coset construction. The displayed $T,J,G^{\pm}$ commute with $J^3$ and close into the $\mathcal{N}=2$ superconformal algebra.

To determine the representations of the algebra, one decomposes the tensor product of an affine $SL(2,R)_{k+2}$ module with the fermion Fock space into modules of the coset algebra and of the gauged $U(1)$ algebra. At the level of characters, this decomposition defines the branching functions. For convenience, we work below in NS sector, and the formulas for other spin structures are obtained via half-integer spectral flow. According to the standard treatment for the coset CFT, the branching functions are defined as:
\begin{equation}
    \chi_{\xi}(\tau,u)\frac{\theta_3(\tau,v)}{\eta(\tau)}=\sum_{m}\chi_{\xi,m}^{NS}(\tau,z)\frac{q^{-\frac{m^2}{k}}e^{2\pi im\omega}}{\eta(\tau)}
\end{equation}
Here $\xi$ labels a representation in $\widehat{SL(2,\mathbb{R})}_{k+2}$, while $\chi_{\xi,m}^{NS}$ is the corresponding NS-sector coset branching function. The factor on the right is the character of the gauged $U(1)$ sector, with $m$ as its $J^3_0$ charge. The angle variables $u,v,z,\omega$ are associated to $U(1)$-currents $j^3,\psi^+\psi^-,J,J^3$. The relations among them are:
\begin{equation}
    u=\frac{2}{k}z+\omega,\quad v=u+z\equiv\frac{k+2}{k}z+\omega
\end{equation}
The label $m$ specifies the $U(1)$-charge of the current. Therefore, it receives contributions from both the bosonic $SL(2,\mathbb{R})$ charge and the fermion charge. If a state has bosonic charge $m_{\text{bos}}$ as the eigenvalue for $j_0^3$ and the total charge is denoted by $m$ as the eigenvalue of $J_0^3$, the fermionic charge is $m-m_{\text{bos}}\in\mathbb{Z}$. This is the selection rule in the coset construction: the total charge is characterized by the $SL(2,\mathbb{R})$ module, because fermion sectors can only shift integer charges. For continuous representations $\hat{\mathcal{C}}_{p,\alpha}$ the bosonic charge lies in $\alpha+\mathbb{Z}$, and hence total charge $m\in\alpha+\mathbb{Z}$. For discrete representations $\hat{\mathcal{D}}_j^+$ they begin at $j$ and differ by integers, giving $m=j+n$ with $n\in\mathbb{Z}$.

We can now summarize the resulting branching functions derived in \cite{Eguchi_2004} and match them with the massive, massless, and graviton representations of the $\mathcal{N}=2$ superconformal algebra introduced in Section~\ref{subsec: spectrum}.
\begin{enumerate}
    \item \textbf{Continuous representation} 

    For a continuous representation $\hat{\mathcal{C}}_{p,\alpha}$, one has $j=\frac{1}{2}+ip$, with $p\in\mathbf{R}_{\geq0}$ and $0\leq\alpha<1$. The vacuum states have the $J_0^3$-spectrum: $J_0^3=\alpha+n$, $n\in\mathbf{Z}$, $\alpha\equiv m\mod1$.

    Due to the absence of null states, the branching functions for continuous series are easily obtained:
    \begin{equation}
        \chi_{\text{con},p,m}^{NS}(\tau,z)=q^{\frac{p^2+m^2}{k}}y^{\frac{2m}{k}}\frac{\theta_3(\tau,z)}{\eta(\tau)^3}
    \end{equation}
    This is precisely the character of a \emph{massive representation} in $\mathcal{N}=2$ superconformal algebra with:
    \begin{equation}
        h=\frac{p^2}{k}+\frac{1}{4k}+\frac{m^2}{k}, \quad Q=\frac{2m}{k}
    \end{equation}
    The unitary condition $p^2+(\alpha-\frac{1}{2})^2\geq0$ in the coset model also agrees with the unitary condition $g_n^{NS}>0$ in $\mathcal{N}=2$ superconformal algebra.

    \item \textbf{Discrete representation}

    $\hat{\mathcal{D}}_j^{\pm}$: $j$ takes discrete values and the vacua have the $J_0^3-$spectrum; $J_0^3=\pm(j+n)$, $n\in\mathbf{Z}_{\geq0}$ for $\hat{\mathcal{D}}_{j}^{\pm}$ respectively. We focus on $\hat{\mathcal{D}}^+_{j}$ because $\hat{\mathcal{D}}^-_j$ is obtained by the spectral flow:
    \begin{equation}
        q^{-\frac{k+2}{4}}e^{-2\pi i\frac{k+2}{2}u}\chi_j^+(\tau,u+\tau)=\chi_{\frac{k+2}{2}-j}^-(\tau,u)
    \end{equation}
The lowest branch, $m=j$, is chiral and is matched with the \emph{massless character} of $\mathcal{N}=2$ algebra, with $h=Q/2=j/k$, $Q=2j/k$:
    \begin{equation}
        \chi_{\text{dis},j,j}^{NS}(\tau,z)=\frac{q^{\frac{j}{k}-\frac{1}{4k}}y^{\frac{2j}{k}}}{1+yq^{1/2}}\frac{\theta_3(\tau,z)}{\eta(\tau)^3}
    \end{equation}
    
The unitary bound\footnote{With spectral flow, the unitary bound can be further restricted to $\frac{1}{2}<j<\frac{k+1}{2}$.} is $0<j<\frac{k+2}{2}$, and states in discrete series within this bound are degenerate.

The branching relation of discrete series is given by:
\begin{equation}
    \chi_j^+(\tau,u)\frac{\theta_3(\tau,v)}{\eta(\tau)}=\sum_{n\in\mathbf{Z}}\frac{q^{\frac{j+n^2+2nj}{k}-\frac{1}{4k}}y^{\frac{2(j+n)}{k}}}{1+yq^{n+1/2}}\frac{\theta_3(\tau,z)}{\eta(\tau)^3}\cdot\frac{q^{-\frac{(j+n)^2}{k}}e^{2\pi i(j+n)\omega}}{\eta(\tau)}
\end{equation}
The other allowed values $m=j+n$ implement the discrete-series selection rule. Their branching functions are identified with the \emph{spectral flowed massless character} of the $\mathcal{N}=2$ superconformal algebra:
    \begin{equation}
        \chi_{\text{dis},j,j+n}^{NS}(\tau,z)=\frac{q^{\frac{j+n^2+2nj}{k}-\frac{1}{4k}}y^{\frac{2(j+n)}{k}}}{1+yq^{n+1/2}}\frac{\theta_3(\tau,z)}{\eta(\tau)^3}\equiv \text{ch}^{NS}(Q=2j/k,n;\tau,z)
    \end{equation}
\begin{enumerate}
    \item $n\geq0$: The vacuum state of $\chi_{\text{dis},j,j+n}^{NS}(\tau,z)$ is $(j_0^+)^n|j,j\rangle
    \otimes|0\rangle_{\psi}$ with:
    \begin{equation}
        h=\frac{2j(n+\frac{1}{2})+n^2}{k}, \quad Q=\frac{2(j+n)}{k}
    \end{equation}

    \item $n<0$: The vacuum state is $(j_{-1}^-)^{|n|-1}|j,j\rangle\otimes\psi_{-1/2}^-|0\rangle_{\psi}$ with:
    \begin{equation}
        h=\frac{-(k-2j)(n+\frac{1}{2})+n^2}{k}, \quad Q=\frac{2(j+n)}{k}-1
    \end{equation}
\end{enumerate}

    \item \textbf{Identity representation}

    Identity representation is generated by the vacuum $j=0$. Branching function of the identity representation $\chi_{\text{dis},0,0}^{NS}(\tau,z)$ is given by:
    \begin{equation}
        \chi_{\text{dis},0,0}^{NS}(\tau,z)=\text{ch}^{NS}(\tau,z)=q^{-\frac{1}{4k}}\frac{1-q}{(1+yq^{1/2})(1+y^{-1}q^{1/2})}\frac{\theta_3(\tau,z)}{\eta(\tau)^3}
    \end{equation}
    which can be identified with the \emph{graviton representation} in $\mathcal{N}=2$ superconformal algebra.

\end{enumerate}

There are also non-unitary representations in $SL(2,\mathbb{R})$ WZW model, which constitute the degenerate states of non-unitary series in the coset model according to the selection rules. These degenerate states (null vectors) are labeled by two integers $r,l$:
\begin{equation}
\begin{aligned}
    j_{r,l}^+&=\frac{1}{2}(1+r+lk) \qquad \text{with} \quad r,l=1,2,3,\cdots\\
    j_{r,l}^-&=\frac{1}{2}(1-r-lk) \qquad \text{with} \quad r=1,2,3,\cdots \quad l=0,1,2,\cdots
\end{aligned}
\end{equation}
The corresponding null vectors have been collected in \cite{Dei:2021yom}. By plugging the relations $h=-\frac{j(j-1)}{k}+\frac{m^2}{k}$ and $\tilde{c}-1=\frac{2}{k}$ for $SL(2,\mathbb{R})$ model into the solution~\eqref{eq:non-unitary repr N=2} for vanishing surface in $\mathcal{N}=2$ algebra, we can easily check that the solutions for spins $j$ are exactly the null vectors we listed above with $l=s/2$ for $l\geq 1$; the $l=0$ series corresponds to the $g_n=0$ (BPS-type) degenerations. Hence the non-unitary degenerate branches reproduce the degenerate $\mathcal{N}=2$ representations discussed previously. This final correspondence will be further explored, giving an exact match of null vectors in Section~\ref{sec:Identification of duality}.

\subsection*{Identification of the Observables}
\label{subsec:identification of parameter and observables}

Recalling the vertex observables in $\mathcal{N}=2$ Liouville defined in section~\ref{subsec:vertex observables} with a pair of parameters $(\alpha,\tilde \alpha)$, one can also introduce the momentum $P$:
\begin{equation}
    \alpha+\tilde{\alpha}=\frac{1}{b}+2iP
\end{equation}
besides the $U(1)$ charge $Q\equiv\frac{1}{b}(\alpha-\tilde \alpha)$. Then we can rewrite the conformal dimension of the primary field in terms of $Q$ and $P$ as:
\begin{equation}
    \Delta_N=P^2+\frac{1}{4b^2}+\frac{b^2 Q^2}{4}, \quad \Delta_R=\Delta_N+\frac{1}{8}
\end{equation}
Comparing with the conformal dimension of continuous representation $\hat{\mathcal{C}}_{p,m}$ in $SL(2,\mathbb{R})/U(1)$ model $h=\frac{p^2}{k}+\frac{1}{4k}+\frac{m^2}{k}$ with $k=b^2$, we will get:
\begin{equation}
\begin{split}
    P^2&=\frac{p^2}{k}=\frac{p^2}{b^2} \quad \Rightarrow \quad p=bP\\
    \frac{b^2 Q^2}{4}&=\frac{m^2}{k}=\frac{m^2}{b^2} \quad \Rightarrow \quad m=\frac{b^2 Q}{2}=\frac{b(\alpha-\tilde{\alpha})}{2}
\end{split}
\end{equation}
More generally, we can match the spin $j$ with the parameters in $\mathcal{N}=2$ algebra by:
\begin{equation}
    j=\frac{b}{2}(\alpha+\tilde \alpha)
    \label{eq:spin-alpha}
\end{equation}

Plugging these relations into~\eqref{eq:NS and R operators}, the NS vertex operators can be identified with the primary states in $SL(2,\mathbb{R})$ model $|j,m\rangle$ via:
\begin{equation}
   N_{\alpha,\tilde \alpha}\quad \Rightarrow \quad e^{\sqrt{\frac{2}{k}}(j\phi+imY)}= |j,m\rangle 
   \label{eq:op-dictionary}
\end{equation}
Correspondingly, one can construct the R vertex operator using a tensor product of the primary states $|j,m\rangle$ and spin $|+\rangle$ or $|-\rangle$. For NS primaries the fermions are in their vacuum, and $m=\frac{b^2Q}{2}$ is the $j_0^3$ eigenvalue of the $SL(2,\mathbb{R})$ factor. For the Ramond fields $R^{\pm}_{\alpha,\tilde\alpha}$ the spin state $|\pm\rangle$ carries fermion charge $\pm\frac{1}{2}$, and the $j_0^3$ eigenvalue is $m=\frac{b^2Q}{2}\mp\frac{1}{2}$, with $Q=\frac{1}{b}(\alpha-\tilde\alpha)=Q_R^{\pm}\mp\frac{1}{2}$. We will make use of this correspondence to show that the null vectors in $\mathcal{N}=2$ Liouville and $SL(2,\mathbb{R})$ model exactly match in Section~\ref{sec:Identification of duality}.

\section{Matching Null Vectors}
\label{sec:Identification of duality}

The dictionary~\eqref{eq:op-dictionary} maps primary operators in $\mathcal{N}=2$ Liouville theory to primary states $\ket{j,m}$ of the coset description. We now extend this comparison to the full degenerate representations, for which particular combinations of descendants become null states. The equivalence of the two theories requires these null states to agree under the identification we showed above.

We will first review the relevant null vectors of the $\mathcal{N}=2$ superconformal algebra and then figure out the dual of two simplest cases in $SL(2,\mathbb{R})$ null vectors, $j_{2,0}^-=-\frac{1}{2}$ and $j^-_{1,1}=-\frac{k}{2}$. Decoupling a null state inside a four-point correlation function produces a differential equation, and solving the resulting equations in these two simplest cases determines the conformal blocks of four-point functions containing a degenerate field.

\subsection{Null Vectors in \texorpdfstring{$\mathcal{N}=2$}{N=2} Superconformal Algebra}

As reviewed in Section~\ref{subsec: spectrum}, a highest-weight module becomes degenerate when the Kac determinant of some descendant states vanishes at a fixed level. On the vanishing surface, a descendant generated by certain combinations of negative modes is itself annihilated by all positive modes and therefore generates a null submodule. In $\mathcal N=2$ Liouville theory these conditions can be expressed directly by the parameters $(\alpha,\tilde\alpha)$ for the vertex observables. We summarize below the two relevant families, conventionally called Class I and Class II which has been derived in \cite{Ahn_2004_ZZ}, and display their null vectors at the lowest level in NS and R sectors.

Among the primary fields there is a series of degenerate fields $N_{r,l}\equiv N_{\alpha_{r,l},\tilde{\alpha}_{r,l}}$ and $R^{\pm}_{r,l}\equiv R^{\pm}_{\alpha_{r,l},\tilde{\alpha}_{r,l}}$, whose subscripts are parameterized by:
\begin{equation}
    \alpha_{r,l}=\frac{1-r}{2b}+\frac{Q}{2}b-\frac{l}{2}b, \quad \tilde{\alpha}_{r,l}=\frac{1-r}{2b}-\frac{Q}{2}b-\frac{l}{2}b
\end{equation}
where $r,l\in\mathbb{Z}_{+}$ and $Q\neq\pm l$, these are called \emph{Class I} degenerate fields at level $rl$; for $Q=\pm l$ the field coincides with a Class-II field. Using the dictionary between spin $j$ and $(\alpha,\tilde \alpha)$~\eqref{eq:spin-alpha}, we see that the Class I degenerate states correspond to those degenerate series in $SL(2,\mathbb{R})$ model with spin $j_{r,l}^-$ excluding the case when $l=0$. In particular, the simplest degenerate state in this series corresponds to that with spin $j_{1,1}^-=-\frac{k}{2}$ and $U(1)$ charge $m=\frac{b^2Q}{2}$. Conformal dimension of the degenerate states is given by\footnote{Comparing with the conformal dimension derived from Kac determinant:
\begin{equation*}
    \Delta_{(r,s)}^{N}=\frac{1}{4b^2}(1-r^2)-\frac{1}{4}sr+\frac{b^2}{4}(Q^2-\frac{s^2}{4})
\end{equation*}
we have $s=2l$.}
\begin{equation}
    \Delta_{(r,l)}^{N}=\frac{1}{4b^2}(1-r^2)-\frac{1}{2}rl+\frac{b^2}{4}(Q^2-l^2)
\end{equation}
The simplest degenerate state in this series is $N_{1,1}$ with $\alpha_{1,1}=-\frac{b}{2}$ for $Q=0$, denoted by $N_{-\frac{b}{2}}$ for simplicity. At level 1, for $r=l=1$, one can check that
\begin{equation}
    \left[\frac{b^2}{2}(1-Q^2)J_{-1}+G^+_{-1/2}G^-_{-1/2}-(1-Q)L_{-1}\right]|N_{-\frac{b}{2}}\rangle
    \label{eq:(1,1) degenerate}
\end{equation}
is annihilated by all the positive modes of the $\mathcal{N}=2$ superconformal algebra. 

The second class of degenerate fields is denoted by $N_r^Q\equiv N_{\alpha_r,\tilde{\alpha}_r}$ and $R^{+,Q}_{r}\equiv R_{\alpha_r,\tilde{\alpha}_r}^{+}$, called \emph{Class-II}\footnote{We restrict to $Q>0$ here, for $Q<0$, $(\alpha_r,\tilde\alpha_r)$ takes the value $(\frac{1-r}{2b},\frac{1-r}{2b}-Qb)$ for $N_r^Q$ and $R^{-,Q}_{r}$.}:
\begin{equation}
    \alpha_r\equiv\frac{1-r}{2b}+Q b, \quad \tilde{\alpha}_{r}\equiv \frac{1-r}{2b}
\end{equation}
with $r$ a positive odd integer for NS sector and even for R sector. These fields have null states at level $r/2$. Similarly, we can translate these parameters into spin $j$ and find that their neutral members ($Q=0$) have spin $j_{r,0}^-$. For $r=1$, these fields becomes either chiral or anti-chiral fields which are annihilated by $G^{\pm}_{-1/2}$ respectively. For $r=3$, the null state at level $\frac{3}{2}$ takes the form:
\begin{equation}
    \left[\left(Q-\frac{2}{b^2}+1\right)G^{+}_{-3/2}-G^+_{-1/2}L_{-1}+G^+_{-1/2}J_{-1}\right]|N_{-\frac{1}{b}}^Q\rangle
    \label{eq:level 3/2 degen NS}
\end{equation}
It's easily observed that this state has $U(1)$ charge $Q+1$ and dimension $3/2$ higher than that of $N_{-\frac{1}{b}}^Q$, therefore it's straightforward to identify it as $N_{-\frac{1}{b}}^{Q+1}$ up to a normalization constant. We can generate infinite null states for any odd integer $r$ in this way: $N_{-\frac{1}{b}}^Q\rightarrow N_{-\frac{1}{b}}^{Q+1}\rightarrow N_{-\frac{1}{b}}^{Q+2}\rightarrow\cdots$.\footnote{For $Q<0$, there are null states generated by $G^-_{-3/2},G^-_{-1/2}$ similarly, and this state can be identified with null state $N^{Q-1}_{-\frac{1}{b}}$. Therefore there is another sequence: $N^{Q}_{r}\rightarrow N^{Q-1}_{r}\rightarrow N^{Q-2}_r\rightarrow\cdots$.} Another simplest null state is given by $R^{+,Q}_{-\frac{1}{2b}}$ with $\alpha_{2}=-\frac{1}{2b}+Q b$. Especially for the null state in neutral sector, we can see that $R^+_{-\frac{1}{2b}}$ corresponds to the null state with spin $j_{2,0}^-=-\frac{1}{2}$. At level $r/2=1$ in R sector, the state satisfying the null state condition is simply:
\begin{equation}
    G_{-1}^+|R^{+,Q}_{-\frac{1}{2b}}\rangle
    \label{eq:(2,0) degenerate}
\end{equation}
which can be identified as $|R^{+,Q+1}_{-\frac{1}{2b}}\rangle$ up to a normalization constant.

To summarize, the dictionary~\eqref{eq:spin-alpha} already identifies the two simplest fields as
\begin{equation}
    R^+_{-\frac{1}{2b}}\ \longleftrightarrow\ j^-_{2,0}=-\frac12,
    \qquad
    N_{-\frac{b}{2}}\ \longleftrightarrow\ j^-_{1,1}=-\frac{k}{2}.
\end{equation}
At this stage the correspondence follows only from matching the primary-field parameters. In the next subsection, we check that their null descendants agree as well by substituting the coset realization of the $\mathcal N=2$ generators and applying the affine-current and fermion (anti-)commutation relations.

\subsection{Matching Null Vectors}

To verify the correspondence between null vector conditions, we express each $\mathcal{N}=2$ null vector in terms of the bosonic currents $j^a$ and the free fermions $\psi^{\pm}$. Acting on the tensor product primary state in $\mathcal{N}=2$ algebra, we then compare the bosonic component with the null vector of the affine $SL(2,\mathbb{R})$ module and simplify the fermionic component using commutation relations of $SL(2,\mathbb{R})$ current algebra. We carry this out for two simplest null vectors listed above.

\subsubsection*{Matching null vector $j_{2,0}^-=-\frac{1}{2}$}

For the $SL(2,\mathbb{R})$ representation with $j=-1/2$, we pick the lowest weight state in the $\mathcal{D}^+$ representation whose relevant null vector is entirely generated by zero modes:
\begin{equation}
    |\mathcal{N}\rangle=j_0^+j_0^+|j=-1/2,m=-1/2\rangle
    \label{eq:null-1/2}
\end{equation}

On the Liouville side, we take the null vector for $R_{-\frac{1}{2b}}^{-}$, which is simply:
\begin{equation}
    \ket{\mathcal{N}}=G_{-1}^-\ket{R_{-\frac{1}{2b}}^{-}}
\end{equation}
The corresponding test state for this null vector can be written as a tensor product state with both bosonic and fermionic sector, where we rewrite the bosonic sector by translating the Liouville parameters into $j,m$ labeling states in $SL(2,\mathbb{R})$:
\begin{equation}
    \ket{R_{-\frac{1}{2b}}^-}\equiv|j=-\frac{1}{2},m=\frac{1}{2}\rangle\otimes|-\rangle
\end{equation}
Here the total $J^3_0$ charge is 0. From $G^-(z)=\frac{\sqrt{2}}{b}\psi^-j^+(z)$, we have $G_{-1}^-=\frac{\sqrt{2}}{b}\sum_{r\in\mathbb{Z}}\psi_{-1-r}^-j_{r}^+$. Plugging in the test state gives:
\begin{equation}
\begin{aligned}
    |\mathcal{N}\rangle&=\frac{\sqrt{2}}{b}j_{0}^+|j=-\frac{1}{2},m=\frac{1}{2}\rangle\otimes \psi_{-1}^-|-\rangle\\
    &\propto\frac{\sqrt{2}}{b}j_0^+j_0^+|j=-\frac{1}{2},m=-\frac{1}{2}\rangle\otimes\psi_{-1}^-|-\rangle
\end{aligned}
\end{equation}
where we can read out that the bosonic part matches exactly with null vector for $\ket{j=-\tfrac{1}{2},m=-\tfrac{1}{2}}$ in $SL(2,\mathbb{R})$.

Here we pick $\ket{R^{-}_{-\frac{1}{2b}}}$ to match the null vector in $\mathcal{D}^+$ representations. We can do the same for $\ket{R^{+}_{-\frac{1}{2b}}}$ to match the null vector in $\mathcal{D}^-$ representations.

\subsubsection*{Matching null vector $j_{1,1}^-=-\frac{k}{2}$}

The second example is less obvious, because the level-$1$ $\mathcal{N}=2$ null vector contains three different generators. We first record the null vector that we aim to recover in $SL(2,R)_{k+2}$ WZW model for $j_{1,1}^-=-\frac{k}{2}$:
\begin{equation}
     \label{eq:null-vector--k/2} |\mathcal{N}\rangle=\left[(\frac{k}{2}+1-m)(\frac{k}{2}-m)j_{-1}^+ + (k-2m)j_{-1}^3j_0^+ + j_{-1}^-j_0^+j_0^+\right]|j=-\frac{k}{2},m-1\rangle
\end{equation}

On the Liouville side, for $m\neq\pm\frac{k}{2}$ (i.e. $Q\neq\pm1$), the corresponding null vector is~\eqref{eq:(1,1) degenerate} with $b^2=k$ and $Q=\tfrac{2m}{k}$, acting on $\ket{N^m_{-b/2}}$. The NS primary $\ket{N^{m}_{-b/2}}$ has a fermionic vacuum in the tensor product state. By substituting the Liouville parameters as before, we can rewrite the test NS primary state as:
\begin{equation}
    \ket{N_{-b/2}^m}\equiv|j=-k/2,m\rangle\otimes|0\rangle_{f}\propto j_0^+|j=-k/2,m-1\rangle\otimes|0\rangle_f
\end{equation}
We now substitute the three $\mathcal{N}=2$ generators with the coset realization:
\begin{equation}
\begin{aligned}
    J_{-1}&=(1+\frac{2}{k})\sum_{r\in \mathbb{Z}+1/2}\psi^+_{-1-r}\psi^-_{r}+\frac{2}{k}j_{-1}^3\\
    G^+_{-1/2}G_{-1/2}^-&=\frac{2}{k}\sum_{r,s\in\mathbb{Z}+1/2}j_{-1/2-r}^-j_{-1/2-s}^+\psi_{r}^+\psi_{s}^-\\
    L_{-1}&=\frac{1}{k}(j_{-1}^+j_0^-+j_{-1}^-j_0^+)+\frac{2}{k}\sum_{r\in\mathbb{Z}+1/2}\psi^+_{-1-r}\psi^-_{r}j_0^3\\
    &+\frac{1}{2}\sum_{s\in\mathbb{Z}+1/2}(2s+1)\psi_{-1-s}^+\psi_s^-
\end{aligned}
\end{equation}

Plugging the mode expansions into the null vector for $\ket{N^m_{-b/2}}$ and using the fact that the generators with positive modes annihilate the primary state, the $\mathcal{N}=2$ null vector separates into two components, which have separately level-1 bosonic sector and level-0 bosonic sector (with level-0 fermionic sector and level-1 fermionic sector)\footnote{These two sectors can be considered as lowest component with 1 and the highest component with $\theta^+\theta^-$ in superspace.}:
\begin{equation}
\begin{aligned}
    \relax[(\frac{k}{2}+1)(1-\frac{4m^2}{k^2})&j_0^++\frac{2}{k}j_0^-j_0^+j_0^+-\frac{2}{k}(1-\frac{2m}{k})j_0^3j_0^+]|j,m-1\rangle\otimes\psi_{-1/2}^+\psi_{-1/2}^-|0\rangle_f\\
    (1-\frac{4m^2}{k^2})j_{-1}^3j_0^++\frac{2}{k}&j_{-1}^-j_0^+j_0^+-(1-\frac{2m}{k})\frac{1}{k}(j_{-1}^+j_0^-j_0^++j_{-1}^-j_0^+j_0^+)|j,m-1\rangle\otimes|0\rangle_f
    \label{eq:raw level 1 null}
\end{aligned}
\end{equation}
Using commutation relations for zero modes of $j^{\pm,0}$, the first line can be rewritten as:
\begin{equation}
    j_0^-j_0^+|j,m\rangle=(-\frac{k}{2}-\frac{k^2}{4}+m+m^2)|j,m\rangle
\end{equation}
which is in accordance with the $SL(2,\mathbb{R})$ current algebra commutation relations. Simplifying the bosonic sector in the second line of equation~\eqref{eq:raw level 1 null} gives:
\begin{equation}
    \left[(1-\frac{4m^2}{k^2})j_{-1}^3j_0^++(\frac{1}{k}+\frac{2m}{k^2})j_{-1}^-j_0^+j_0^+-(\frac{1}{k}-\frac{2m}{k^2})j_{-1}^+j_0^-j_0^+\right]|j,m-1\rangle
\end{equation}
We can rewrite the last term using:
\begin{equation}
\begin{split}
    j_0^+|j,m\rangle=(m+j)|j,m+1\rangle\\
    j_0^-|j,m\rangle=(m-j)|j,m-1\rangle
    \label{eq:up-down-shift-current}
\end{split} 
\end{equation}
and we get:
\begin{equation}
    \left[(1-\frac{4m^2}{k^2})j_{-1}^3j_0^++(\frac{1}{k}+\frac{2m}{k^2})j_{-1}^-j_0^+j_0^+-(\frac{1}{k}-\frac{2m}{k^2})(m-1-\frac{k}{2})(m+\frac{k}{2})j_{-1}^+\right]|j,m-1\rangle
\end{equation}
which can be simplified by multiplying $\frac{k^2}{k+2m}$, is exactly the null vector~\eqref{eq:null-vector--k/2}. We have therefore matched not only the spins of the two degenerate primaries but also their explicit null vectors. This agreement helps with writing down the BPZ-type equations for $\mathcal{N}=2$ Liouville theory in the following subsection. In addition, we can also match the null vectors of $\mathcal{N}=2$ algebra with those in $SL(2,\mathbb{R})$ model in higher levels. We collect the results for level-2 null vectors in Appendix~\ref{app:null vectors level k}.

\subsection{BPZ-type Equations as Constraints on Structure Constant}
\label{subsec:BPZ equations}

Null vectors give rise to differential equations for correlators containing a degenerate insertion. Since the null vectors of the two theories coincide under the coset realization, we formulate these constraints in the parent $SL(2,\mathbb{R})_{k+2}$ WZW model, where they are well understood. Correlators of the supercoset, through the duality for those of $\mathcal{N}=2$ Liouville theory, are related to $m$-basis WZW correlators by stripping off the free-field factor of the gauged $U(1)$ symmetry \cite{Ribault:2005ms}. In this subsection we review the resulting decoupling equations and the Knizhnik--Zamolodchikov (KZ) equation, written with the labels $(j,m)$ of the WZW model.

We start with a primary field $\ket{j,m}$ associated to the operator $\Phi_{j,m}$ via:
\begin{equation}
    \ket{j,m}\equiv \Phi_{j,m}(0)\ket{0}
\end{equation}
A primary field associated to $\Phi_{j,m}(z)$ depends on the worldsheet coordinate $z$:
\begin{equation}
    \Phi_{j,m}(z)\equiv e^{zL_{-1}}\Phi_{j,m}(0)e^{-zL_{-1}}
\end{equation}
and $\Phi_{j,m}(z)$ is a $m$-basis field. There is another set of basis called $x$-basis, whose corresponding fields are labeled only with spin $j$. They are related to $m$-basis fields by\footnote{For continuous $m,\bar m$ the sum is replaced by a Mellin transform, given in the conventions of the literature in~\eqref{eq:m-basis function}.}:
\begin{equation}
    \Phi_j(x;z)\equiv \sum_{m}x^{m-j}\frac{\Gamma(m+j)}{\Gamma(m+1-j)}\Phi_{j,m}(z)
\end{equation}
where the normalization factor $\frac{\Gamma(m+j)}{\Gamma(m+1-j)}$ follows the convention~\eqref{eq:up-down-shift-current}, and the polynomial of $x$ is fixed by the $\widehat{\mathfrak{sl}(2,R)}_{k+2}$ current algebra:
\begin{equation}
    [j^3_m,j^3_n]=-\frac{\kappa}{2}m\,\delta_{m+n,0}\ ,\quad
 [j^3_m,j^\pm_n]=\pm j^\pm_{m+n}\ ,\quad
 [j^+_m,j^-_n]=-2j^3_{m+n}+\kappa\,m\,\delta_{m+n,0}
 \label{eq:affine} 
\end{equation}
with $\kappa\equiv k+2$. These operators also depend on $z$ via $j^a(z)\equiv e^{zL_{-1}}j^a(0)e^{-zL_{-1}}$, and their OPEs with primary fields are:
\begin{equation}
    j^a(z)\Phi_{j}(x;w)=\frac{D^a}{z-w}\Phi_j(x;w)
\end{equation}
with $D^a$ defined by:
\begin{equation}
D^+=\partial_{x}\ ,\quad
D^3=x\partial_x+j\ ,\quad
D^-=x^2\partial_x+2jx\
 \label{eq:Dhat}
\end{equation}
The null vector~\eqref{eq:null-vector--k/2} can be simplified using~\eqref{eq:up-down-shift-current}:
\begin{equation}
    A(m)\left[j_{-1}^-\ket{j,m+1}-2j_{-1}^3\ket{j,m}+j_{-1}^+\ket{j,m-1}\right], \quad A(m)\equiv (m-\frac{k}{2})(m-1-\frac{k}{2})
    \label{eq:m-null vector}
\end{equation}
for $j=-\frac{k}{2}$. The bracket carries no $m$-dependence, since the same null vector condition applies for all eigenstates of $j_0^3$. In particular, for the lowest-weight state $\ket{j,j}$, the null vector becomes:
\begin{equation}
     \ket{\mathcal{N}}=\Big[j^-_{-1}j^+_0j^+_0+2(k-1)\,j^3_{-1}j^+_0+k(k-1)\,j^+_{-1}\Big]\ket{j,j}\ .
     \label{eq:lowest weight null vector}
\end{equation}

Now we take the four-point function in both $m$-basis and $x$-basis:
\begin{equation}
\begin{aligned}
    G_4(m_1,m_2,m_3,m_4)=\langle\Phi_{j_1,m_1}(z_1)&\Phi_{j_2,m_2}(z_2)\Phi_{j_3,m_3}(z_3)\Phi_{j_4,m_4}(z_4)\rangle\\
    G_4(x_1,x_2,x_3,x_4;z_1,z_2,z_3,z_4)=\langle\Phi_{j_1}(x_1;z_1)&\Phi_{j_2}(x_2;z_2)\Phi_{j_3}(x_3;z_3)\Phi_{j_4}(x_4;z_4)\rangle
\end{aligned}
\end{equation}
If we set $j_2=-\frac{k}{2}$ and plug the relation~\eqref{eq:m-null vector} in the $m$-basis four-point function, we will get:
\begin{equation}
\begin{aligned}
    \sum_{i=1,3,4}\frac{1}{z_2-z_i}\Big[(m_i-j_i)G_4(m_i-1,&m_2+1,\{m_{j\neq i,2}\})-2m_iG_4(m_1,m_2,m_3,m_4)\\
    \qquad \quad &+(m_i+j_i)G_4(m_i+1,m_2-1,\{m_{j\neq i,2}\})\Big]=0
\end{aligned}
\end{equation}
Each term preserves $\sum_i m_i=0$ from charge conservation. But the equation above is not a differential equation within a fixed $U(1)$-charge sector, instead it relates the correlators whose individual charges differ by 1 unit while the number of total charges is fixed. The way to extract the information of conformal blocks is to check the $x$-basis correlators. Still setting $j_2=-\frac{k}{2}$ and conjugating~\eqref{eq:lowest weight null vector} with $e^{x_2j_0^+}$, the $x$-basis four-point function satisfies the relation:
\begin{multline}
    0=\sum_{i=1,3,4}\frac{1}{z_i-z_2}\Big[(D_i^--2x_2D_i^3+x_2^2D_i^+)\partial^2_{x_2}+2(k-1)(D_i^3-x_2D_i^+)\partial_{x_2}\\
    +k(k-1)D_i^+\Big]G_4(\{x_{i=1,\cdots,4}\};\{z_{i=1,\cdots,4}\})
    \label{eq:null diff no gauge fixing}
\end{multline}
where we have used:
\begin{equation}
    e^{x_2j_0^+}j_{-1}^-e^{-x_2j_0^+}=j_{-1}^--2x_2j_{-1}^3+x_2^2j_{-1}^+, \quad e^{x_2j_0^+}j_{-1}^3e^{-x_2j_0^+}=j_{-1}^3-x_2j_{-1}^+
\end{equation}
With the global Ward identities, we pick a gauge $z_1=x_1=0$, $z_3=x_3=1$, $z_4=x_4=\infty$, then the four-point function depends on the cross-ratio:
\begin{equation}
    z_2=z=\frac{z_{21}z_{43}}{z_{31}z_{42}}, \qquad x_2=x=\frac{x_{21}x_{43}}{x_{31}x_{42}},\
\end{equation}
and the eq.~\eqref{eq:null diff no gauge fixing} can be rewritten as the ordinary third-order ODE for $x$:
\begin{multline}
  \Big[-2(x-1)x(x-z)\partial^3_{x}
  +\Big(x^2\big(-2j_1-2j_3+2j_4+5\kappa-18\big)\\
  -2x\big(2j_4z+\kappa z-6z-2j_1+2\kappa-6\big)
  +z\big(-2j_1+2j_3+2j_4+\kappa-6\big)\Big)\partial^2_{x}\\
  -2(\kappa-3)\Big(2x\big(-j_1-j_3+j_4+\kappa-3\big)+2j_1-2j_4z-\kappa+2z+2\Big)\partial_{x}\\
  +(\kappa^2-5\kappa+6)\big(-2j_1-2j_3+2j_4+\kappa-2\big)\Big]\,G(x,z)=0
  \label{eq:ODE-2}
\end{multline}
with $\kappa=k+2$. Another differential equation can be obtained by setting $j_2=-\frac{1}{2}$ and plugging~\eqref{eq:null-1/2} in the four-point function:
\begin{equation}
    \partial_x^2G(x,z)=0
    \label{eq:ODE-1}
\end{equation}
And together with the KZ equation derived from $L_{-1}=\frac{1}{k}(j_{-1}^-j_0^++j_{-1}^+j_0^--2j_{-1}^3j_{0}^3)$:
\begin{equation}
  \partial_z G(x,z)=\frac{1}{k}\left(\frac{\mathcal P}{z}+\frac{\mathcal Q}{z-1}\right)G(x,z)
  \label{eq:KZ}
\end{equation}
where $\mathcal P$ and $\mathcal Q$ are differential operators in $x$,
\begin{align}
  \mathcal P&=x^2(x-1)\,\partial_x^2-\Big((\sigma-1)x^2+2j_1x-2j_2x(x-1)\Big)\partial_x-2\sigma j_2x-2j_1j_2\ ,
  \label{eq:KZP}\\
  \mathcal Q&=-(1-x)^2x\,\partial_x^2+(1-x)\Big((\sigma-1)(1-x)+2j_3+2j_2x\Big)\partial_x-2\sigma j_2(1-x)-2j_3j_2\ ,
  \label{eq:KZQ}
\end{align}
with $\sigma\equiv j_4-j_1-j_2-j_3$. For $j_2=-\frac{1}{2}$ and $j_2=-\frac{k}{2}$, there are two corresponding differential equations. By solving these three differential equations~\eqref{eq:ODE-1}, \eqref{eq:ODE-2} and \eqref{eq:KZ}, we will get the solution of conformal blocks in $SL(2,\mathbb{R})$ WZW model.

Our algebraic analysis suggests that the structure constants of $\mathcal{N}=2$ Liouville theory encode the same data as those of the $SL(2,\mathbb{R})$ WZW model. This correspondence alone, however, does not establish their precise identification. In particular, three continuous representations of $SL(2,\mathbb{R})$ admit two linearly independent Clebsch–Gordan coefficients, leaving open which linear combination is selected in $\mathcal{N}=2$ Liouville theory. As discussed in \cite{Ribault:2005wp,Ribault:2009ui}, the $SL(2,\mathbb{R})/U(1)$ coset model selects a unique structure constant through Wick rotation of the $m$-basis structure constant of the $H_3^+$ model. The fermionic FZZ duality therefore provides a candidate for the corresponding Liouville structure constant, although the algebraic constraints considered so far do not suffice to establish this identification independently. A further question concerns the structure constants with spectral-flow violation $\sum w=\pm1$ in the $SL(2,\mathbb{R})$ WZW model\footnote{More general spectral-flowed correlators have been studied in \cite{Dei:2021xgh,Dei:2021yom}.}, which also contribute to coset correlators: their interpretation in $\mathcal{N}=2$ Liouville theory remains to be clarified. In the next section, we address these questions by examining Liouville correlators in superspace.

\section{\texorpdfstring{$\mathcal{N}=2$}{N=2} Liouville Structure Constants in Superspace}
\label{sec:correlators in superspace}

Having checked the algebraic duality between $\mathcal{N}=2$ Liouville theory and the $SL(2,\mathbb{R})$ WZW model, we now check how the proposed structure constants fit in the superspace configuration. We will give the results directly and show that they satisfy the constraints given by superconformal Ward identities. We also check the crossing symmetry numerically.

\subsection{Structure Constants of \texorpdfstring{$\mathcal{N}=2$}{N=2} Liouville Theory}
\label{subsec:N=2 Liouville structure constant}

To determine the full structure constant, we must consider both the holomorphic and anti-holomorphic sectors, each of which should independently satisfy the conditions~\eqref{eq:C_t sol} derived from superconformal Ward identities in Appendix~\ref{app:superconformal Ward identities}. We introduce another set of parameters $(\bar\alpha,\bar{\tilde\alpha},\bar\eta)$ to describe the primary vertex operators with conformal weight $\bar{h}$ and $U(1)$ charge $\bar Q$ in the anti-holomorphic sector:
\begin{equation}
    V^{\bar\eta}_{\bar\alpha,\bar{\tilde\alpha}}(\bar z,\bar\theta^+,\bar\theta^-)=e^{\frac{1}{\sqrt{2}}(\bar\alpha+\bar{\tilde\alpha})\phi(\bar z)+\frac{i}{\sqrt{2}}(\bar\alpha-\bar{\tilde\alpha})Y(\bar z)+i\bar\eta K(\bar z)}
\end{equation}
Similar to the holomorphic sector, the conformal weight and $U(1)$ charge are given by:
\begin{equation}
    \bar h=-\bar\alpha\bar{\tilde\alpha}+\frac{1}{2b}(\bar\alpha+\bar{\tilde\alpha})+\frac{\bar\eta^2}{2}, \qquad \bar Q=\frac{1}{b}(\bar\alpha-\bar{\tilde\alpha})+\bar\eta
\end{equation}
Combination of both sectors should obey the selection rule $\sum_i(Q_i-\bar Q_i)\in\frac{2}{b^2}\mathbb{Z}$ for generic $b$, which follows from the selection rule of $SL(2,\mathbb{R})/U(1)$ coset model $m-\bar{m}\in\mathbb{Z}$\footnote{$\bar{m}$ denotes the $U(1)$ charge in the anti-holomorphic sector. The selection rule of $m$-basis wave function in coset model has been derived in \cite{Ribault:2009ui} and reviewed in Appendix~\ref{app: structure constant of SL(2,R)}.}. Since $\sum_iQ_i$ and $\sum_i\bar Q_i$ can only be $0,\pm1$, the only way of satisfying the selection rule is that $\sum_iQ_i=\sum_i\bar Q_i$. Therefore, for a gauge fixed three-point function in superspace, combining both sectors gives:
\begin{equation}
    G_3(Z,\bar Z)=\sum_{a,\bar  a}P_a(Z)P_{\bar a}(\bar Z)C_{a\bar a}(\alpha_i,\bar\alpha_i,\tilde\alpha_i,\bar{\tilde\alpha}_i), \quad a,\bar a\in\{0,+,-,t\}
\end{equation}
The only possible combinations of the structure constants would be:
\begin{equation}
\begin{aligned}
    C_{00}, \quad C_{0t}, \quad C_{t0}, \quad C_{tt} &\qquad \text{for the neutral sector}\\
    C_{++}, \quad C_{--} &\qquad \text{for the charged sector}
\end{aligned}
\end{equation}

A subtlety arises from the spin convention used in the literature on coset structure constants. There, the spin $j_H$ corresponds to the conformal weight $-\frac{1}{k}j_H(j_H+1)$, whereas our convention gives $-\frac{1}{k}j(j-1)$. Matching these expressions admits two possible identifications, $j_H=-j$ and $j_H=j-1$, related by the reflection $j_H\mapsto-j_H-1$. As we show below, our consistency conditions select one of these two identifications. The neutral sector structure constant $C_{00}(\alpha_i, \bar\alpha_i,\tilde\alpha_i,\bar{\tilde\alpha}_i)$ is the spectral-flow-preserving structure constant~\eqref{eq:SFO-structure constant} in $SL(2,\mathbb{R})/U(1)$ model under the dictionary $j=\frac{b}{2}(\alpha+\tilde\alpha)=\frac{b}{2}(\bar\alpha+\bar{\tilde\alpha})$ and $m=\frac{b}{2}(\alpha-\tilde\alpha), \bar m=\frac{b}{2}(\bar\alpha-\bar{\tilde\alpha})$ with the identification $j_H=-j$ \cite{Hosomichi:2004ph}:
\begin{multline}
    C_{00}(\alpha_i, \bar\alpha_i,\tilde\alpha_i,\bar{\tilde{\alpha}}_i)=\delta(\sum_i\alpha_i=\sum_i\tilde\alpha_i)\delta(\sum_i\bar\alpha_i=\sum_i\bar{\tilde\alpha}_i)F(\alpha_i, \bar\alpha_i,\tilde\alpha_i,\bar{\tilde{\alpha}}_i)D_0(\alpha_i, \bar\alpha_i,\tilde\alpha_i,\bar{\tilde{\alpha}}_i)\\
    \text{with} \quad  D_0(\alpha_i, \bar\alpha_i,\tilde\alpha_i,\bar{\tilde{\alpha}}_i)=\frac{\tilde{r}(b)\Upsilon^{'}(0)\prod_{i=1}^3\Upsilon(\alpha_i+\tilde\alpha_i+b)}{\Upsilon
    (b+\frac{1}{2}\sum_i^3(\alpha_i+\tilde\alpha_i)-b^{-1})\prod_{(i,j,k)}\Upsilon(\frac{1}{2}(\alpha_{ij}^k+\tilde\alpha_{ij}^k)+b)}
    \label{eq:neutral-structure-constant}
\end{multline}
where $\alpha_{ij}^k\equiv\alpha_i+\alpha_j-\alpha_k$, $\tilde\alpha_{ij}^k\equiv\tilde\alpha_i+\tilde\alpha_j-\tilde\alpha_k$ and $(i,j,k)$ represents all possible cyclic combinations. $\Upsilon\equiv\Upsilon_b$ is defined in~\eqref{eq:Upsilon func}. The normalization factor follows $\tilde{r}(b)=\frac{(\mu^{2b^{-2}} b^{2b^{-2}})^{-\frac{b}{2}\sum_{i=1}^3(\alpha_i+\tilde\alpha_i)+1}}{\sqrt{2}b^{-2}}$ and $F(\alpha_i, \bar\alpha_i,\tilde\alpha_i,\bar{\tilde{\alpha}}_i)$ has the expression:
\begin{multline}
    F(\alpha_i, \bar\alpha_i,\tilde\alpha_i,\bar{\tilde{\alpha}}_i)\equiv \pi^{-2}\int d^2zd^2w z^{-b\alpha_1}\bar z^{-b\bar\alpha_1}(1-z)^{-b\alpha_2}(1-\bar z)^{-b\bar\alpha_2}\\
    \times w^{-b\tilde\alpha_1}\bar w^{-b\bar{\tilde\alpha}_1}(1-w)^{-b\tilde\alpha_2}(1-\bar w)^{-b\bar{\tilde{\alpha}}_2}|z-w|^{-4+b\sum_{i=1}^3(\alpha_i+\tilde\alpha_i)}
    \label{eq:F-neutral}
\end{multline}
where $\bar\alpha_i$ and $\bar{\tilde\alpha}_i$ are the parameters of vertex operators in the anti-holomorphic sector. The charged sector structure constants $C_{\pm\pm}(\alpha_i, \bar\alpha_i,\tilde\alpha_i,\bar{\tilde{\alpha}}_i)$ are given by the spectral-flow-violating structure constant with $\sum w=\pm1$ \cite{Fukuda:2001jd,Hosomichi:2004ph}, evaluated on the shifted labels~\eqref{eq:shifted-argu C} and multiplied by the factors of~\eqref{eq:OPE-G^pm}: 
\begin{multline}
    C_{\pm\pm}(\alpha_i, \bar\alpha_i,\tilde\alpha_i,\bar{\tilde{\alpha}}_i)=\frac{2}{b^2}\delta(\sum_i(\alpha_i-\tilde\alpha_i)\pm b)\delta(\sum_i(\bar\alpha_i-\bar{\tilde\alpha}_i)\pm b)F_{\pm}(\alpha_i, \bar\alpha_i,\tilde\alpha_i,\bar{\tilde{\alpha}}_i)\\
    \times \frac{r(b)\Upsilon'(0)\prod_{i=1}^3\Upsilon(\alpha_i+\tilde\alpha_i+b)}{\Upsilon(\frac{b}{2}+\frac{1}{2}\sum_{i=1}^3(\alpha_i+\tilde\alpha_i)-b^{-1})\prod_{(i,j,k)}\Upsilon(\frac{1}{2}(\alpha_{ij}^k+\tilde\alpha_{ij}^k)+\frac{b}{2})}
    \label{eq:charged-structure-constant}
\end{multline}
where the normalization factor $r(b)=\frac{(\mu^{2b^{-2}} b^{-2+2b^{-2}})^{-\frac{b}{2}\sum_{i=1}^3(\alpha_i+\tilde\alpha_i)+1}}{\sqrt{2}b^{-1-b^2}}$ and $\tilde r(b)$ are those of \cite{Hosomichi:2004ph} with $b\to b^{-1}$, and $F_{\pm}(\alpha_i, \bar\alpha_i,\tilde\alpha_i,\bar{\tilde{\alpha}}_i)$ are defined by:
\begin{equation}
\begin{aligned}
   F_+(\alpha_i, \bar\alpha_i,\tilde\alpha_i,\bar{\tilde{\alpha}}_i)&=(-1)^{\frac{b}{2}(\alpha_2-\tilde\alpha_2-\bar\alpha_2+\bar{\tilde\alpha}_2)}\frac{\prod_{i=1}^3\Gamma(1-b\tilde\alpha_i)}{\prod_{i=1}^3\Gamma(b\bar{\tilde{\alpha}}_i)}\\
   F_-(\alpha_i, \bar\alpha_i,\tilde\alpha_i,\bar{\tilde{\alpha}}_i)&=(-1)^{\frac{b}{2}(\alpha_2-\tilde\alpha_2-\bar\alpha_2+\bar{\tilde\alpha}_2)}\frac{\prod_{i=1}^3\Gamma(1-b\alpha_i)}{\prod_{i=1}^3\Gamma(b\bar{\alpha}_i)}
   \label{eq:Fpm}
\end{aligned}
\end{equation}

Eliminating $C_t$ from~\eqref{eq:C_t sol} gives, after gauge fixing the worldsheet coordinates $z_i$'s:
\begin{equation}
\begin{aligned}
    \alpha_a\tilde\alpha_bC_{0}(\alpha_a+\frac{1}{b},&\tilde\alpha_a-\frac{1}{b},\alpha_b-\frac{1}{b},\tilde\alpha_b+\frac{1}{b},\alpha_c,\tilde\alpha_c)+\alpha_b\tilde\alpha_aC_{0}(\alpha_a-\frac{1}{b},\tilde\alpha_a+\frac{1}{b},\alpha_b+\frac{1}{b},\tilde\alpha_b-\frac{1}{b},\alpha_c,\tilde\alpha_c)\\
    &=(h_a+h_b-h_c)C_0(\alpha_a,\tilde\alpha_a,\alpha_b,\tilde\alpha_b,\alpha_c,\tilde\alpha_c)\\
    (b\tilde\alpha_c-1)C_0&(\alpha_a,\tilde\alpha_a,\alpha_b,\tilde\alpha_b,\alpha_c,\tilde\alpha_c)+b\tilde\alpha_aC_0(\alpha_a-\frac{1}{b},\tilde\alpha_a+\frac{1}{b},\alpha_b,\tilde\alpha_b,\alpha_c+\frac{1}{b},\tilde\alpha_c-\frac{1}{b})\\
    &+b\tilde\alpha_bC_0(\alpha_a,\tilde\alpha_a,\alpha_b-\frac{1}{b},\tilde\alpha_b+\frac{1}{b},\alpha_c+\frac{1}{b},\tilde\alpha_c-\frac{1}{b})=0\\
    (b\alpha_c-1)C_0&(\alpha_a,\tilde\alpha_a,\alpha_b,\tilde\alpha_b,\alpha_c,\tilde\alpha_c)+b\alpha_aC_0(\alpha_a+\frac{1}{b},\tilde\alpha_a-\frac{1}{b},\alpha_b,\tilde\alpha_b,\alpha_c-\frac{1}{b},\tilde\alpha_c+\frac{1}{b})\\
    &+b\alpha_bC_0(\alpha_a,\tilde\alpha_a,\alpha_b+\frac{1}{b},\tilde\alpha_b-\frac{1}{b},\alpha_c-\frac{1}{b},\tilde\alpha_c+\frac{1}{b})=0
    \label{eq:consistency check}
\end{aligned}
\end{equation}
where we have used the action of $G^{\pm}_{-1/2}$ in~\eqref{eq:OPE-G^pm}. The shifted arguments arise because, for $\sum_i\eta_i=0$, the structure constants of spectral flowed operators are those of the unflowed ones with shifted labels (see Appendix~\ref{app:superconformal Ward identities}):
\begin{equation}
    \langle V_{\alpha_3,\tilde\alpha_3}^{\eta_3}V_{\alpha_2,\tilde\alpha_2}^{\eta_2}V_{\alpha_1,\tilde\alpha_1}^{\eta_1}\rangle\sim C_0(\alpha_1-\frac{\eta_1}{b},\tilde\alpha_1+\frac{\eta_1}{b},\alpha_2-\frac{\eta_2}{b},\tilde\alpha_2+\frac{\eta_2}{b},\alpha_3-\frac{\eta_3}{b},\tilde\alpha_3+\frac{\eta_3}{b})
    \label{eq:shifted-argu C}
\end{equation} 
Since the shifts induced by $\eta$ and $\bar{\eta}$ leave the spins $j_i$ unchanged, $D_0(\alpha_i, \bar\alpha_i,\tilde\alpha_i,\bar{\tilde{\alpha}}_i)$ acts as an overall constant in the consistency check~\eqref{eq:consistency check}. The nontrivial requirement is therefore that the integrand of $F(\alpha_i, \bar\alpha_i,\tilde\alpha_i,\bar{\tilde{\alpha}}_i)$ from equation~\eqref{eq:F-neutral} satisfies these conditions with the appropriate polynomial factors. In the same way, the remaining neutral components follow from $C_{00}$ by shifting the holomorphic and anti-holomorphic labels independently. With $C_t$ defined by the first line of~\eqref{eq:C_t sol},
\begin{equation}
    C_{t0}=-2\alpha_2\tilde\alpha_1\,C_{00}\big|_{S},\qquad C_{0t}=-2\bar\alpha_2\bar{\tilde\alpha}_1\,C_{00}\big|_{\bar S},\qquad C_{tt}=4\alpha_2\tilde\alpha_1\bar\alpha_2\bar{\tilde\alpha}_1\,C_{00}\big|_{S,\bar S}
\end{equation}
where $S:(\alpha_1,\tilde\alpha_1,\alpha_2,\tilde\alpha_2)\to(\alpha_1-\frac{1}{b},\tilde\alpha_1+\frac{1}{b},\alpha_2+\frac{1}{b},\tilde\alpha_2-\frac{1}{b})$ and $\bar S$ is the same shift of the barred labels. Equivalently, $S$ inserts $\frac{z(1-w)}{w(1-z)}$ in the integrand of $F$~\eqref{eq:F-neutral}, and $\bar S$ its anti-holomorphic counterpart. 

The integrand of $F(\alpha_i,\bar\alpha_i,\tilde\alpha_i,\bar{\tilde{\alpha}}_i)$ factorizes into holomorphic and anti-holomorphic parts of the same functional form. It therefore suffices to verify the consistency condition in one sector; the corresponding condition in the other follows by the appropriate replacement of parameters. We set $F(\alpha_i,\bar\alpha_i,\tilde\alpha_i,\bar{\tilde{\alpha}}_i)=\pi^{-2}\int d^2z d^2w I\cdot \bar I$ with $I\equiv z^{-b\alpha_1}(1-z)^{-b\alpha_2}w^{-b\tilde\alpha_1}(1-w)^{-b\tilde\alpha_2}(z-w)^{-2+\frac{b}{2}\sum_{i=1}^3(\alpha_i+\tilde\alpha_i)}$ and $\bar I$ the anti-holomorphic counterpart. Then the first equation in~\eqref{eq:consistency check} can be written in terms of integrand $I$ as:
\begin{equation}
\begin{aligned}
   (a,b)=(1,2) \quad & \int I\big(\tilde\alpha_1\alpha_2\frac{z}{w}\frac{1-w}{1-z}+\alpha_1\tilde\alpha_2\frac{w}{z}\frac{1-z}{1-w}-(h_1+h_2-h_3)\big)=0\\
   (a,b)=(1,3) \quad & \int I\big(\tilde\alpha_1\alpha_3\frac{z}{w}+\alpha_1\tilde\alpha_3\frac{w}{z}-(h_1+h_3-h_2)\big)=0\\
   (a,b)=(2,3) \quad & \int I(\tilde\alpha_2\alpha_3\frac{1-z}{1-w}+\alpha_2\tilde\alpha_3\frac{1-w}{1-z}-(h_2+h_3-h_1))=0
   \label{eq:1-consistency}
\end{aligned}
\end{equation}
The second equation of~\eqref{eq:consistency check} can be written as:
\begin{equation}
\begin{aligned}
    &\int I\big((b\tilde\alpha_1-1)+b\tilde\alpha_2\frac{w}{z}\frac{1-z}{1-w}+b\tilde\alpha_3\frac{w}{z}\big)=0\\
    &\int I\big((b\tilde\alpha_2-1)+b\tilde\alpha_1\frac{z}{w}\frac{1-w}{1-z}+b\tilde\alpha_3\frac{1-w}{1-z}\big)=0\\
    & \int I\big((b\tilde\alpha_3-1)+b\tilde\alpha_1\frac{z}{w}+b\tilde\alpha_2\frac{1-z}{1-w}\big)=0
    \label{eq:2-consistency}
\end{aligned}
\end{equation}
And the third equation in~\eqref{eq:consistency check} can be rewritten in the similar form as above but with $z\leftrightarrow w$ and $\tilde\alpha\leftrightarrow \alpha$. Furthermore, the equations~\eqref{eq:1-consistency} can be obtained from~\eqref{eq:2-consistency} and its dual equations with $\alpha$'s, which means that we only need to check whether $I$ satisfies the equations~\eqref{eq:2-consistency} and its dual. And it turns out that the integrands in~\eqref{eq:2-consistency} can be written as total derivative of $w$ for~\eqref{eq:2-consistency} and of $z$ for its dual. More explicitly, the first line in~\eqref{eq:2-consistency} is proved to be equivalent as:
\begin{equation}
    \int \partial_w\Phi_1(z,w)=\int \partial_w (I\frac{w(w-z)}{z})=0
\end{equation}
for the second and third equations in~\eqref{eq:2-consistency}, the generating functions $\Phi_i(z,w)$ are:
\begin{equation}
    \Phi_2(z,w)=I\frac{(1-w)(w-z)}{1-z}, \qquad \Phi_3(z,w)=I(w-z)
\end{equation}
The dual equations of~\eqref{eq:2-consistency} follow in the same way, with $z\leftrightarrow w$ and $\alpha_i\leftrightarrow\tilde\alpha_i$. This shows that $I$, with the identification $j_H=-j$, satisfies the Ward identities of the $\mathcal{N}=2$ superconformal algebra; with the other identification, $j_H=j-1$, the integrand is obtained by $b\tilde\alpha_i\to1-b\alpha_i$, $b\alpha_i\to1-b\tilde\alpha_i$ and does not satisfy~\eqref{eq:2-consistency}. The Ward identities do not constrain the factor $D_0$, nor the relative normalization between the neutral and charged sectors. We test both numerically through crossing symmetry in the next subsection.

\subsection{Numerical Check for Crossing Symmetry}
\label{subsec:4-pt and crossing symmetry}

To check the crossing symmetry, we start from the four-point function:
\begin{equation}
    G_4=\bra{h_4}\mathcal{V}^{\eta=0}_{\alpha_3,\tilde\alpha_3}(1,\theta_3^{\pm})\mathcal{V}^{\eta=0}_{\alpha_2,\tilde\alpha_2}(z,\theta_2^{\pm})\ket{h_1}
\end{equation}
with out-state $\bra{h_4}\equiv\lim_{\omega\rightarrow\infty}\omega^{2h_4}\bra{0}\phi_4(\omega)$ and in-state $\ket{h_1}\equiv \phi_1(0)\ket{0}$. We have fixed the gauge at $z_2=z,z_3=1$ and $\theta_1^{\pm}=\theta_4^{\pm}=0$ using Ward identities. Combining with the anti-holomorphic part, the four-point function in the bosonic sector reads:
\begin{equation}
    G_4(z,\bar z)=\bra{h_4,\bar h_4}\phi_3(1,1)\phi_2(z,\bar z)\ket{h_1,\bar h_1}
\end{equation}

To compute the four-point function, we insert an identity operator and introduce intermediate states, where we can make use of the structure constants data. The complete spectrum of the theory, as we reviewed in Section~\ref{sec:N=2 Liouville} and Section~\ref{sec:Relation to SL(2,R) WZW}, contains both massive representations and massless representations. The identity operator is written as:
\begin{multline}
    \mathbf{1}=\sum_{w,\lambda\in\mathbb{Z}}\bigg[\int_0^{\infty}\frac{dP}{B(P)}\sum_{A,\bar A,B,\bar B}A\bar A\ket{h_P,\bar h_P}_{w,\lambda}(K^{-1})_{AB}(\bar K^{-1})_{\bar A \bar B}\,{}_{w,\lambda}\bra{h_P,\bar h_P}B^{\dagger}\bar B^{\dagger}\\
    +\sum_{\substack{n\ge 0\\ \frac12<j_n<\frac{1+b^2}{2}}}\frac{1}{B_{P    _n;w,\lambda}}\Pi_{P_n;\,w,\lambda}\bigg],
\qquad P_n=\frac{i\,(j_n-\tfrac12)}{b}
\label{eq:general-identity-op}
\end{multline}
where the second term is given by the residue of poles $P_n$ sitting in the window $\frac12<j_n<\frac{1+b^2}{2}$, constituting the massless series in the spectrum. Here $w,\lambda$ are integers parameterizing $U(1)$ charge of the primary state via $Q=\frac{w}{b^2}+\lambda$ and $\bar Q=\lambda-\frac{w}{b^2}$. And physical states in massless series require that the $U(1)$ charges in holomorphic and anti-holomorphic sectors have the same sign, giving the condition for $(w,\lambda)$ as $b^2|\lambda|>|w|$\footnote{In $SL(2,\mathbb{R})$ WZW model, the zero-mode Hilbert space is $L^2(SL(2,\mathbb{R}))$, and the Plancherel theorem tells that its decomposition is a direct integral of $\mathcal{C}_j\otimes\mathcal{C}_j$ over the continuous series plus a sum over the discrete series of $\mathcal{D}_j^+\otimes \mathcal{D}_j^+$ and $\mathcal{D}^-_{j}\otimes \mathcal{D}_j^-$. A state in $\mathcal{D}^+\otimes\mathcal{D}^-$ would not satisfy square-integrability and is not a bound state, thus does not belong to the spectrum.}. The configurations are chosen such that no pole $P_n$ lies in this window for any internal state $\ket{h_P,\bar h_P}_{w,\lambda}$ in either channel, i.e. whenever $m$ and $\bar m$ have the same sign, $\min(|m|,|\bar m|)-n\notin(\frac12,\frac{1+b^2}{2})$ for all integers $n\geq0$, with $m=\frac{1}{2}(w+b^2\lambda)$ and $\bar m=\frac{1}{2}(b^2\lambda-w)$. And we only insert the first term of~\eqref{eq:general-identity-op} into the four-point function $G_4(z,\bar z)$, where $A,B$ and $\bar A,\bar B$ are monomials of descendant operators of $\mathcal{N}=2$ superconformal algebra. $K_{AB}$ and $K_{\bar A\bar B}$ are the Gram matrices which satisfy:
\begin{equation*}
   K_{AB}=\bra{h_P}A^{\dagger}B\ket{h_P}, \quad \bar K_{\bar A\bar B}=\bra{\bar h_P}\bar A^{\dagger}\bar B\ket{\bar h_P} 
\end{equation*}
A complete spectrum sums over the monomials $A,B\in \mathcal{B}(N,Q)$ and $\bar A,\bar B\in\mathcal{B}(\bar N,\bar Q)$ at all levels $N,\bar N$ and charges $Q,\bar Q$, but for numerical test we set a truncation for the level $N$ and $\bar N$. Plugging in the identity operator, the four-point function becomes:
\begin{multline}
    G_4(z,\bar z)=\sum_{\delta=0,\pm1}\int \frac{dP}{B(P)} \sum_{N,\bar N}\sum_{A,\bar A,B,\bar B}\bra{h_4,\bar h_4}\phi_3(1,1)A\bar A\ket{h_P,\bar h_P}_{w_P,\lambda_P}(K^{-1})_{AB}(\bar K^{-1})_{\bar A\bar B}\\
    \times{}_{w_P,\lambda_P}\bra{h_P,\bar h_P}B^{\dagger}\bar B^{\dagger}\phi_2(z,\bar z)\ket{h_1,\bar h_1}
\end{multline}
with $w_P=w_1+w_2$ and $\lambda_P=\lambda_1+\lambda_2+\delta$ following the charge conservation of three-point functions. The measure $B(P)$ can be obtained from the structure constant $C_{00}$ by:
\begin{equation*}
    B(P)\delta(P-P')=\lim_{\epsilon\rightarrow0^+}C_{00}(\{\alpha_P\},\{\alpha_{P'}\},\{\alpha_{\epsilon}\})
\end{equation*}
with $\alpha_P+\tilde\alpha_P=\bar\alpha_P+\bar{\tilde\alpha}_P=b^{-1}+2iP$, $\alpha_{P'}+\tilde\alpha_{P'}=\bar\alpha_{P'}+\bar{\tilde\alpha}_{P'}=b^{-1}-2iP'$ and $\alpha_{\epsilon}+\tilde\alpha_{\epsilon}=\bar\alpha_{\epsilon}+\bar{\tilde\alpha}_{\epsilon}=2\epsilon$. Plugging these parameters into the structure constant $C_{00}$~\eqref{eq:neutral-structure-constant} and~\eqref{eq:F-neutral}, the measure is simply $B(P)=\sqrt{2}\pi$ if we set $P'=P-\Delta P$ and take the limit $\Delta P\rightarrow 0,\,\epsilon\rightarrow 0^+$. The matrix elements factorize into a linear combination of the vectors, whose elements correspond to the four components of three-point function in superspace fixed by Ward identities. The three-point functions can also be written as a linear combination of these vectors with coefficients given by the structure constants derived before. Then the four-point function can be rewritten as\footnote{The three-point functions can be written as:
\begin{equation*}
\begin{aligned}
    \bra{h_4,\bar h_4}\phi_3(1,1)A\bar A\ket{h_P,\bar h_P}_{w_P,\lambda_P}&=\sum_{a,\bar a}u_A^a \bar u_{\bar A}^{\bar a}D_s^{a\bar a}\\
    {}_{w_P,\lambda_P}\bra{h_P,\bar h_P}B^{\dagger}\bar B^{\dagger}\phi_2(z,\bar z)\ket{h_1,\bar h_1}&=z^{h_P+N-h_1-h_2}\bar z^{\bar h_P+\bar N-\bar h_1-\bar h_2}\sum_{b,\bar b}v_B^b\bar v_{\bar B}^{\bar b}C_s^{b\bar b}
\end{aligned}
\end{equation*}}:
\begin{equation}
\begin{aligned}
    G_s(z,\bar z)&=\sum_{\delta=0,\pm1}\int_0^{\infty}dP\sum_{a,\bar a,b,\bar b}D_s^{a\bar a}M_{s,ab}^{\delta}(z)C_s^{b\bar b}\bar M_{s,\bar a\bar b}^{\delta}(\bar z)\\
    \text{with} \quad M_{s,ab}^{\delta}&=z^{h_P-h_1-h_2}\sum_N z^N\sum_{A,B\in\mathcal{B(N,-\delta)}}u_A^a(K_{N}^{-1})_{AB}v_{B}^b, \quad \{a,\bar a,b,\bar b\}=\{0,+,-,t\}
\end{aligned}    
\end{equation}

A four-point function can be decomposed in two ways: fuse the fields at 0 and $z$ ($s$-channel), or the fields at $z$ and 1 ($t$-channel), obtained from the $s$-channel by exchanging the fields $1\leftrightarrow3$ and $z\leftrightarrow 1-z$ in our case. The four-point function in $t$-channel is given by:
\begin{equation}
    G_t(1-z,1-\bar z)=\sum_{\delta=0,\pm1}\int_0^{\infty}dP\sum_{a,\bar a,b,\bar b}D_t^{a\bar a}M_{t,ab}^{\delta}(1-z)C_t^{b\bar b}\bar M_{t,\bar a\bar b}^{\delta}(1-\bar z)
\end{equation}
In the numerical tests, we sum over the intermediate primary fields with momentum $P$ up to $P_{m}$ with step $h_s$ and the basis of Gram matrix $K^{-1}$ up to level $N_c=\bar N_c$. We also set numerical data for external fields momentum $(P_1,P_2,P_3,P_4)$ and charges given by $(w_i,\lambda_i)$. Charge conservation gives a constraint $\sum w_i=\sum\lambda_i=0$. With these data, we set the configuration $(b,P_i,w_i,\lambda_i)$, the cross-ratio $z$, the block level $N_c$, the momentum step $h_s$ and cutoff $P_m$. With crossing symmetry, one can fit the relative normalization $\kappa_f$ of the charged sector. Schematically, the four-point function can be written as a sum of neutral part and charge part by:
\begin{equation}
\begin{aligned}
    G_s&=S_0+\kappa_f S_c\\
    G_t&=T_0+\kappa_f T_c
\end{aligned}
\end{equation}
where $S_0\, (T_0)$ are given by the sum over 16 neutral terms $D^{a\bar a}M_{ab}^{0}C_t^{b\bar b}\bar M_{\bar a\bar b}^{0}$ with $(a,\bar a,b,\bar b)=(0,t)$, and $S_c\, (T_c)$ are given by the sum over two possible charged terms $D^{++}M_{+-}^{-1}C^{--}\bar M_{+-}^{-1}$ and $D^{--}M_{-+}^{1}C_t^{++}\bar M_{-+}^{1}$. With the structure constants given by~\eqref{eq:neutral-structure-constant} and~\eqref{eq:charged-structure-constant}, we solve for $\kappa_f$ independently in each of 100 data sets (five configurations of external momenta and charges, three of them with spinning fields, ten values of $b$ each, and $z=\frac12,\frac25$; $N_c=6$ or $7$, $P_m=5$). We find $|\kappa_f-1|<10^{-3}$ for 79 of them, while the neutral sector alone typically violates crossing at the level of $10\%$ or more. The remaining 21 sets, all at $z=\frac25$, give $|\kappa_f-1|$ up to $2.1\times 10^{-3}$, but this deviation decreases with each additional level of the block expansion and extrapolates to $\kappa_f=1$ within $10^{-4}$. The code and the full data are provided as ancillary files.

\section{Discussion}
\label{sec:discussion}

In this paper we have proposed the structure constants of spacelike $\mathcal N=2$ Liouville theory. Our results hold in the full quantum theory, beyond the semiclassical regime. Motivated by its duality with the supersymmetric coset $SL(2,\mathbb R)/U(1)$, we derived the structure constants on the Liouville side algebraically. We then assembled these data into superspace and verified that they are consistent with crossing symmetry numerically.

The main structural difference from the DOZZ formula and its $\mathcal N=1$ analogue stems from the $U(1)_R$ symmetry. Superconformal primaries are labeled by both a momentum $P$ and a $U(1)_R$ charge $Q$, and the structure constants derived in Section~\ref{sec:correlators in superspace} are functions of both labels. Moreover, the $U(1)_R$ current, realized in part by the compact boson $Y$, generates a continuous spectral flow relating the NS and R sectors, so that the structure constants in the two sectors are not independent. This contrasts with $\mathcal N=1$ Liouville theory, where the NS and R structure constants must be determined separately.

Given the wide range of applications of the structure constants of $\mathcal N=0$ and $\mathcal N=1$ Liouville theory, we conclude by discussing several directions for future work.

\paragraph{Timelike $\mathcal{N}=2$ Liouville theory}

A natural extension is to develop timelike $\mathcal{N}=2$ Liouville theory with $c < 3$, corresponding to $\hat Q_L^2<0$ (equivalently $b\rightarrow i\hat b$, $\hat Q_L=1/\hat b$ and $c=3+6\hat{Q}_L^2=3-6/b^2$). The bosonic case shows that this cannot be achieved by analytically continuing the structure constants themselves because DOZZ formula has a natural boundary at imaginary $b$. However, the shift equations that determine it continue to $c < 1$, where they admit a different solution containing crossing symmetry \cite{Zamolodchikov:2005fy,Kostov:2005kk,Ribault:2015sxa}. The analogue of $\mathcal{N}=1$ Liouville theory was addressed in~\cite{Muhlmann:2025ngz}. In the $\mathcal{N}=2$ case, we therefore expect the timelike structure constants to arise as solutions of the bootstrap equations obtained by the analytical continuation. Our derivation for spacelike structure constants also suggests an independent approach from the coset perspective. Just as the spacelike theory is dual to the supercoset $SL(2,\mathbb{R})/U(1)$, its $c_t<3$ timelike counterpart may be related to the compact supercoset $SU(2)/U(1)$ at non-rational level, analogous to the realization of timelike Liouville theory through non-rational $SU(2)/U(1)$ coset \cite{Jaskolski:2015uya}. One can try to derive the structure constants for timelike $\mathcal{N}=2$ Liouville theory.

\paragraph{$\mathcal{N}=2$ $\widehat{\textrm{VMS}}$}

With timelike and spacelike $\mathcal{N}=2$ Liouville theory, one can explore a generalization of $\mathcal{N}=2$ Virasoro minimal string. The bosonic Virasoro minimal string (VMS) is a critical string theory in two dimensions, whose worldsheet CFT consists of a spacelike Liouville theory with $c\geq 25$ and timelike Liouville theory with $c\leq 1$ together with $bc$ ghosts \cite{Collier:2023cyw}. Its duality to a double-scaled matrix integral, which holds for any value of continuous parameter $c$, reduces to the JT gravity/matrix integral duality in the large central charge limit. Its $\mathcal{N}=1$ supersymmetric generalization was initiated in~\cite{Rangamani:2025wfa} by coupling $\mathcal{N}=1$ spacelike Liouville theory with $c>27/2$ and an analogous timelike Liouville with $c<3/2$ to two-dimensional $\mathcal{N}=(1,1)$ supergravity on the worldsheet, see also \cite{Collier:2026bhh} for subsequent developments. Such worldsheet strings are expected to be related to deformations of matrix models dual to JT supergravity analyzed in detail in~\cite{Stanford:2019vob}. The $\mathcal{N}=2$ analogue would couple spacelike $\mathcal{N}=2$ Liouville theory and its timelike counterpart, with central charges summing to 6, to $\mathcal{N}=2$ supergravity on the worldsheet, where the $U(1)_R$ symmetry is gauged and spectral flow relates the NS and R sectors. The structure constants obtained here, together with their timelike counterpart, could provide the ingredients for a worldsheet description of these models. In the semiclassical limit, one would expect such theories to reduce to $\mathcal{N}=2$ JT supergravity, whose matrix-model description was developed in \cite{Turiaci:2023jfa}. A worldsheet description would provide a one-parameter deformation of $\mathcal N=2$ JT supergravity that remains exactly solvable beyond the semiclassical regime.

\paragraph{3d (super)gravity}

Another direction concerns the three-dimensional gravity. Worldsheet theories built from pairs of Liouville CFTs have been related to three-dimensional gravity: the VMS is closely related to 3d gravity \cite{Collier:2023cyw,Collier:2023fwi} through Virasoro TQFT, and the complex Liouville string to de Sitter quantum gravity in 2+1 dimensions \cite{Collier:2025lux}. More recently, the $\mathcal{N}=1$ super Virasoro minimal string was shown to arise from the quantization of 3d supergravity \cite{Eberhardt:2026hfh}. It is natural to ask whether this picture extends to $\mathcal{N}=2$. The structure constants obtained in this work would provide ingredients for such a construction. Three-dimensional $\mathcal{N}=(2,2)$ supergravity also admits near-BPS black holes. It would be interesting to see whether such a framework would shed light on them.

\paragraph{NS5-branes and little string theory}

Another setting is the type II superstring, where $\mathcal{N}=2$ Liouville theory is coupled to $\mathcal{N}=1$ worldsheet supergravity and appears as part of the target space. The main example is the near-horizon geometry of $k$ NS5-branes arranged on a circle, described by the worldsheet theory $\mathbb{R}^{5,1}\times\big(SL(2,\mathbb{R})_k/U(1)\times SU(2)_k/U(1)\big)/\mathbb{Z}_k$, which is holographically dual to little string theory in a double scaling limit \cite{Giveon:1999px,Giveon:1999tq,Israel:2004ir}. Here $SU(2)_k/U(1)$ is the $\mathcal{N}=2$ minimal model, and the supercoset $SL(2,\mathbb{R})_k/U(1)$, equivalently $\mathcal{N}=2$ Liouville theory with $b^2=k$, caps off the throat. Our structure constants therefore enter directly into the correlation functions of this background. Similar backgrounds describe string theory near Calabi-Yau singularities \cite{Ooguri:1995wj,Giveon:1999zm}, and it would be interesting to explore possible applications of our results in such settings.

\paragraph{2d black holes}

Through the mirror symmetry, $\mathcal{N}=2$ Liouville theory is related to the fermionic 2d black hole \cite{Witten:1991yr,Hori:2001ax}. This exactly solvable background has served as a useful setting for studying black hole thermodynamics, Hagedorn behavior and the transition between black holes and strings \cite{Kutasov:2000jp,Chen:2021dsw}. It would be interesting to explore whether our work can provide further insights into these questions.

\acknowledgments

The author would like to give special thanks to Lorenz Eberhardt for his guidance throughout this project and helpful suggestions on the presentation of this work.
We would like to thank Erik Verlinde for helpful discussions and comments on the draft.
JT is supported by the Spinoza grant and the Delta ITP consortium, both of which are programs of the Netherlands Organisation for Scientific Research (NWO) that is funded by the Dutch Ministry of Education, Culture and Science (OCW).
We acknowledge use of Opus 5 and Fable 5 (Anthropic) and Gemini Pro (Google) for assistance with calculations, the numerical crossing test and cross-checking the draft. The author takes full responsibility for the content of the paper.

\appendix
\addtocontents{toc}{\protect\setcounter{tocdepth}{1}}

\section{Structure Constants of \texorpdfstring{$H_3^+$}{H3} and \texorpdfstring{$SL(2,\mathbb{R})/U(1)$}{SL(2,R)/U(1)} Model}
\label{app: structure constant of SL(2,R)}

In this Appendix, we briefly review the $H_3^+$, $SL(2,\mathbb{R})$, and $SL(2,\mathbb{R})/U(1)$ models. We outline the essential steps in determining the structure constants of the $H_3^+$ WZNW model and show the solutions of the conformal blocks explicitly, referring to \cite{Teschner:1997ft} for the detailed derivation. We explain how the structure constants of the $SL(2,\mathbb{R})/U(1)$ coset model are obtained from those of the $H_3^+$ model. Through the parameter dictionary relating the coset CFT to $\mathcal{N}=2$ Liouville theory, these results yield the proposed Liouville structure constants whose consistency is examined in section~\ref{sec:correlators in superspace}.

\subsection{Structure Constant of \texorpdfstring{$H_3^+$}{H3} WZNW Model}
\label{app:H_3}

The coset $H_3^+\equiv SL(2,\mathbb{C})/SU(2)$ is the set of all hermitian 2$\times$2 matrices with determinant 1. We focus on the $x$-basis primary fields denoted by $\Phi_{j}(x,\bar x;z,\bar z)$. The global Ward identities associated with the symmetries generated by $j_0^a, \bar j_0^a$ and $L_n,\bar L_n$ for $n=-1,0,1$ completely fix the dependence of the two- and three-point correlation functions on the worldsheet and spacetime coordinates, up to overall function of the spins $j_i$\footnote{In this review, we follow Teschner's convention with conformal weight given by $-\frac{1}{k}j_H(j_H+1)$, but we drop the subscript $H$ for brevity. And $\alpha$'s appearing in this section are not relevant to $\alpha$'s defined in the main context.}:
\begin{multline}
    \langle\Phi_{j_2}(x_2|z_2)\Phi_{j_1}(x_1|z_1)\rangle=N(j_1)|z_1-z_2|^{4h_1}\delta^{(2)}(x_1-x_2)\delta(j_1,-1-j_2)+B(j_1)|z_1-z_2|^{4h_1}|x_1-x_2|^{4j_1}\delta(j_1,j_2)\\
    \langle\Phi_{j_3}(x_3|z_3)\Phi_{j_2}(x_2|z_2)\Phi_{j_1}(x_1|z_1)\rangle=|x_1-x_2|^{2(j_1+j_2-j_3)}|x_1-x_3|^{2(j_1+j_3-j_2)}|x_2-x_3|^{2(j_2+j_3-j_1)}\\
    \times|z_1-z_2|^{2(h_3-h_1-h_2)}|z_1-z_3|^{2(h_2-h_1-h_3)}|z_2-z_3|^{2(h_1-h_2-h_3)}C(j_1,j_2,j_3)
    \label{eq:3-pt function x-basis}
\end{multline}
The two terms in the two-point function arise due to the equivalence of representations with spin $j$ and spin $-1-j$. Assume the operators to be normalized by 
\begin{equation}
    N(j)\equiv1 \quad \text{such that}\quad R(j)=B(j)
\end{equation}
where $R(j)$ is the reflection amplitude and can be fixed from the structure constant. Furthermore, we assume $C(j_1,j_2,j_3)$ to be symmetric in its variables as is necessary for the primary fields to be mutually local. Four-point functions can be expanded in terms of three-point functions by using the OPE of operators $\Phi_{j_1}(x_1|z_1)$ and $\Phi_{j_2}(x_2|z_2)$. One arrives at:
\begin{multline}
    \langle\Phi_{j_1}(x_1|z_1)\Phi_{j_2}(x_2|z_2)\Phi_{j_3}(x_3|z_3)\Phi_{j_4}(x_4|z_4)\rangle\\
    =\int_{\mathcal{S}_{s}}d\mu(j_{21})C(j_4,j_3,j_{21})D(j_{21};j_2,j_1)|\mathcal{F}_{s,j_{21}}[\begin{smallmatrix}
        j_4 &j_3& j_2 &j_1\\
        x_4 &x_3 &x_2 &x_1
    \end{smallmatrix}](z_4,\cdots,z_1)|^2
\end{multline}
where $D(j;j_2,j_1)=C(-1-j,j_2,j_1)$ is determined by the structure constant if one assumes that only one of the two linearly dependent operators $\Phi_j(x|z_1)$ and $\Phi_{-1-j}(x|z_1)$ appears in the OPE of $\Phi_{j_1}(x_1|z_1)$ and $\Phi_{j_2}(x_2|z_2)$. The information of a four-point function is therefore split into the part determined directly by current algebra symmetries (the conformal blocks), the structure constant and the set $\mathcal{S}_s$ of intermediate representations. The latter is equivalent to knowledge of the fusion rules. An alternative representation of four-point function is obtained by using the OPE of $\Phi_{j_3}$ and $\Phi_{j_2}$ to get an expansion in terms of ``$t$-channel" conformal blocks:
\begin{multline}
    \langle\Phi_{j_1}(x_1|z_1)\Phi_{j_2}(x_2|z_2)\Phi_{j_3}(x_3|z_3)\Phi_{j_4}(x_4|z_4)\rangle\\
    =\int_{\mathcal{S}_{t}}d\mu(j_{32})C(j_4,j_1,j_{32})D(j_{32};j_2,j_3)|\mathcal{F}_{t,j_{32}}[\begin{smallmatrix}
        j_4 &j_3& j_2 &j_1\\
        x_4 &x_3 &x_2 &x_1
    \end{smallmatrix}](z_4,\cdots,z_1)|^2
\end{multline}
The expansion in $s$-channel and $t$-channel should produce the same four-point function, which is crossing symmetry. With the fusion relation:
\begin{equation}
    \mathcal{F}_{s,j_{21}}[\begin{smallmatrix}
        j_4 &j_3& j_2 &j_1\\
        x_4 &x_3 &x_2 &x_1
    \end{smallmatrix}](z_4,\cdots,z_1)=\int d\mu(j_{32})F_{j_{21}j_{32}}[\begin{smallmatrix}
        j_3&j_2\\
        j_4&j_1
    \end{smallmatrix}]\mathcal{F}_{t,j_{32}}[\begin{smallmatrix}
        j_4 &j_3& j_2 &j_1\\
        x_4 &x_3 &x_2 &x_1
    \end{smallmatrix}](z_4,\cdots,z_1),
\end{equation}
the structure constant follows the relation:
\begin{equation}
\begin{split}
    \int_{\mathcal{S}_s}d\mu(j_{21})F_{j_{21}j_{32}}[\begin{smallmatrix}
        j_3&j_2\\
        j_4&j_1
    \end{smallmatrix}]&\bar{F}_{j_{21}j'_{32}}[\begin{smallmatrix}
        j_3&j_2\\
        j_4&j_1
    \end{smallmatrix}]C(j_4,j_3,j_{21})D(j_{21};j_2,j_1)\\
    &=\delta(j_{32},j'_{32})C(j_4,j_{32},j_1)D(j_{32};j_2,j_3)
    \label{eq:crossing symmetry}
\end{split}
\end{equation}

Under the gauge $x_1=z_1=0$, $x_4=z_4=\infty$, $x_2=x,z_2=z$, $x_3=z_3=1$, the $s$-channel conformal block can be written as a function of cross-ratios $x,z$:
\begin{multline}
    \mathcal{F}^s_{j_{21}}[\begin{smallmatrix}
        j_4 & j_3 & j_2 & j_1\\
        x_4 & x_3 & x_2 & x_1
    \end{smallmatrix}](z_1,z_2,z_3,z_4)=|z_{42}|^{-4\Delta_2}|z_{41}|^{2(\Delta_3+\Delta_2-\Delta_1-\Delta_4)}|z_{43}|^{2(\Delta_1+\Delta_2-\Delta_3-\Delta_4)}\\
    \times|z_{31}|^{2(\Delta_4-\Delta_2-\Delta_3-\Delta_1)}|x_{43}|^{2(j_4+j_3-j_1-j_2)}|x_{42}|^{4j_2}|x_{41}|^{2(j_4+j_1-j_3-j_2)}|x_{31}|^{2(j_3+j_2+j_1-j_4)}\mathcal{F}(x,\bar x|z, \bar z)
    \label{eq:xz-conformal block}
\end{multline}
The holomorphic part of the conformal block $\mathcal{F}$ satisfies the differential equations~\eqref{eq:ODE-2},~\eqref{eq:ODE-1} and~\eqref{eq:KZ} given in section~\ref{subsec:BPZ equations}\footnote{The ODEs are the same up to different conventions between $j$ and $j_H$.}. Solving the equation~\eqref{eq:ODE-1} and KZ equation gives a set of solutions for $s$- and $t$-channel:
\begin{equation}
    \{\mathcal{F}^{2,0}_{s+},\mathcal{F}^{2,0}_{s-}\} \qquad \text{and} \qquad \{\mathcal{F}^{2,0}_{t+},\mathcal{F}^{2,0}_{t-}\}
\end{equation}
Solving another equation~\eqref{eq:ODE-2} gives three independent solutions:
\begin{equation}
    \{\mathcal{F}^{1,1}_{s+},\mathcal{F}^{1,1}_{s-},\mathcal{F}^{1,1}_{s\times}\} \qquad \text{and} \qquad \{\mathcal{F}^{1,1}_{t+},\mathcal{F}^{1,1}_{t-},\mathcal{F}^{1,1}_{t\times}\}
\end{equation}
These solutions are of hypergeometric type. With the solution of conformal blocks, we can use crossing symmetry and single-valuedness of the four-point function to uniquely fix the structure constant. Concretely, the property that the off-diagonal part of crossing symmetry equation~\eqref{eq:crossing symmetry} for $j_{2,0}=\frac{1}{2}$ reads:
\begin{equation}
    \frac{E^{2,0}_{s+}}{E^{2,0}_{s-}}=-\frac{F^{2,0}_{-+}F^{2,0}_{--}}{F^{2,0}_{+-}F^{2,0}_{++}}
\end{equation}
where $E^{2,0}_{s\sigma}\equiv C(j_4,j_3,j_1-\frac{\sigma}{2})D(j_1-\frac{\sigma}{2};\frac{1}{2},j_1)$ and RHS can be read out from the solutions of conformal blocks $\mathcal{F}^{2,0}_{s\sigma}$ and $\mathcal{F}^{2,0}_{t\sigma}$. However, to get a functional equation for $C(j_4,j_3,j_1)$, we have to further fix $C_-^{2,0}(j_1)\equiv D^{2,0}(j_1-\frac{\sigma}{2};\frac{1}{2},j_1)$ on the LHS after setting $C_+^{2,0}(\alpha_1)\equiv1$. This step requires to choose a normalization of the structure constant and depends on the consistency for dual cases $j_{2,0}$ and $j_{1,1}$, which has been explained in detail in \cite{ribault2022conformalfieldtheoryplane}. Eventually, one gets the functional equation:
\begin{equation}
    \frac{C(\alpha_4,\alpha_3,\alpha_1+b^{-1})}{C(\alpha_4,\alpha_3,\alpha_1)}=\frac{(\nu(b^{-1}))^{-1}\gamma(b^{-1}(2\alpha_1))\gamma(b^{-1}(2\alpha_1+b^{-1}))\gamma(b^{-1}(\alpha_3+\alpha_4-\alpha_1-b^{-1}))}{\gamma(b^{-1}(\alpha_1+\alpha_3-\alpha_4))\gamma(b^{-1}(\alpha_1+\alpha_4-\alpha_3))\gamma(b^{-1}(\alpha_3+\alpha_4+\alpha_1-b^{-1}))}
\end{equation}
$\nu(b^{-1})$ is the normalization factor. Following the same method, one can solve the differential equations in the case $j_{1,1}$ and get a dual functional equation from the conformal blocks $\mathcal{F}^{1,1}_{s\sigma}$ and $\mathcal{F}^{1,1}_{t\sigma}$:
\begin{equation}
    \frac{C(\alpha_4,\alpha_3,\alpha_1+b)}{C(\alpha_4,\alpha_3,\alpha_1)}=\frac{(\tilde{\nu}(b))^{-1}\gamma(b(2\alpha_1))\gamma(b(2\alpha_1+b))\gamma(b(\alpha_3+\alpha_4-\alpha_1-b))}{\gamma(b(\alpha_1+\alpha_3-\alpha_4))\gamma(b(\alpha_1+\alpha_4-\alpha_3))\gamma(b(\alpha_3+\alpha_4+\alpha_1-b^{-1}))}
\end{equation}
where $\tilde\nu(b)$ is the normalization factor. The subtlety between $SL(2,\mathbb{R})$ model and $H_3^+$ model comes from the additional independent solution $\mathcal{F}^{1,1}_{s\times},\mathcal{F}^{1,1}_{t\times}$. A four-point function $\Psi$ can be written as a linear combination of the $s$- or $t$-channel conformal blocks as follows, for simplicity we drop the subscript $(1,1)$:
\begin{equation}
    \Psi=\sum_{\sigma=+,-}E_{\sigma}\mathcal{G}_{\sigma}(x,z)\bar{\mathcal{G}}_{\sigma}(\bar x,\bar z)+E^D\mathcal{G}_{\times}(x,z)\bar{\mathcal{G}}_{\times}(\bar x,\bar z)
\end{equation}
where $\mathcal{G}_{+}\equiv\mathcal{F}_{+}$ and $\mathcal{G}_{-},\mathcal{G}_{\times}$ are the linear combinations of $\mathcal{F}_-$ and $\mathcal{F}_{\times}$ so that $\mathcal{G}_-$ has the same monodromy as the corresponding bosonic Liouville conformal blocks $\mathcal{F}_{-}^L$ and $\mathcal{G}_{\times}$ has one dimensional monodromy representation. The ratio $E_+/E_-$ is fixed by crossing symmetry. In the $H_3^+$ model, $\bar x=x^*$ and $\bar z=z^*$, and single-valuedness of $\Psi$ requires the off-diagonal terms to vanish, which fixes $E^D$ as well. In the $SL(2,\mathbb{R})$ model, $x$ and $\bar x$ are independent coordinates, so $E^D$ is not fixed by this argument and the structure constant is not unique; we return to this point in the next subsection. For $H_3^+$, the two functional equations above shift $\alpha_1$ by $b^{-1}$ and by $b$. For irrational $b^2$ these shifts are incommensurate, so a solution that is continuous in $\alpha_1$ is unique up to a factor independent of $\alpha_1$, and by the symmetry of $C$ in its arguments up to an overall constant. The result extends to rational $b^2$ by continuity in $b$. The solution is~\cite{Teschner:1997ft}:
\begin{equation}
    C^H(j_3,j_2,j_1)=\frac{C_0(b^{-1})(\nu(b^{-1}))^{j_{123}}\Upsilon(-2b^{-1}j_1)\Upsilon(-2b^{-1}j_2)\Upsilon(-2b^{-1}j_3)}{\Upsilon(-b^{-1}(j_{123}+1))\Upsilon(-b^{-1}j_{12}^3)\Upsilon(-b^{-1}j_{13}^2)\Upsilon(-b^{-1}j_{23}^1)}
    \label{eq:H3 structure constant}
\end{equation}
with $j_{123}\equiv j_1+j_2+j_3$ and $j_{ab}^c\equiv j_a+j_b-j_c$. And $\Upsilon(x)$ is defined by:
\begin{equation}
    \ln\Upsilon(x)=\int_0^{\infty}\frac{dt}{t}\left[(\frac{Q}{2}-x)^2e^{-t}-\frac{\sinh^2[\left(\frac{Q}{2}-x\right)\frac{t}{2}]}{\sinh{\frac{bt}{2}}\sinh{\frac{t}{2b}}}\right], \qquad Q=b+b^{-1}
    \label{eq:Upsilon func}
\end{equation}

\subsection{Structure Constants of \texorpdfstring{$SL(2,\mathbb{R})/U(1)$}{SL(2,R)/U(1)} Coset CFT}

We now introduce the $m$-basis functions $\Phi_{m,\bar{m}}^j(h)$ and recall how the correlators of the $SL(2,\mathbb{R})/U(1)$ model are obtained from $H_3^+$ correlators by Wick rotation.

The group element of $H_3^+$ can be parameterized using three real coordinates $(\rho,\theta,\tau)$ as:
\begin{equation}
    h\equiv\left(\begin{matrix}
        e^{\tau}\cosh\rho & e^{i\theta}\sinh\rho \\
        e^{-i\theta}\sinh\rho & e^{-\tau}\cosh\rho   \end{matrix}\right)
\end{equation}
The analytic functions on $H_3^+$ are mapped to those on $AdS_3$, the universal cover of $SL(2,\mathbb{R})$, via the map $\Phi(\rho,\theta,\tau)\in H_3^+\rightarrow \Phi(\rho,\theta,i\tau)\in AdS_3$, which is also called the Wick rotation.

There are series of functions in ``$m$-bases'', which are well suited to the Wick rotation we mentioned above. In the case of $H_3^+$, they are conventionally defined as:
\begin{equation}
    \Phi_{j,m,\bar{m}}(h)\equiv\int d^2x x^{-j-1+m}\bar{x}^{-j-1+\bar{m}}\Phi^j_x(h) \quad \text{with} \quad m-\bar{m}\in\mathbb{Z}
    \label{eq:m-basis function}
\end{equation}
The numbers $m,\bar{m}$ can be parameterized in terms of an integer $n\in\mathbb{Z}$ and a momentum $p\in\mathbb{R}$ in the $H_3^+$ model:
\begin{equation}
    m=\frac{n+ip}{2},\quad \bar{m}=\frac{-n+ip}{2}
\end{equation}
These $m$-basis functions obey the reflection property:
\begin{equation}
    \Phi_{m,\bar{m}}^j=R_{m,\bar{m}}^j\Phi^{-j-1}_{m,\bar{m}}, \quad R_{m,\bar{m}}^j=\frac{\Gamma(2j+1)\Gamma(-j+m)\Gamma(-j-\bar{m})}{\Gamma(-2j-1)\Gamma(j+1+m)\Gamma(j+1-\bar{m})}
\end{equation}
where $R_{m,\bar{m}}^j=R_{\bar{m},m}^j$ due to $m-\bar{m}\in\mathbb{Z}$.
The functions on $AdS_3$ are obtained from the above functions $\Phi_{m,\bar{m}}^j(h)$ by the Wick rotation $\tau\rightarrow i\tau$. In order for the resulting functions to be delta-function normalizable, we now need to also apply Wick rotation to the conjugate momentum $p\rightarrow -ip$, therefore the resulting $m$-basis functions on $AdS_3$ remain the same.

These $m$-base functions are complete on the space $H_3^+,SL(2,\mathbb{R})$ and $AdS_3$. While $H_3^+$ only has functions in continuous representations $\mathcal{C}^{j,\alpha}$ with $j\in -\frac{1}{2}+i\mathbb{R}_+$ and $\alpha\in [0,1)$, $SL(2,\mathbb{R})$ and $AdS_3$ also have functions in discrete representations $\mathcal{D}^{j,\pm}$. For $SL(2,\mathbb{R})$, they are $\mathcal{D}^{j,\pm}$ with $j\in\frac{1}{2}\mathbb{N}$ and $ m,\bar{m}\in\pm(j+1+\mathbb{N})$. For $AdS_3$, they are $\mathcal{D}^{j,\pm}$ with $j\in(-\frac{1}{2},\infty)$ and $m,\bar{m}\in\pm(j+1+\mathbb{N})$. The minisuperspace limit of the spectrum of the $AdS_3$ WZNW model is the space of delta-function normalizable functions on $AdS_3$. To the correspondence of interest, now we focus on the $SL(2,R)$ and $H_3^+$ cases. According to \cite{Ribault:2009ui}, the tensor product laws for $sl(2,R)$ representations under minisuperspace limit can be derived by studying the Clebsch-Gordan coefficients. These coefficients are the three-point invariants, viewed as functions $C(j_1,j_2,j_3|m_1,m_2,m_3)$ subject to equations:
\begin{equation}
    \sum_{i=1}^3(m_i+j_i+1)C(m_i+1)=\sum_{i=1}^3(m_i-j_i-1)C(m_i-1)=\sum_{i=1}^3m_iC=0
    \label{eq:sol-tensor sl(2,R)}
\end{equation}
which comes from the zero-mode Ward identity. In terms of the spin of the main text, obtained by $j_i\to j_i-1$, the first two equations are the last two lines of~\eqref{eq:consistency check} with $b\alpha_i=j_i+m_i$ and $b\tilde\alpha_i=j_i-m_i$. In the case of three continuous representations, there are two linearly independent solutions for the equation~\eqref{eq:sol-tensor sl(2,R)}. In the case when at least one representation is discrete, say $\mathcal{D}^{j_1,+}$, there is only 1 linearly independent solution because of the constraint $m_1\in j_1+1+\mathbb{N}$.

Now we turn to the $SL(2,\mathbb{R})/U(1)$ coset model. It was shown in \cite{Ribault:2005ms} that the $H_3^+$ correlators in the $m$-basis are, up to normalization, $SL(2,\mathbb{R})/U(1)$ coset correlators with three continuous representations. Applying the transformation~\eqref{eq:m-basis function} to the $H_3^+$ three-point function gives the spectral flow preserving structure constant. The spectral flow violating ones, with $\sum_i w_i=\pm1$, do not follow from $H_3^+$ and require a separate computation \cite{Fukuda:2001jd,Maldacena:2001km}; we use the expressions and normalizations of \cite{Hosomichi:2004ph} with $b\to b^{-1}$. From here on we follow the conventions of~\cite{Hosomichi:2004ph}, in which the spin is related to the one used in Section~\ref{app:H_3} by $j\to-j-1$. In the spectral flow preserving case the structure constant is\footnote{The cosmological constant $\mu,\bar\mu$ in the $\mathcal{N}=2$ Liouville action is related to $\nu$ via $\mu=\bar\mu=\nu^{b^2/2}$.}:
\begin{multline}
  \delta^{(2)}(\sum_im_i)F(j_i,m_i,\bar{m}_i)D_0(j_1,j_2,j_3)  \quad \text{with}\\
  D_0(j_1,j_2,j_3)=\frac{\tilde{r}(b)\Upsilon'(0)\Upsilon(b^{-1}(2j_1+1))\Upsilon(b^{-1}(2j_2+1))\Upsilon(b^{-1}(2j_3+1))}{\Upsilon(b-b^{-1}(j_{123}+1))\Upsilon(b-b^{-1}j_{23}^1)\Upsilon(b-b^{-1}j_{31}^2)\Upsilon(b-b^{-1}j_{12}^3)}
  \label{eq:SFO-structure constant}
\end{multline}
where $j_{123}\equiv j_1+j_2+j_3$, $j_{ab}^c\equiv j_a+j_b-j_c$, $\tilde{r}(b)=\frac{(\nu b^{2b^{-2}})^{j_{123}+1}}{\sqrt{2}b^{-2}}$ and $F(j_i,m_i,\bar m_i)$ is defined by:
\begin{multline}
    F(j_i,m_i,\bar{m}_i)=\pi^{-2}\int d^2zd^2\omega z^{j_1-m_1}\bar{z}^{j_1-\bar{m}_1}(1-z)^{j_2-m_2}(1-\bar{z})^{j_2-\bar{m}_2}\\
\times \omega^{j_1+m_1}\bar{\omega}^{j_1+\bar{m}_1}(1-\omega)^{j_2+m_2}(1-\bar{\omega})^{j_2+\bar{m}_2}|z-\omega|^{-4-2j_{123}}
\end{multline}
which is equivalent to applying the transformation~\eqref{eq:m-basis function} to $x$ coordinate dependence in the three-point function after $j_i\to-j_i-1$. For spectral flow violating case, the structure constant is non-vanishing when $\sum_{i=1}^{3}w_i=\pm1$, and the results are:
\begin{multline}
    \delta^{(2)}(\sum_{i}m_i\pm(\frac{k}{2}+1))F_{\pm}(j_i,m_i,\bar{m}_i)D_{\pm}(j_1,j_2,j_3) \qquad \text{with}\\
    D_{\pm}(j_1,j_2,j_3)=\frac{r(b)\Upsilon'(0)\Upsilon(b^{-1}(2j_1+1))\Upsilon(b^{-1}(2j_2+1))\Upsilon(b^{-1}(2j_3+1))}{\Upsilon(\frac{b}{2}-b^{-1}(j_{123}+1))\Upsilon(\frac{b}{2}-b^{-1}j_{23}^1)\Upsilon(\frac{b}{2}-b^{-1}j_{31}^2)\Upsilon(\frac{b}{2}-b^{-1}j_{12}^3)}
    \label{eq:SFV-structure constant}
\end{multline}
where $r(b)=\frac{(\nu b^{-2+2b^{-2}})^{j_{123}+1}}{\sqrt{2}b^{-1-b^2}}$ and $F_{\pm}(j_i,m_i,\bar{m}_i)$ are defined by:
\begin{equation}
    F_{\pm}(j_i,m_i,\bar{m}_i)=(-1)^{m_2-\bar{m}_2}\frac{\Gamma(1+j_1\pm m_1)\Gamma(1+j_2\pm m_2)\Gamma(1+j_3\pm m_3)}{\Gamma(-j_1\mp \bar{m}_1)\Gamma(-j_2\mp\bar{m}_2)\Gamma(-j_3\mp\bar{m}_3)}
\end{equation}

\section{Matching Null Vectors at Level 2}
\label{app:null vectors level k}

Having matched the simplest two examples of the null vectors in $\mathcal{N}=2$ Liouville and $SL(2,\mathbb{R})$ WZW model, we can generalize them to higher levels following the same mechanism. Here we collect the null vectors at level 2 for $SL(2,\mathbb{R})$ model and $\mathcal{N}=2$ superconformal algebra. We
work in the NS sector and denote the test state for constructing the null vectors by $\ket{j,m}$. The test state can be ``translated" to the corresponding primary state in $\mathcal{N}=2$ algebra by plugging in the duality of parameters.

For $SL(2,\mathbb{R})$ WZW model, a neutral descendant state at level 2 is expanded by a set of basis:
\begingroup
\small
\begin{align}
 \mathcal B_{SL(2,R)}={}&\bigl\{
 j^-_{-2},j^3_{-2},j^+_{-2},(j^-_{-1})^2,
 j^-_{-1}j^3_{-1},j^-_{-1}j^+_{-1},(j^3_{-1})^2,
 j^3_{-1}j^+_{-1},(j^+_{-1})^2\bigr\}.             \label{eq:affine-level-two-basis}
\end{align}
\endgroup
We can write the ansatz of the null vector at level 2 as a linear combination of these basis states. Requiring that acting $j_1^{\pm},j_1^3$ all annihilate this state, we get a solution for all null vectors at level 2. We used Mathematica to perform the computation and get the solutions for $j$ to be:
\begin{equation}
    j_1=-\frac{k+1}{2}, \quad j_2=-k, \quad j_3=k+1, \quad j_4=\frac{k+3}{2}
\end{equation}
For $j_1=-\frac{k+1}{2}$, the null vector is:
\begingroup
\small
\begin{equation}
\begin{aligned}
    \ket{\mathcal{N}_{j_1}}&=-\frac{1}{2}j_{-2}^-\ket{j,m+1}+\frac{1}{2}j_{-2}^3\ket{j,m}-\frac{1}{2}j_{-2}^+\ket{j,m-1}-\frac{1}{4}j_{-1}^-j_{-1}^-\ket{j,m+2}\\
    &-\frac{1}{2}j_{-1}^-j_{-1}^+\ket{j,m}+j_{-1}^-j_{-1}^3\ket{j,m+1}-j_{-1}^3j_{-1}^3\ket{j,m}+j_{-1}^3j_{-1}^+\ket{j,m-1}\\
    &-\frac{1}{4}j_{-1}^+j_{-1}^+\ket{j,m-2}
\end{aligned}
\end{equation}
\endgroup
For $j_2=-k$, the null vector is:
\begingroup
\small
\begin{equation}
\begin{aligned}
    \ket{\mathcal{N}_{j_2}}&=\frac{k^2-k+2m}{2d}j_{-2}^-\ket{j,m+1}-\frac{k^2+m-1}{d}j_{-2}^3\ket{j,m}+\frac{k^2+k+2m-2}{2d}j_{-2}^+\ket{j,m-1}\\
    &-\frac{k-m-1}{2d}j_{-1}^-j_{-1}^-\ket{j,m+2}+j_{-1}^-j_{-1}^3\ket{j,m+1}+\frac{m}{d}j_{-1}^-j_{-1}^+\ket{j,m}\\
    &+\frac{2m}{d}j_{-1}^3j_{-1}^3\ket{j,m}-\frac{k+2m-1}{d}j_{-1}^3j_{-1}^+\ket{j,m-1}+\frac{k+m-1}{2d}j_{-1}^+j_{-1}^+\ket{j,m-2}
\end{aligned}
\end{equation}
\endgroup
For $j_3=k+1$, the null vector is:
\begingroup
\small
\begin{equation}
\begin{aligned}
    \ket{\mathcal{N}_{j_3}}&=\frac{k^2-k+2m}{2d}j_{-2}^-\ket{j,m+1}+\frac{(k-m)(k^2+m-1)}{de}j_{-2}^3\ket{j,m}\\
    &+\frac{(k-m)(k-m+1)(k^2+k+2m-2)}{2de(k+m)}j_{-2}^+\ket{j,m-1}+\frac{k+m+2}{2d}j_{-1}^-j_{-1}^-\ket{j,m+2}\\
    &+j_{-1}^-j_{-1}^3\ket{j,m+1}-\frac{m(k-m)}{de}j_{-1}^-j_{-1}^+\ket{j,m}-\frac{2m(k-m)}{de}j_{-1}^3j_{-1}^3\ket{j,m}\\
    &-\frac{(k-m)(k-m+1)(k+2m-1)}{de(k+m)}j_{-1}^3j_{-1}^+\ket{j,m-1}\\
    &-\frac{(k-m)(k-m+1)(k-m+2)}{2de(k+m)}j_{-1}^+j_{-1}^+\ket{j,m-2}
\end{aligned}
\end{equation}
\endgroup
For $j_4=\frac{k+3}{2}$, the null vector is:
\begingroup
\small
\begin{equation}
\begin{aligned}
    \ket{\mathcal{N}_{j_4}}&=-\frac{1}{2}j_{-2}^-\ket{j,m+1}-\frac{k-2m+1}{2(k+2m+3)}j_{-2}^3\ket{j,m}-\frac{(k-2m+3)(k-2m+1)}{2(k+2m+1)(k+2m+3)}j_{-2}^+\ket{j,m-1}\\
    &+\frac{k+2m+5}{4(k-2m-1)}j_{-1}^-j_{-1}^-\ket{j,m+2}+j_{-1}^-j_{-1}^3\ket{j,m+1}+\frac{k-2m+1}{2(k+2m+3)}j_{-1}^-j_{-1}^+\ket{j,m}\\
    &+\frac{k-2m+1}{k+2m+3}j_{-1}^3j_{-1}^3\ket{j,m}+\frac{(k-2m+3)(k-2m+1)}{(k+2m+1)(k+2m+3)}j_{-1}^3j_{-1}^+\ket{j,m-1}\\
    &+\frac{(k-2m+5)(k-2m+3)(k-2m+1)}{4(k+2m-1)(k+2m+1)(k+2m+3)}j_{-1}^+j_{-1}^+\ket{j,m-2}
\end{aligned}
\end{equation}
\endgroup
where $d=k-2m-1,\quad e=k+m+1$.

From the Liouville side, a level-2 null vector is expanded by the neutral level-two basis in NS sector:
\begingroup
\small
\begin{align}
 \mathcal B_{\mathcal N=2}={}&\bigl\{
 J_{-2},L_{-2},J_{-1}^{2},
 G^+_{-3/2}G^-_{-1/2},G^+_{-1/2}G^-_{-3/2},
 J_{-1}G^+_{-1/2}G^-_{-1/2}, \notag\\
 &\hspace{3.1cm}L_{-1}G^+_{-1/2}G^-_{-1/2},
 L_{-1}J_{-1},L_{-1}^{2}\bigr\}.                    \label{eq:N2-level-two-basis}
\end{align}
\endgroup
The primary states $\ket{h,Q}$ in $\mathcal{N}=2$ Liouville can also be labelled by $j,m$ via:
\begin{equation}
 h=-\frac{j(j-1)}{k}+\frac{m^2}{k},\qquad
 Q=\frac{2m}{k},\qquad c=3+\frac{6}{k}
\end{equation}
We can write the ansatz of the level-2 null vector as a linear combination of the descendant operators in~\eqref{eq:N2-level-two-basis} acting on the primary state $\ket{h,Q}$. Requiring that acting $L_1,G^{\pm}_{1/2},J_1$ all annihilate this state, we get a solution for all null vectors at level 2. We find that the solutions for $j$ agrees with those in $SL(2,\mathbb{R})$ model. The corresponding null vectors are as follows. For $j_1=-\frac{k+1}{2}$ and $j_4=\frac{k+3}{2}$, the null vector is:
\begingroup
\small
\begin{equation}
\begin{aligned}
    \ket{\mathcal{N}_j^L}&=\frac{k^2-2k+2m(k-1)}{2k}J_{-2}\ket{h,Q}+\frac{1}{k}L_{-2}\ket{h,Q}+\frac{(k+2m)^2-1}{4}J_{-1}J_{-1}\ket{h,Q}\\
    &-\frac{k^3+2k^2-(4m^2+1)k-4m-2}{2[(k-2m)^2-1]}G_{-3/2}^+G_{-1/2}^-\ket{h,Q}+\frac{k^2+(2m+1)k-2}{2(k-2m-1)}G_{-1/2}^+G_{-3/2}^-\ket{h,Q}\\
    &+\frac{k[k^2-(4m^2+1)]}{(k-2m)^2-1}J_{-1}G_{-1/2}^+G_{-1/2}^-\ket{h,Q}+\frac{4mk}{(k-2m)^2-1}L_{-1}G_{-1/2}^+G_{-1/2}^-\ket{h,Q}\\
    &-(k+2m)L_{-1}J_{-1}\ket{h,Q}+L_{-1}L_{-1}\ket{h,Q}
\end{aligned}
\end{equation}
\endgroup
And for $j_2=-k$ and $j_3=k+1$, the null vector is:
\begingroup
\small
\begin{equation}
\begin{aligned}
    \ket{\mathcal{N}_j^L}&=\frac{k-2k^2+2m-2km}{2}J_{-2}\ket{h,Q}+kL_{-2}\ket{h,Q}+m(k+m)J_{-1}J_{-1}\ket{h,Q}\\
    &-\frac{k(k+m)}{2(k-m)}G_{-3/2}^+G_{-1/2}^-\ket{h,Q}-\frac{k}{2}G_{-1/2}^+G_{-3/2}^-\ket{h,Q}+\frac{km}{k-m}J_{-1}G_{-1/2}^+G_{-1/2}^-\ket{h,Q}\\
    &-\frac{k}{k-m}L_{-1}G_{-1/2}^+G_{-1/2}^-\ket{h,Q}-(k+2m)L_{-1}J_{-1}\ket{h,Q}+L_{-1}L_{-1}\ket{h,Q}
\end{aligned}
\end{equation}
\endgroup

Finally, for each of the four branches we substitute the coset expressions for $L_n,J_n,G_r^\pm$ into the corresponding $\mathcal N=2$ null vectors and normal order the affine currents and fermions.  The fermionic components turn out to vanish according to the current algebra, and the remaining bosonic components satisfies
\begin{equation}
 \left.\ket{\mathcal N^L_{j_a}}\right|_{\rm coset}
   =C_a(k,m)\ket{\mathcal N_{j_a}},
 \qquad a=1,2,3,4,                                   \label{eq:level-two-null-proportionality}
\end{equation}
With the normalizations displayed above (coefficient of $L_{-1}L_{-1}$ equal to one in $\ket{\mathcal N^L_{j_a}}$ and coefficient of $j^-_{-1}j^3_{-1}$ equal to one in $\ket{\mathcal N_{j_a}}$), the constants are:
\begingroup
\small
\begin{equation}
\begin{aligned}
    C_1(k,m)&=-\frac{(k+2m)^2-1}{k^2}, & C_2(k,m)&=\frac{2(k-2m-1)(k+m)}{k^2},\\
    C_3(k,m)&=-\frac{2(k-2m-1)(k+m)(k+m+1)}{k^2(k-m)}, & C_4(k,m)&=\frac{(k+2m-1)(k+2m+1)(k+2m+3)}{k^2(k-2m+1)}
\end{aligned}
\end{equation}
\endgroup
Our Mathematica implementation works for generic level $k$ and can therefore be used to systematically extend the analysis to higher levels.

\section{Superconformal Ward Identities}
\label{app:superconformal Ward identities}

In this section, we will first derive the coordinate dependence of the three-point functions in superspace from superconformal Ward identities, and then extract the constraints for structure constants in neutral sector.

Consider a three-point function $G_3=\langle \mathcal{V}_{\alpha_1,\tilde\alpha_1}^{\eta=0}(z_1;\theta_1^{\pm})\mathcal{V}_{\alpha_2,\tilde\alpha_2}^{\eta=0}(z_2;\theta_2^{\pm})\mathcal{V}_{\alpha_3,\tilde\alpha_3}^{\eta=0}(z_3;\theta_3^{\pm}) \rangle$, the lowest bosonic component of each superfield is a NS vertex operator with conformal dimension $h_i$ and $U(1)$ charge $Q_i$ for $i=1,2,3$. We focus on the holomorphic part first, and the anti-holomorphic part can be solved in the same way. For the structure constant, we have to combine both sectors in a consistent way, which will be clarified later.

Ward identities correspond to the operators $L_{-1},L_0,L_1,J_0,G^+_{\pm1/2},G^-_{\pm1/2}$, which act on a superfield of weight $h$ and charge $Q$ as:
\begin{multline}
    L_{-1}=\partial_z, \quad L_0=z\partial_z+\frac{1}{2}(\theta^+\partial_{\theta^+}+\theta^-\partial_{\theta^-})+h, \quad J_0=(\theta^+\partial_{\theta^+}-\theta^-\partial_{\theta^-})+Q,\\
    L_1=z^2\partial_z+z(\theta^+\partial_{\theta^+}+\theta^-\partial_{\theta^-})+2hz-Q\theta^+\theta^- \quad G^+_{-1/2}=\partial_{\theta^+}-\theta^-\partial_z, \quad G^-_{-1/2}=\partial_{\theta^-}-\theta^+\partial_z, \\
    G^+_{1/2}=z(\partial_{\theta^+}-\theta^-\partial_z)+\theta^+\theta^-\partial_{\theta^+}-(2h+Q)\theta^-, \quad G^-_{1/2}=z(\partial_{\theta^-}-\theta^+\partial_z)-\theta^+\theta^-\partial_{\theta^-}-(2h-Q)\theta^+
\end{multline}
They give a constraint which can be written concisely as:
\begin{equation}
    \sum_i D_X^{(i)}G_3=\sum_i(V^{(i)}_X+m_X^{(i)})G_3=0
\end{equation}
where $i=1,2,3$ representing the coordinates these vector fields act on, and $X$ denotes eight vector fields giving constraints for the coordinates following the relation $[X,\Phi]=D_X\Phi$. We decompose the matrix of differential operators into first-order derivative part $V_X^{(i)}$ and the multiplier $m_X^{(i)}$. The supertranslation operators $D_{L_{-1}},D_{G^+_{-1/2}},D_{G^-_{-1/2}}$ contains only first-order derivative part. Similar to bosonic case, we introduce supertranslation invariant coordinates $Z_{ij}$ and $\theta_{ij}^{\pm}$:
\begin{equation}
    Z_{ij}\equiv z_{ij}-\theta^+_i\theta_j^--\theta_i^-\theta_j^+, \quad \theta_{ij}^{\pm}\equiv \theta_i^{\pm}-\theta_j^{\pm}
\end{equation}
It's straightforward to check that:
\begin{equation}
    \sum_kD_{L_{-1}}^{(k)}Z_{ij}=\sum_kV_{L_{-1}}^{(k)}Z_{ij}=0,  \quad \sum_kD_{G_{-1/2}^{\pm}}^{(k)}Z_{ij}=\sum_kV_{G_{-1/2}^{\pm}}^{(k)}Z_{ij}=0
\end{equation}
and $\theta_{ij}^{\pm}$ is trivially invariant under supertranslations. Therefore, each solution of the Ward identities should depend on the coordinate combinations $Z_{ij}$ and $\theta_{ij}^{\pm}$. Now we turn to other five constraints given by $G^{\pm}_{1/2},L_0,L_1,J_0$. We first check how the corresponding generators $V_X$ act on $Z_{ij}$:
\begin{equation}
\begin{aligned} 
    \sum_kV_{L_0}^{(k)}Z_{ij}=Z_{ij}, \quad   \sum_kV_{L_1}^{(k)}Z_{ij}&=(z_i+z_j)Z_{ij}, \quad \sum_kV_{J_0}^{(k)}Z_{ij}=0,\\
    \sum_kV_{G_{1/2}^{\pm}}^{(k)}Z_{ij}&=-(\theta_j^{\mp}+\theta_i^{\mp})Z_{ij}
\end{aligned}
\end{equation}
Motivated by bosonic case, the possible dependence on $Z_{ij}$ for the three-point function could be $P_0(Z)\equiv Z_{21}^{-h_{12}^3}Z_{32}^{-h_{23}^1}Z_{31}^{-h_{13}^2}$ with $h_{ab}^c\equiv h_a+h_b-h_c$. Since $Z_{ij}$ only contains the term with even number of Grassmann variable, we can use the chain rule, i.e., $V_X(Z_{ij}^a)=aZ_{ij}^{a-1}V_XZ_{ij}$. Then we can directly prove that $P_0(Z)$ is translation invariant. For five other constraints, with the relation we have derived above, we can write out:
\begin{equation}
\begin{aligned}
    \sum_kD_{L_0}^{(k)}P_0(Z)&=-(h_1+h_2+h_3)P_0(Z)+\sum_{k=1}^3h_kP_0(Z)=0, \quad \sum_kD_{J_0}^{(k)}P_0(Z)=\sum_kQ_kP_0(Z), 
    \\
\sum_kD_{L_1}^{(k)}P_0(Z)&=-\sum_kQ_k\theta_k^+\theta_k^-P_0(Z), \quad \sum_kD_{G^{\pm}_{1/2}}^{(k)}P_0(Z)=\sum_k\mp Q_k\theta_k^{\mp}P_0(Z)
\end{aligned}
\end{equation}
We find out that all conformal weight dependence cancels, and $P_0(Z)$ is the solution of Ward identities if all $Q_k$'s are zero. To loosen the condition, we introduce another factor $e^{B_Q}$ to cancel the additional charge terms appearing in $D_{L_1}P_0(Z)$ and $D_{G^{\pm}_{1/2}}P_0(Z)$. The constraints for the supertranslation invariant factor $B_Q$ are as follows:
\begin{align*}
    \sum_kV_{L_0}^{(k)}B_Q=0, \quad \sum_kV_{J_0}^{(k)}B_Q=0,\quad  \sum_kV_{L_{-1}}^{(k)}B_Q=0, \quad \sum_kV_{G^{\pm}_{-1/2}}^{(k)}B_Q=0,\\
    \sum_kV_{L_1}^{(k)}B_Q=\sum_kQ_k\theta_k^+\theta_k^-, \quad \sum_kV_{G^{\pm}_{1/2}}^{(k)}B_Q=\sum_k\pm Q_k\theta_k^{\mp} \qquad
\end{align*}
Setting the ansatz for $B_Q$ as a linear combination of all quadratic terms $B_Q=\sum_{i,j}M_{ij}\theta_i^+\theta_j^-$, solving the constraints above turns to solving a first-order differential equation for matrix $M$\footnote{The conditions for $B_Q$ above can be rewritten as some equations for $M$:
\begin{equation*}
    M\cdot\mathbf{e}=0, \quad M^T\cdot \mathbf{e}=0, \quad M\cdot \mathbf{z}=\mathbf{Q}, \quad M^T\cdot \mathbf{z}=\mathbf{Q}
\end{equation*}
These conditions give eight constraints for nine entries of the matrix $M_{ij}$. In particular, $\mathbf{e}^TM\mathbf{z}=\sum Q_i=0$. The solution for $M$ is:
\begin{equation*}
    M=M_Q+cM_B
\end{equation*}
where $M_B$ corresponds to $B\equiv \sum_{ij}M_{B,ij}\theta_i^+\theta_j^-$ satisfying $M_B\cdot\mathbf{e}=M_B^T\cdot\mathbf{e}=M_B\cdot\mathbf{z}=M_B^T\cdot\mathbf{z}=0$.}. A solution for three-point function in superspace can be figured out by choosing the gauge $M_{Q,12}=M_{Q,21}=0$ for the charged part of $M$:
\begin{equation}
    G_3=P_0(Z)e^{B_Q}(C_0+KB)
    \label{eq:neutral charge sol}
\end{equation}
with constants $C_0,K$ and $B_Q,B$ defined as follows\footnote{We already checked that $B_Q^2\neq0$ and $B^2=0$.}:
\begingroup
\small
\begin{equation}
\begin{aligned}
B_Q&=\frac{Q_1}{z_{13}}\,\theta_1^+\theta_1^-+\frac{Q_2}{z_{23}}\,\theta_2^+\theta_2^-
+\Big(\frac{Q_1}{z_{13}}+\frac{Q_2}{z_{23}}\Big)\theta_3^+\theta_3^-
-\frac{Q_1}{z_{13}}\big(\theta_1^+\theta_3^-+\theta_3^+\theta_1^-\big)
-\frac{Q_2}{z_{23}}\big(\theta_2^+\theta_3^-+\theta_3^+\theta_2^-\big)\\
B&=\frac{\theta_2^{+}\theta_1^{-}-\theta_2^{-}\theta_1^{+}}{z_{21}}
 +\frac{\theta_3^{+}\theta_2^{-}-\theta_3^{-}\theta_2^{+}}{z_{32}}
 +\frac{\theta_1^{+}\theta_3^{-}-\theta_1^{-}\theta_3^{+}}{z_{13}}
 +\Big(\frac{1}{z_{12}}-\frac{1}{z_{13}}\Big)\theta_1^{+}\theta_1^{-}\\
 &+\Big(\frac{1}{z_{23}}-\frac{1}{z_{21}}\Big)\theta_2^{+}\theta_2^{-}
 +\Big(\frac{1}{z_{31}}-\frac{1}{z_{32}}\Big)\theta_3^{+}\theta_3^{-}
\end{aligned}
\label{eq:Bq-B}
\end{equation}
\endgroup

In the gauge $\theta_1=\theta_3=0,z_1=0,z_3\rightarrow\infty$, the three-point function becomes:
\begin{equation}
    G_3=\bra{h_3}\mathcal{V}^{\eta=0}_{\alpha_2,\tilde\alpha_2}(z,\theta_2^{\pm})\ket{h_1}=z^{h_3-h_1-h_2}(C_0-\frac{K}{z}\theta_2^+\theta_2^-)
\end{equation}
where the out-state $\bra{h_3}\equiv \lim_{\omega\rightarrow\infty}\omega^{2h_3}\bra{0}\phi_3(\omega)$ and the in-state $\ket{h_1}\equiv\phi_1(0)\ket{0}$\footnote{$\phi$ denotes the lowest component of the superfield multiplet.}. $C_0$ is the structure constant of the three lowest components of the superfield multiplets, given in Section~\ref{sec:correlators in superspace}. This solution is a two-dimensional vector in the neutral sector due to the charge conservation condition $\sum_iQ_i=0$. In this gauge there are two further one-dimensional solutions in the charged sectors $\sum_iQ_i=\mp1$, which appear below as the terms linear in $\theta_2^{\pm}$.

The constant $K$ is not fixed by the Ward identities. $L_0$ covariance relates it to the three-point function of the descendant $G^+_{-1/2}G^-_{-1/2}\phi_2$ as follows. Recalling the superfield multiplet defined in~\eqref{eq: NS multiplet}, we can write the intermediate state in the three-point function as:
\begin{equation}
  \mathcal{V}^{\eta=0}_{\alpha_2,\tilde\alpha_2}(z,\theta_2^{\pm})=\phi_2(z)+\theta_2^+\psi_2^+(z)+\theta_2^-\psi_2^-(z)+\theta_2^+\theta_2^-\tilde{\phi}_2(z)
\end{equation}
where $\phi_2(z)$ is the NS vertex operator with conformal dimension $h_2$ and charge $Q_2$ as defined above. Comparing with~\eqref{eq: NS multiplet}, we can read out that $\psi_2^+(z)=G^{+}_{-1/2}\phi_2(z),\psi_2^-(z)=G^-_{-1/2}\phi_2(z)$ and $\tilde{\phi}_2(z)=(L_{-1}-G_{-1/2}^+G^-_{-1/2})\phi_2(z)$. Plugging it into the three-point function in the gauge we chose above, we get four components for the three-point function:
\begin{multline*}
    \bra{h_3}\mathcal{V}^{\eta=0}_{\alpha_2,\tilde\alpha_2}(z,\theta_2^{\pm})\ket{h_1}=\bra{h_3}\phi_2(z)\ket{h_1}+\bra{h_3}\psi_2^+(z)\ket{h_1}\theta_2^+\\
    +\bra{h_3}\psi_2^-(z)\ket{h_1}\theta_2^-+\bra{h_3}\tilde{\phi}_2(z)\ket{h_1}\theta_2^+\theta_2^-
\end{multline*}
For each component, the state $\chi(z)=e^{zL_{-1}}\chi(0)e^{-zL_{-1}}$ is generated by the translation operator $L_{-1}$, where $\chi=\phi_2,\psi_2^{\pm},\tilde{\phi}_2$. The conformal dimension of these states $\phi,\psi^+,\psi^-,\tilde{\phi}$ are given by the eigenvalues of $L_0$:
\begin{equation}
    (h_2,Q_2), \quad (h_2+\frac{1}{2},Q_2+1), \quad (h_2+\frac{1}{2},Q_2-1) \quad (h_2+1,Q_2)
\end{equation}
And the three-point function takes the form\footnote{We can compute the commutators of $L_0$ and these states,i.e., $[L_0,\chi(z)]$, and sandwich them into $\bra{h_3}$ and $\ket{h_1}$. Using $[L_0,L_{-1}]=L_{-1}$, we have:
\begin{equation*}
    [L_0,\chi(z)]=(h_{\chi}+z\partial_z)\chi(z)
\end{equation*}
Therefore, we can write down the relation:
\begin{equation*}
    \bra{h_3}[L_0,\chi(z)]\ket{h_1}=(h_3-h_1)\bra{h_3}\chi(z)\ket{h_1}=(h_{\chi}+z\partial_z)\bra{h_3}\chi(z)\ket{h_1}
\end{equation*}
leading to $\bra{h_3}\chi(z)\ket{h_1}=C_{\chi}z^{h_3-h_1-h_{\chi}}$. For the top component, since $\tilde{\phi}_2(z)=(L_{-1}-G_{-1/2}^+G_{-1/2}^-)\phi_2(z)$, we have:
\begin{equation*}
\begin{aligned}
    \bra{h_3}\tilde{\phi}_2(z)\ket{h_1}&=\partial_z\bra{h_3}\phi_2(z)\ket{h_1}-\bra{h_3}G_{-1/2}^+G_{-1/2}^-\phi_2(z)\ket{h_1}\\
    &=((h_3-h_2-h_1)C_0-C_t)z^{h_3-h_2-h_1-1}
\end{aligned}
\end{equation*}}:
\begin{multline}
    C_0z^{h_3-h_1-h_2}+C_+z^{h_3-h_1-h_2-1/2}\theta_2^+ +C_-z^{h_3-h_1-h_2-1/2}\theta_2^-\\
    +((h_3-h_2-h_1)C_0-C_t)z^{h_3-h_1-h_2-1}\theta_2^+\theta_2^-
\end{multline}
with $C_t$ denoting the constant coefficient of the field $G_{-1/2}^+G_{-1/2}^-\phi_2(z)$.

The charge of the three-point function can be extracted from the eigenvalue of $J_0$, which gives the selection rules if we consider $[J_0,\chi(z)]=Q_{\chi}\chi(z)$ sandwiched by in-state $\ket{h_1}$ and out-state $\bra{h_3}$. $\phi_2(z)$ and $\tilde{\phi}_2(z)$ are in the \emph{neutral sector} since\footnote{We used $\bra{h_3}J_0=\lim_{\omega\rightarrow\infty}\omega^{2h_3}\bra{0}\phi_3(\omega)J_0=-\lim_{\omega\rightarrow\infty}\omega^{2h_3}\bra{0}[J_0,\phi_3(\omega)]=-Q_3\bra{h_3}$ and $\bra{0}J_0=0$.}:
\begin{equation}
\begin{aligned}
    \bra{h_3}[J_0,\tilde{\phi}_2(z)]\ket{h_1}&=Q_{2}((h_3-h_2-h_1)C_0-C_t)z^{h_3-h_1-h_2-1}\\
    &=-(Q_3+Q_1)((h_3-h_2-h_1)C_0-C_t)z^{h_3-h_1-h_2-1}\\
    \bra{h_3}[J_0,\phi_2(z)]\ket{h_1}&=Q_{2}C_0z^{h_3-h_1-h_2}=-(Q_3+Q_1)C_0z^{h_3-h_1-h_2}
\end{aligned}
\end{equation}
giving $Q_1+Q_2+Q_3=0$. $\psi^+_2(z)$ and $\psi^-_2(z)$ are in the \emph{charged sector} since:
\begin{equation}
\begin{aligned}
   \bra{h_3}[J_0,\psi_2^+(z)]\ket{h_1}&=(Q_{2}+1)C_+z^{h_3-h_1-h_2-1/2}=-(Q_3+Q_1)C_+z^{h_3-h_1-h_2-1/2}\\
   \bra{h_3}[J_0,\psi_2^-(z)]\ket{h_1}&=(Q_{2}-1)C_-z^{h_3-h_1-h_2-1/2}=-(Q_3+Q_1)C_-z^{h_3-h_1-h_2-1/2}
\end{aligned}
\end{equation}
giving $Q_1+Q_2+Q_3=-1$ for the structure constant $C_+$ and $Q_1+Q_2+Q_3=1$ for the structure constant $C_-$.

Having fixed the structure constant of the top component $K$ via a specific gauge fixing state, we can write down the structure constant for all possible two-fermion states in terms of $C_0$ and $C_t$ from the general formula~\eqref{eq:neutral charge sol} and~\eqref{eq:Bq-B} for the three-point function. There are six possible combinations for the two-fermion states associated to $\theta_i^+\theta_j^-,i\neq j$ after gauge fixing the Grassmann variables. With $P_0(z)\equiv z_{21}^{h_3-h_1-h_2}z_{32}^{h_1-h_2-h_3}z_{31}^{h_2-h_1-h_3}$, the three-point function can be written as $G_3=\langle\phi_3\phi_2\phi_1\rangle+\theta_i^+\theta_j^-\langle\phi_k\psi_i^+\psi_j^-\rangle$ with $k\neq i,j$\footnote{$\langle\phi_3\phi_2\phi_1\rangle=C_0P_0(z)$.}, and the top components for the correlator are given by:
\begin{equation}
\begin{aligned}
    \langle\phi_3\psi_2^-\psi_1^+\rangle&=\frac{P_0(z)}{z_{21}}C_t\\
    \langle\phi_3\psi_2^+\psi_1^-\rangle&=\frac{P_0(z)}{z_{21}}(2(h_3-h_1-h_2)C_0-C_t)\\
    \langle\psi_3^-\phi_2\psi_1^+\rangle&=-\frac{P_0(z)}{z_{31}}(C_t+(2h_1-Q_1)C_0)\\
    \langle\psi_3^+\phi_2\psi_1^-\rangle&=\frac{P_0(z)}{z_{31}}((2h_2-2h_3-Q_1)C_0+C_t)\\
    \langle\psi_3^-\psi_2^+\phi_1\rangle&=\frac{P_0(z)}{z_{32}}((2h_1-2h_3+Q_2)C_0+C_t)\\
    \langle\psi_3^+\psi_2^-\phi_1\rangle&=-\frac{P_0(z)}{z_{32}}(C_t+(2h_2+Q_2)C_0)
    \label{eq:C_t sol}
\end{aligned}
\end{equation}
Recall that $\psi_i^{\pm}(z)=G_{-1/2}^{\pm}\phi_i(z)$, the LHS can be independently obtained from the bosonic component $\langle\phi_3\phi_2\phi_1\rangle$ using the OPE $G_{-1/2}^{\pm}$ on the NS vertex operators, which leads to the spectral-flowed NS vertex operator by $\pm1$ unit. Using the free field realization of $\mathcal{N}=2$ generators, we have:
\begin{equation}
    G_{-1/2}^+V^{\eta=0}_{\alpha,\tilde\alpha}(z)=-i\sqrt{2}\tilde\alpha V^{\eta=+1}_{\alpha,\tilde\alpha}(z), \quad G^-_{-1/2}V^{\eta=0}_{\alpha,\tilde\alpha}(z)=-i\sqrt{2}\alpha V^{\eta=-1}_{\alpha,\tilde\alpha}(z)
    \label{eq:OPE-G^pm}
\end{equation}
where the spectral-flowed operator $V^{\eta}_{\alpha,\tilde\alpha}(z)$ carries the $SL(2,\mathbb{R})$ labels of $V^{\eta=0}_{\alpha-\eta/b,\tilde\alpha+\eta/b}(z)$. As for the RHS, once $C_0$ is fixed, the relations above should lead to the unique $C_t$. This provides the additional constraint for the structure constant of $\mathcal{N}=2$ Liouville theory apart from the algebraic check.

\bibliographystyle{JHEP}
\bibliography{biblio.bib}

@article{Teschner:1997ft,
    author = "Teschner, J.",
    title = "{On structure constants and fusion rules in the SL(2,C) / SU(2) WZNW model}",
    eprint = "hep-th/9712256",
    archivePrefix = "arXiv",
    doi = "10.1016/S0550-3213(99)00072-3",
    journal = "Nucl. Phys. B",
    volume = "546",
    pages = "390--422",
    year = "1999"
}

@article{Teschner_1995,
    author = {Teschner, J{\"o}rg},
    title = "{On the Liouville three-point function}",
    eprint = "hep-th/9507109",
    archivePrefix = "arXiv",
    doi = "10.1016/0370-2693(95)01200-A",
    journal = "Phys. Lett. B",
    volume = "363",
    pages = "65--70",
    year = "1995"
}

@article{ribault2022conformalfieldtheoryplane,
    author = "Ribault, Sylvain",
    title = "{Conformal field theory on the plane}",
    eprint = "1406.4290",
    archivePrefix = "arXiv",
    primaryClass = "hep-th",
    month = "6",
    year = "2014"
}

@article{Maldacena:2000hw,
    author = "Maldacena, Juan Martin and Ooguri, Hirosi",
    title = "{Strings in AdS(3) and SL(2,R) WZW model 1.: The Spectrum}",
    eprint = "hep-th/0001053",
    archivePrefix = "arXiv",
    reportNumber = "CALT-68-2245, CITUSC-99-010, HUTP-99-A027, LBNL-44375, UCB-PTH-99-48, LBL-44375",
    doi = "10.1063/1.1377273",
    journal = "J. Math. Phys.",
    volume = "42",
    pages = "2929--2960",
    year = "2001"
}

@article{Maldacena:2000kv,
    author = "Maldacena, Juan Martin and Ooguri, Hirosi and Son, John",
    title = "{Strings in AdS(3) and the SL(2,R) WZW model. Part 2. Euclidean black hole}",
    eprint = "hep-th/0005183",
    archivePrefix = "arXiv",
    reportNumber = "CALT-68-2266, CITUSC-00-021, HUTP-00-A009, UCB-PTH-00-10",
    doi = "10.1063/1.1377039",
    journal = "J. Math. Phys.",
    volume = "42",
    pages = "2961--2977",
    year = "2001"
}

@article{Ribault:2005wp,
    author = "Ribault, Sylvain and Teschner, Joerg",
    title = "{H+(3)-WZNW correlators from Liouville theory}",
    eprint = "hep-th/0502048",
    archivePrefix = "arXiv",
    reportNumber = "KCL-MTH-05-01",
    doi = "10.1088/1126-6708/2005/06/014",
    journal = "JHEP",
    volume = "06",
    pages = "014",
    year = "2005"
}

@article{Ribault:2005ms,
    author = "Ribault, Sylvain",
    title = "{Knizhnik-Zamolodchikov equations and spectral flow in AdS(3) string theory}",
    eprint = "hep-th/0507114",
    archivePrefix = "arXiv",
    reportNumber = "KCL-MTH-05-07",
    doi = "10.1088/1126-6708/2005/09/045",
    journal = "JHEP",
    volume = "09",
    pages = "045",
    year = "2005"
}

@article{Boucher:1986bh,
    author = "Boucher, Wayne and Friedan, Daniel and Kent, Adrian",
    title = "{Determinant Formulae and Unitarity for the N=2 Superconformal Algebras in Two-Dimensions or Exact Results on String Compactification}",
    reportNumber = "EFI-86-14-CHICAGO",
    doi = "10.1016/0370-2693(86)90260-1",
    journal = "Phys. Lett. B",
    volume = "172",
    pages = "316",
    year = "1986"
}

@article{Eguchi:1988af,
    author = "Eguchi, Tohru and Taormina, Anne",
    title = "{On the Unitary Representations of $N=2$ and $N=4$ Superconformal Algebras}",
    reportNumber = "CERN-TH-5044/88",
    doi = "10.1016/0370-2693(88)90360-7",
    journal = "Phys. Lett. B",
    volume = "210",
    pages = "125--132",
    year = "1988"
}

@article{Eguchi_2004,
    author = "Eguchi, Tohru and Sugawara, Yuji",
    title = "{SL(2;R)/U(1) supercoset and elliptic genera of non-compact Calabi-Yau manifolds}",
    eprint = "hep-th/0403193",
    archivePrefix = "arXiv",
    reportNumber = "UT-04-12",
    doi = "10.1088/1126-6708/2004/05/014",
    journal = "JHEP",
    volume = "05",
    pages = "014",
    year = "2004"
}

@article{NAKAYAMA_2004,
    author = "Nakayama, Yu",
    title = "{Liouville field theory: A decade after the revolution}",
    eprint = "hep-th/0402009",
    archivePrefix = "arXiv",
    reportNumber = "UT-04-02",
    doi = "10.1142/S0217751X04019500",
    journal = "Int. J. Mod. Phys. A",
    volume = "19",
    pages = "2771--2930",
    year = "2004"
}

@article{Ahn_2004_ZZ,
    author = "Ahn, Changrim and Stanishkov, Marian and Yamamoto, Masayoshi",
    title = "{ZZ-branes of $N=2$ super-Liouville theory}",
    eprint = "hep-th/0405274",
    archivePrefix = "arXiv",
    doi = "10.1088/1126-6708/2004/07/057",
    journal = "JHEP",
    volume = "07",
    pages = "057",
    year = "2004"
}

@article{Ahn_2004,
    author = "Ahn, Changrim and Kim, Chanju and Rim, Chaiho and Stanishkov, Marian",
    title = "{Duality in $N=2$ super-Liouville theory}",
    eprint = "hep-th/0210208",
    archivePrefix = "arXiv",
    doi = "10.1103/PhysRevD.69.106011",
    journal = "Phys. Rev. D",
    volume = "69",
    pages = "106011",
    year = "2004"
}

@article{Hosomichi:2004ph,
    author = "Hosomichi, Kazuo",
    title = "{N=2 Liouville theory with boundary}",
    eprint = "hep-th/0408172",
    archivePrefix = "arXiv",
    doi = "10.1088/1126-6708/2006/12/061",
    journal = "JHEP",
    volume = "12",
    pages = "061",
    year = "2006"
}

@article{Fukuda:2001jd,
    author = "Fukuda, Takeshi and Hosomichi, Kazuo",
    title = "{Three point functions in sine-Liouville theory}",
    eprint = "hep-th/0105217",
    archivePrefix = "arXiv",
    reportNumber = "YITP-01-43",
    doi = "10.1088/1126-6708/2001/09/003",
    journal = "JHEP",
    volume = "09",
    pages = "003",
    year = "2001"
}

@article{Maldacena:2001km,
    author = "Maldacena, Juan Martin and Ooguri, Hirosi",
    title = "{Strings in AdS(3) and the SL(2,R) WZW model. Part 3. Correlation functions}",
    eprint = "hep-th/0111180",
    archivePrefix = "arXiv",
    reportNumber = "CALT-68-2360, CITUSC-01-042",
    doi = "10.1103/PhysRevD.65.106006",
    journal = "Phys. Rev. D",
    volume = "65",
    pages = "106006",
    year = "2002"
}

@article{Ribault:2009ui,
    author = "Ribault, Sylvain",
    title = "{Minisuperspace limit of the AdS$_3$ WZNW model}",
    eprint = "0912.4481",
    archivePrefix = "arXiv",
    primaryClass = "hep-th",
    reportNumber = "LPTA-09-100, LPTA:09-100",
    doi = "10.1007/JHEP04(2010)096",
    journal = "JHEP",
    volume = "04",
    pages = "096",
    year = "2010"
}

@book{blumenhagen2009introduction,
  author = {Blumenhagen, Ralph and Plauschinn, Erik},
  title = {Introduction to Conformal Field Theory: With Applications to String Theory},
  publisher = {Springer},
  year = {2009},
  volume = {779},
  series = {Lecture Notes in Physics}
}

@article{Dorn:1994xn,
    author = "Dorn, Harald and Otto, H. J.",
    title = "{Two and three point functions in Liouville theory}",
    eprint = "hep-th/9403141",
    archivePrefix = "arXiv",
    reportNumber = "HU-BERLIN-IEP-94-2, DESY-94-066",
    doi = "10.1016/0550-3213(94)00352-1",
    journal = "Nucl. Phys. B",
    volume = "429",
    pages = "375--388",
    year = "1994"
}

@article{Zamolodchikov_1996,
    author = "Zamolodchikov, Alexander B. and Zamolodchikov, Alexei B.",
    title = "{Structure constants and conformal bootstrap in Liouville field theory}",
    eprint = "hep-th/9506136",
    archivePrefix = "arXiv",
    reportNumber = "LPM-95-24, RU-95-39",
    doi = "10.1016/0550-3213(96)00351-3",
    journal = "Nucl. Phys. B",
    volume = "477",
    pages = "577--605",
    year = "1996"
}

@article{Kazakov:2000pm,
    author = "Kazakov, Vladimir and Kostov, Ivan K. and Kutasov, David",
    title = "{A Matrix model for the two-dimensional black hole}",
    eprint = "hep-th/0101011",
    archivePrefix = "arXiv",
    reportNumber = "SACLAY-SPH-T-00-123, LPTHENS-00-32, EFI-2000-29",
    doi = "10.1016/S0550-3213(01)00606-X",
    journal = "Nucl. Phys. B",
    volume = "622",
    pages = "141--188",
    year = "2002"
}

@article{Hori:2001ax,
    author = "Hori, Kentaro and Kapustin, Anton",
    title = "{Duality of the fermionic 2d black hole and $N=2$ Liouville theory as mirror symmetry}",
    eprint = "hep-th/0104202",
    archivePrefix = "arXiv",
    reportNumber = "HUTP-01-A019",
    doi = "10.1088/1126-6708/2001/08/045",
    journal = "JHEP",
    volume = "08",
    pages = "045",
    year = "2001"
}

@article{Dijkgraaf:1991ba,
    author = "Dijkgraaf, Robbert and Verlinde, Herman L. and Verlinde, Erik P.",
    title = "{String propagation in a black hole geometry}",
    reportNumber = "PUPT-1252, IASSNS-HEP-91-22",
    doi = "10.1016/0550-3213(92)90237-6",
    journal = "Nucl. Phys. B",
    volume = "371",
    pages = "269--314",
    year = "1992"
}

@article{Witten:1991yr,
    author = "Witten, Edward",
    title = "{On string theory and black holes}",
    reportNumber = "IASSNS-HEP-91-12",
    doi = "10.1103/PhysRevD.44.314",
    journal = "Phys. Rev. D",
    volume = "44",
    pages = "314--324",
    year = "1991"
}

@article{Dei:2021yom,
    author = "Dei, Andrea and Eberhardt, Lorenz",
    title = "{String correlators on AdS$_{3}$: four-point functions}",
    eprint = "2107.01481",
    archivePrefix = "arXiv",
    primaryClass = "hep-th",
    doi = "10.1007/JHEP09(2021)209",
    journal = "JHEP",
    volume = "09",
    pages = "209",
    year = "2021"
}

@article{Dei:2021xgh,
    author = "Dei, Andrea and Eberhardt, Lorenz",
    title = "{String correlators on AdS$_{3}$: three-point functions}",
    eprint = "2105.12130",
    archivePrefix = "arXiv",
    primaryClass = "hep-th",
    doi = "10.1007/JHEP08(2021)025",
    journal = "JHEP",
    volume = "08",
    pages = "025",
    year = "2021"
}

@article{Polyakov:1981rd,
    author = "Polyakov, Alexander M.",
    editor = "Khalatnikov, I. M. and Mineev, V. P.",
    title = "{Quantum Geometry of Bosonic Strings}",
    reportNumber = "Print-81-0351 (LANDAU INST)",
    doi = "10.1016/0370-2693(81)90743-7",
    journal = "Phys. Lett. B",
    volume = "103",
    pages = "207--210",
    year = "1981"
}

@article{David:1988hj,
    author = "David, F.",
    title = "{Conformal Field Theories Coupled to 2D Gravity in the Conformal Gauge}",
    reportNumber = "SACLAY-SPHT-88-132",
    doi = "10.1142/S0217732388001975",
    journal = "Mod. Phys. Lett. A",
    volume = "3",
    pages = "1651",
    year = "1988"
}

@article{Distler:1988jt,
    author = "Distler, Jacques and Kawai, Hikaru",
    title = "{Conformal Field Theory and 2D Quantum Gravity}",
    reportNumber = "CLNS-88-853",
    doi = "10.1016/0550-3213(89)90354-4",
    journal = "Nucl. Phys. B",
    volume = "321",
    pages = "509--527",
    year = "1989"
}

@inproceedings{Ginsparg:1993is,
    author = "Ginsparg, Paul H. and Moore, Gregory W.",
    title = "{Lectures on 2-D gravity and 2-D string theory}",
    booktitle = "{Theoretical Advanced Study Institute (TASI 92): From Black Holes and Strings to Particles}",
    eprint = "hep-th/9304011",
    archivePrefix = "arXiv",
    reportNumber = "YCTP-P23-92, LA-UR-92-3479",
    pages = "277--469",
    year = "1993"
}

@article{Knizhnik:1988ak,
    author = "Knizhnik, V. G. and Polyakov, Alexander M. and Zamolodchikov, A. B.",
    editor = "Khalatnikov, I. M. and Mineev, V. P.",
    title = "{Fractal Structure of 2D Quantum Gravity}",
    reportNumber = "PRINT-88-0812 (LANDAU)",
    doi = "10.1142/S0217732388000982",
    journal = "Mod. Phys. Lett. A",
    volume = "3",
    pages = "819",
    year = "1988"
}

@article{Brezin:1990rb,
    author = "Brezin, E. and Kazakov, V. A.",
    title = "{Exactly Solvable Field Theories of Closed Strings}",
    doi = "10.1016/0370-2693(90)90818-Q",
    journal = "Phys. Lett. B",
    volume = "236",
    pages = "144--150",
    year = "1990"
}

@article{Douglas:1989ve,
    author = "Douglas, Michael R. and Shenker, Stephen H.",
    editor = "Brezin, E. and Wadia, S. R.",
    title = "{Strings in Less Than One-Dimension}",
    reportNumber = "RU-89-34",
    doi = "10.1016/0550-3213(90)90522-F",
    journal = "Nucl. Phys. B",
    volume = "335",
    pages = "635",
    year = "1990"
}

@article{Gross:1989vs,
    author = "Gross, David J. and Migdal, Alexander A.",
    editor = "Brezin, E. and Wadia, S. R.",
    title = "{Nonperturbative Two-Dimensional Quantum Gravity}",
    reportNumber = "PUPT-1148",
    doi = "10.1103/PhysRevLett.64.127",
    journal = "Phys. Rev. Lett.",
    volume = "64",
    pages = "127",
    year = "1990"
}

@article{DiFrancesco:1993cyw,
    author = "Di Francesco, P. and Ginsparg, Paul H. and Zinn-Justin, Jean",
    title = "{2-D Gravity and random matrices}",
    eprint = "hep-th/9306153",
    archivePrefix = "arXiv",
    reportNumber = "LA-UR-93-1722, SACLAY-SPH-T-93-061",
    doi = "10.1016/0370-1573(94)00084-G",
    journal = "Phys. Rept.",
    volume = "254",
    pages = "1--133",
    year = "1995"
}

@article{Polyakov:1981re,
    author = "Polyakov, Alexander M.",
    editor = "Khalatnikov, I. M. and Mineev, V. P.",
    title = "{Quantum Geometry of Fermionic Strings}",
    reportNumber = "Print-81-0350 (LANDAU INST)",
    doi = "10.1016/0370-2693(81)90744-9",
    journal = "Phys. Lett. B",
    volume = "103",
    pages = "211--213",
    year = "1981"
}

@article{Murthy:2003es,
    author = "Murthy, Sameer",
    title = "{Notes on noncritical superstrings in various dimensions}",
    eprint = "hep-th/0305197",
    archivePrefix = "arXiv",
    reportNumber = "PUPT-2086",
    doi = "10.1088/1126-6708/2003/11/056",
    journal = "JHEP",
    volume = "11",
    pages = "056",
    year = "2003"
}

@article{Kutasov:1990ua,
    author = "Kutasov, David and Seiberg, Nathan",
    title = "{Noncritical superstrings}",
    reportNumber = "RU-90-41, PUPT-1193",
    doi = "10.1016/0370-2693(90)90233-V",
    journal = "Phys. Lett. B",
    volume = "251",
    pages = "67--72",
    year = "1990"
}

@article{Muhlmann:2025ngz,
    author = {M{\"u}hlmann, Beatrix and Narovlansky, Vladimir and Tsiares, Ioannis},
    title = "{On the three-point functions in timelike $ \mathcal{N} $ = 1 Liouville CFT}",
    eprint = "2505.08890",
    archivePrefix = "arXiv",
    primaryClass = "hep-th",
    doi = "10.1007/JHEP02(2026)236",
    journal = "JHEP",
    volume = "02",
    pages = "236",
    year = "2026"
}

@article{Poghossian:1996agj,
    author = "Poghossian, R. H.",
    title = "{Structure constants in the N=1 superLiouville field theory}",
    eprint = "hep-th/9607120",
    archivePrefix = "arXiv",
    doi = "10.1016/S0550-3213(97)00218-6",
    journal = "Nucl. Phys. B",
    volume = "496",
    pages = "451--464",
    year = "1997"
}

@article{Rashkov:1996np,
    author = "Rashkov, R. C. and Stanishkov, M.",
    title = "{Three point correlation functions in N=1 superLiouville theory}",
    eprint = "hep-th/9602148",
    archivePrefix = "arXiv",
    doi = "10.1016/0370-2693(96)00474-1",
    journal = "Phys. Lett. B",
    volume = "380",
    pages = "49--58",
    year = "1996"
}

@article{Distler:1989nt,
    author = "Distler, Jacques and Hlousek, Zvonimir and Kawai, Hikaru",
    title = "{Superliouville Theory as a Two-Dimensional, Superconformal Supergravity Theory}",
    reportNumber = "CLNS-88/871",
    doi = "10.1142/S0217751X90000180",
    journal = "Int. J. Mod. Phys. A",
    volume = "5",
    pages = "391",
    year = "1990"
}

@article{Muhlmann:2026pdc,
    author = {M{\"u}hlmann, Beatrix and Zikopoulos, Themistocles},
    title = "{Toward the Structure Constants of $\mathcal{N}=2$ Liouville Theory}",
    eprint = "2607.25662",
    archivePrefix = "arXiv",
    primaryClass = "hep-th",
    month = "7",
    year = "2026"
}

@article{Jaskolski:2015uya,
    author = "Jaskolski, Zbigniew and Suchanek, Paulina",
    title = "{Non-rational su(2) cosets and Liouville field theory}",
    eprint = "1510.01773",
    archivePrefix = "arXiv",
    primaryClass = "hep-th",
    month = "10",
    year = "2015"
}

@article{Zamolodchikov:2005fy,
    author = "Zamolodchikov, Alexei B.",
    title = "{Three-point function in the minimal Liouville gravity}",
    eprint = "hep-th/0505063",
    archivePrefix = "arXiv",
    reportNumber = "LPM-04-10",
    doi = "10.1007/s11232-005-0003-3",
    journal = "Theor. Math. Phys.",
    volume = "142",
    pages = "183--196",
    year = "2005"
}

@article{Ribault:2015sxa,
    author = "Ribault, Sylvain and Santachiara, Raoul",
    title = "{Liouville theory with a central charge less than one}",
    eprint = "1503.02067",
    archivePrefix = "arXiv",
    primaryClass = "hep-th",
    doi = "10.1007/JHEP08(2015)109",
    journal = "JHEP",
    volume = "08",
    pages = "109",
    year = "2015"
}

@article{Kostov:2005kk,
    author = "Kostov, I. K. and Petkova, V. B.",
    title = "{Bulk correlation functions in 2-D quantum gravity}",
    eprint = "hep-th/0505078",
    archivePrefix = "arXiv",
    reportNumber = "SPHT-T05-068",
    doi = "10.1007/s11232-006-0011-y",
    journal = "Theor. Math. Phys.",
    volume = "146",
    pages = "108--118",
    year = "2006"
}

@article{Collier:2023cyw,
    author = {Collier, Scott and Eberhardt, Lorenz and M{\"u}hlmann, Beatrix and Rodriguez, Victor A.},
    title = "{The Virasoro minimal string}",
    eprint = "2309.10846",
    archivePrefix = "arXiv",
    primaryClass = "hep-th",
    doi = "10.21468/SciPostPhys.16.2.057",
    journal = "SciPost Phys.",
    volume = "16",
    number = "2",
    pages = "057",
    year = "2024",
    note = "[Erratum: SciPost Phys. 16, 1 (2024)]"
}

@article{Rangamani:2025wfa,
    author = "Rangamani, Mukund and Zheng, Jianming",
    title = "{Towards the super Virasoro minimal string}",
    eprint = "2505.08892",
    archivePrefix = "arXiv",
    primaryClass = "hep-th",
    doi = "10.1007/JHEP09(2025)061",
    journal = "JHEP",
    volume = "09",
    pages = "061",
    year = "2025"
}

@article{Collier:2026bhh,
    author = {Collier, Scott and M{\"u}hlmann, Beatrix and Rangamani, Mukund and Tsiares, Ioannis and Zheng, Jianming},
    title = "{Super Liouville theory on the torus: bootstrap and the super Virasoro minimal string}",
    eprint = "2610.01909",
    archivePrefix = "arXiv",
    primaryClass = "hep-th",
    month = "10",
    year = "2026"
}

@article{Stanford:2019vob,
    author = "Stanford, Douglas and Witten, Edward",
    title = "{JT gravity and the ensembles of random matrix theory}",
    eprint = "1907.03363",
    archivePrefix = "arXiv",
    primaryClass = "hep-th",
    doi = "10.4310/ATMP.2020.v24.n6.a4",
    journal = "Adv. Theor. Math. Phys.",
    volume = "24",
    number = "6",
    pages = "1475--1680",
    year = "2020"
}

@article{Turiaci:2023jfa,
    author = "Turiaci, Gustavo J. and Witten, Edward",
    title = "{$ \mathcal{N} $ = 2 JT supergravity and matrix models}",
    eprint = "2305.19438",
    archivePrefix = "arXiv",
    primaryClass = "hep-th",
    doi = "10.1007/JHEP12(2023)003",
    journal = "JHEP",
    volume = "12",
    pages = "003",
    year = "2023"
}

@article{Collier:2023fwi,
    author = "Collier, Scott and Eberhardt, Lorenz and Zhang, Mengyang",
    title = "{Solving 3d gravity with Virasoro TQFT}",
    eprint = "2304.13650",
    archivePrefix = "arXiv",
    primaryClass = "hep-th",
    doi = "10.21468/SciPostPhys.15.4.151",
    journal = "SciPost Phys.",
    volume = "15",
    number = "4",
    pages = "151",
    year = "2023"
}

@article{Collier:2025lux,
    author = {Collier, Scott and Eberhardt, Lorenz and M{\"u}hlmann, Beatrix},
    title = "{A microscopic realization of dS$_3$}",
    eprint = "2501.01486",
    archivePrefix = "arXiv",
    primaryClass = "hep-th",
    doi = "10.21468/SciPostPhys.18.4.131",
    journal = "SciPost Phys.",
    volume = "18",
    number = "4",
    pages = "131",
    year = "2025"
}

@article{Eberhardt:2026hfh,
    author = "Eberhardt, Lorenz",
    title = "{The super Virasoro minimal string from 3d supergravity}",
    eprint = "2604.26038",
    archivePrefix = "arXiv",
    primaryClass = "hep-th",
    doi = "10.21468/SciPostPhys.21.3.062",
    journal = "SciPost Phys.",
    volume = "21",
    number = "3",
    pages = "062",
    year = "2026"
}

@article{Giveon:1999px,
    author = "Giveon, Amit and Kutasov, David",
    title = "{Little string theory in a double scaling limit}",
    eprint = "hep-th/9909110",
    archivePrefix = "arXiv",
    reportNumber = "NSF-ITP-99-104, RI-8-99, EFI-99-41",
    doi = "10.1088/1126-6708/1999/10/034",
    journal = "JHEP",
    volume = "10",
    pages = "034",
    year = "1999"
}

@article{Giveon:1999tq,
    author = "Giveon, Amit and Kutasov, David",
    title = "{Comments on double scaled little string theory}",
    eprint = "hep-th/9911039",
    archivePrefix = "arXiv",
    reportNumber = "NSF-ITP-99-123, RI-10-99, EFI-99-45",
    doi = "10.1088/1126-6708/2000/01/023",
    journal = "JHEP",
    volume = "01",
    pages = "023",
    year = "2000"
}

@article{Ooguri:1995wj,
    author = "Ooguri, Hirosi and Vafa, Cumrun",
    title = "{Two-dimensional black hole and singularities of CY manifolds}",
    eprint = "hep-th/9511164",
    archivePrefix = "arXiv",
    reportNumber = "HUTP-95-A045, LBL-37996, UCB-PTH-95-41",
    doi = "10.1016/0550-3213(96)00008-9",
    journal = "Nucl. Phys. B",
    volume = "463",
    pages = "55--72",
    year = "1996"
}

@article{Giveon:1999zm,
    author = "Giveon, Amit and Kutasov, David and Pelc, Oskar",
    title = "{Holography for noncritical superstrings}",
    eprint = "hep-th/9907178",
    archivePrefix = "arXiv",
    reportNumber = "RI-4-99, EFI-99-31",
    doi = "10.1088/1126-6708/1999/10/035",
    journal = "JHEP",
    volume = "10",
    pages = "035",
    year = "1999"
}

@article{Kutasov:2000jp,
    author = "Kutasov, D. and Sahakyan, D. A.",
    title = "{Comments on the thermodynamics of little string theory}",
    eprint = "hep-th/0012258",
    archivePrefix = "arXiv",
    reportNumber = "EFI-2000-54",
    doi = "10.1088/1126-6708/2001/02/021",
    journal = "JHEP",
    volume = "02",
    pages = "021",
    year = "2001"
}

@article{Chen:2021dsw,
    author = "Chen, Yiming and Maldacena, Juan and Witten, Edward",
    title = "{On the black hole/string transition}",
    eprint = "2109.08563",
    archivePrefix = "arXiv",
    primaryClass = "hep-th",
    doi = "10.1007/JHEP01(2023)103",
    journal = "JHEP",
    volume = "01",
    pages = "103",
    year = "2023"
}

@article{David:2014aha,
    author = "David, Fran{\c{c}}ois and Kupiainen, Antti and Rhodes, R{\'e}mi and Vargas, Vincent",
    title = "{Liouville Quantum Gravity on the Riemann sphere}",
    eprint = "1410.7318",
    archivePrefix = "arXiv",
    primaryClass = "math.PR",
    doi = "10.1007/s00220-016-2572-4",
    journal = "Commun. Math. Phys.",
    volume = "342",
    number = "3",
    pages = "869--907",
    year = "2016"
}

@article{Israel:2004ir,
    author = "Israel, Dan and Kounnas, Costas and Pakman, Ari and Troost, Jan",
    title = "{The Partition function of the supersymmetric two-dimensional black hole and little string theory}",
    eprint = "hep-th/0403237",
    archivePrefix = "arXiv",
    reportNumber = "LPTENS-04-16, RI-02-04",
    doi = "10.1088/1126-6708/2004/06/033",
    journal = "JHEP",
    volume = "06",
    pages = "033",
    year = "2004"
}

\end{document}